\documentclass[prmaterials,aps,preprint,showkeys,longbibliography,floatfix]{revtex4-2}

\usepackage{epsfig,amsmath,amssymb,txfonts,hyperref,physics}

\makeatletter 
\newcommand*{\ifdraft}[2]{\preprintsty@sw{#1}{#2}}
\makeatother

\newcommand*{\VECIT}[1]{\mathbf{#1}}
\newcommand*{\VECITS}[1]{\boldsymbol{#1}}  

\newcommand*{\DUchem}[2]{\Delta U^{\text{#2}+\text{#1}-}}  
\newcommand*{\DUAB}[0]{\DUchem{A}{B}}  
\newcommand*{\DUBA}[0]{\DUchem{B}{A}}  
\newcommand*{\conc}[1]{x_\text{#1}} 
\newcommand*{\xFe}[0]{\conc{Fe}}
\newcommand*{\xCu}[0]{\conc{Cu}}
\newcommand*{\xNi}[0]{\conc{Ni}}
\newcommand*{\xv}[0]{\VECIT{x}}
\newcommand*{\dxv}[0]{\VECITS{\Delta}\xv}
\newcommand*{\hpv}[0]{\hat{\VECIT{p}}}
\newcommand*{\htv}[0]{\hat{\VECIT{t}}}
\newcommand*{\nv}[0]{\VECIT{n}}

\newcommand*{\muv}[0]{\VECITS{\mu}}
\newcommand*{\hh}[0]{\underline{h}}
\newcommand*{\nablav}[0]{\VECITS{\nabla}}
\newcommand*{\dmuv}[0]{\VECITS{\Delta}\muv}

\newcommand*{\kv}[0]{\VECIT{k}}
\newcommand*{\rv}[0]{\VECIT{r}}
\newcommand*{\rhv}[0]{\VECIT{\hat{r}}}

\newcommand*{\kB}[0]{k_{\text{B}}}

\newlength{\wholefigwidth}
\newlength{\smallfigwidth}
\newlength{\halfsmallfigwidth}
\newlength{\PDfigwidth}
\newlength{\figwidth}
\newcommand{\Fig}[1]{Fig.~\ref{fig:#1}}

\newcommand{\App}[1]{Appendix~\ref{sec:#1}}
\newcommand{\Eqn}[1]{Eqn.~\ref{eqn:#1}}

\begin{document}

\title{Computing ternary liquid phase diagrams: Fe-Cu-Ni}

\author{Dallas R. Trinkle}
\email{dtrinkle@illinois.edu}
\affiliation{Department of Materials Science and Engineering, University of Illinois, Urbana-Champaign, Urbana, Illinois 61801, USA}

\date{\today}
\begin{abstract}
  We compute the phase separation of the immiscible liquid alloy Fe-Cu-Ni. Our computational approach uses a virtual semigrand canonical Widom approach to determine differences in excess chemical potentials between different species. Using an embedded atom potential for Fe-Cu-Ni, we simulate liquid states over a range of compositions and temperatures. This raw data is then fit to Redlich-Kister polynomials for the Gibbs free energy with a simple temperature dependence. Using the analytic form, we can determine the phase diagram for the ternary liquid, compute the miscibility gap and spinodal decomposition as a function of temperature for this EAM potential. In addition, we compute density as a function of composition and temperature, and predict pair correlation functions. We use static structure factors to estimate the second derivative of the Gibbs free energy (the $S^0$ method) and compare with our fit Gibbs free energy. Finally, using a nonequilibrium Hamiltonian integration method, we separately compute absolute Gibbs free energies for the pure liquid states; this shows that our endpoints are accurate to within 1 meV for our ternary Gibbs free energy, as well as the absolute Gibbs free energy for the ternary liquid.
\end{abstract}

\maketitle

\section{Introduction}
Liquid immiscible alloys are a difficult challenge for alloy design, as two chemical elements phase separate while in the liquid before cooling into the solid state\cite{Ratke1995}. Useful microstructures may be possible if cooling is sufficiently rapid; in the past, this has been difficult to achieve at the macroscale. The recent development of additive manufacturing that relies on rapid melting via electron-beam or laser followed by rapid cooling has renewed interest in immiscible alloys, as new microstructures with desirable properties may now be feasible in immiscible systems such as Fe-Cu\cite{Zafari2021a,Zafari2021b,Wei2025}. The determination of phase boundaries for liquid immiscible alloys is difficult and time-consuming\cite{Ratke1995}; not surprisingly, much more phase diagram data is available for liquid alloys that do \textit{not} phase separate on cooling. Complicating matters further, for the case of Fe-Cu, the immiscibility gap is in the undercooled region of the phase diagram\cite{Nakagawa1958, Elder1989, Chen1995}. Being able to predict the liquid phase diagram along with driving forces for phase separation---including the effect of additional elements on phase stability---is an important step to for alloy design in additive manufacturing with immiscible alloys.

Computing liquid phase diagrams can either be accomplished through coexistence simulations or via determination of the Gibbs free energy across composition and temperature space. A wide variety of coexistence simulation methods exist\cite{Norman1969, Frenkel2002, Panagiotopoulos1987, Mehta1994, Hitchcock1999, Kofke1988, Ghazisaeidi2021}; while powerful, these techniques allow for the determination of only a single pair of coexisting phases at one temperature. In a ternary, this amounts to determining a single tie-line; to survey a ternary composition space would require many different compositions for each temperature. Moreover, only the points on the binodal can be found; these methods do not provide information about driving forces away from equilibrium, which can be used in mesoscale solidification simulations. Chemical potential information is difficult to acquire in computational simulations; in non-solid systems, one atomistic computational method is the test-particle approach due to Widom, where an atom is virtually inserted at random locations, and the change in energy computed\cite{Frenkel2002}. This converges slowly in condensed systems, as the insertion energy is often large, giving rise to poor statistics. Methods based on searching for free cavities in the liquid have been proposed to improve the sampling efficiency\cite{Hong2012}, though the method requires significant computational effort per sampling point. Additional weighting methods\cite{Qin2016,Perego2018} have also tried to improve the efficiency of test particle insertion. Free energy methods have been used to get absolute free energies via thermodynamic switching for solids\cite{Freitas2016}, and later for liquids\cite{PaulaLeite2019}. These methods are relatively time-consuming to repeat for a range of compositions, but can produce an absolute Gibbs free energy. Recently, a virtual semigrand canonical Widom approach---based on the difference method from \cite{Sindzingre1987}---was demonstrated to compute accurate chemical potentials for a solid system\cite{Peters2020}. Related to the test particle method, virtual \textit{exchanges} of chemical species are attempted, and the energy change computed. As an atom is replaced instead of inserted, the energy changes are not enormous, and so convergence is relatively straightforward. The computational effort is roughly double that of a molecular-dynamics or Monte Carlo simulation of equilibrium. From this data, chemical potential differences can be computed, even in parts of the phase space that are metastable; this information is sufficient to determine a phase diagram. Finally, an even newer method, the $S^0$-method can extract chemical potential derivatives from static structure factors alone\cite{Cheng2022}. The power of this approach is that it only relies on post-processing of trajectory data; the weakness is that it alone is insufficient to determine the Gibbs free energy, as it gives only second derivative information rather than first derivative information such as chemical potentials.

In this paper, we compute the Gibbs free energy of the ternary liquid Fe-Cu-Ni using an EAM potential by Bonny et al.\cite{Bonny2009}. We choose to compute the Gibbs free energy to provide possible input for future solidification simulations away from equilibrium, as well as assessing the phase diagram. The calculations are done using the virtual semigrand canonical Widom approach to determine chemical potential differences between the three species on a grid of compositions and over a range of temperatures, from 1200K up to 3000K. Molecular dynamics simulations, run with LAMMPS\cite{LAMMPS}, serve to provide samples for thermodynamic averages. The resulting chemical potentials are then used to fit a Gibbs free energy function of composition and temperature in a CALPHAD form. From this, we determine the liquid phase diagram, including binodal and spinodal boundaries as well as the critical points. The molecular dynamics runs also provide information on pair correlation functions, static structure factors, and liquid densities. We compare our results with available experimental data\cite{Chen1995}. In addition, we assess the applicability of the $S^0$ method to accurately capture free energy information, and benchmark our computation against a more computationally intensive nonequilibrium Hamiltonian integration method.

\section{Methodology}
Sampling requires equilibrating a variety of concentrations at multiple temperatures using the EAM potential with a combination of Monte Carlo and molecular dynamics runs. We choose a 2048 atom cell from an initial $8\times8\times8$ FCC Cu cell, and all molecular dynamics runs are integrated with 1 fs timesteps. The molecular dynamics runs use thermostats and barostats to maintain desired temperatures and an equilibrium pressure, as implemented in LAMMPS\cite{Shinoda2004,Martyna1994,Parrinello1981,Tuckerman2006}, using 0.1 ps temperature damping, 1 ps pressure damping, 2 thermostat chains for temperature and 3 for pressure. To find all of our starting configurations, we bootstrap from a single 3000K pure Cu molecular dynamics run at the experimentally measured liquid density at 3000K. This was chosen as the metal with the lowest melting temperature, it can reach a liquid configuration the fastest starting from an initial FCC configuration. Molecular dynamics equilibration for 100 ps produced a liquid initial state. From there, different substitutions were added. First along the binary lines Cu-Fe and Cu-Ni in atomic concentration steps of 1/16, then along Fe-Ni, and finally ternary concentrations in a 1/8 concentration grid. Each was equilibrated using a hybrid MD and Monte Carlo approach where after 100 time steps (0.1 ps) of molecular dynamics, 100 binary swaps of Fe-Cu, Cu-Ni, and Ni-Fe were attempted. This continued for 30 ps, and the state stored as a starting point for thermodynamic sampling and trajectory analysis. The temperature was then lowered by 300K, and equilibration continued. In this way, a grid of 3 unary, 45 binary, and 21 ternary concentrations were equilibrated at 3000K, 2700K, 2400K, 2100K, 1800K, 1500K, and 1200K.

With these initial states, the raw data for free energy modeling comes from molecular dynamics using a virtual semigrand canonical Widom approach. While the virtual approach is compatible with a hybrid MD / MC approach, we chose to perform only molecular dynamics so that we could also have a long trajectory for both pair correlation analysis and structure factor computation for the $S^0$ method using the same data. For these runs, the same thermostat and barostats were used. Every 2048 steps, the virtual semigrand canonical Widom approach determined the energy change for each atom in the cell to change its chemistry into the other chemistries: Fe to Cu or Ni, Cu to Fe or Ni, and Ni to Fe or Cu. This is accomplished via a modification to LAMMPS to compute the virtual swaps. As there are 2048 atoms in the cells, this roughly doubles the run time of the simulation. We ran trajectories for $2^{17}$ fs = 131 ps, which gives 64 separate evaluations of the energy changes. At the same time, we wrote the atomic positions every 128 fs, giving 1024 snapshots for structural analysis.

Chemical potential differences are calculated directly from averages computed with the virtual semigrand canonical Widom method. In the traditional Widom method, the chemical potential is calculated from the energy change from virtually inserting a test particle anywhere in the sample; for condensed systems, this method is slow to converge. A related method, laid out in the same paper\cite{Sindzingre1987} and recently demonstrated for solid systems\cite{Peters2020}, can be instead used to compute the \textit{difference} in chemical potential between two species. Instead of evaluating the energy to insert a particle, we compute the energy to \textit{change the identity} of an atom of type A into one of type B: $\DUAB$ (removing an A atom, inserting a B atom at the same location). We can write the chemical potential difference as the difference in the ideal gas, plus the difference in excess chemical potential
\begin{equation}
  \Delta\mu_{\text{B}-\text{A}} = \mu_\text{B}^\text{ideal}-\mu_\text{A}^\text{ideal}
  +  \mu_\text{B}^\text{excess}-\mu_\text{A}^\text{excess}.
  \label{eqn:dmu}
\end{equation}
The ideal gas difference is straightforward in terms of the masses $m_\text{A}$ and $m_\text{B}$ and number of atoms $N_\text{A}$ and $N_\text{B}$,
\begin{equation}
  \mu_\text{B}^\text{ideal}-\mu_\text{A}^\text{ideal} =
  -\kB T\left[\frac32 \ln\left(\frac{m_\text{B}}{m_\text{A}}\right)
    + \ln\left(\frac{N_\text{A}}{N_\text{B}}\right)\right].
  \label{eqn:dmu-ideal}
\end{equation}
The excess chemical potential difference in \Eqn{dmu} can be written either in terms of $\DUAB$ or $\DUBA$ as
\begin{equation}
  \begin{split}
    \mu_\text{B}^\text{excess}-\mu_\text{A}^\text{excess}
    &= -\kB T \ln\left\langle\exp\left( -\beta \DUAB \right) \right\rangle_{N_\text{A}N_\text{B}} \\
    &= +\kB T \ln\left\langle\exp\left( -\beta \DUBA \right) \right\rangle_{N_\text{A}N_\text{B}}
  \end{split}
  \label{eqn:dmu-excess}
\end{equation}
where $\beta = (\kB T)^{-1}$ and the thermodynamic averages are taken at constant particle numbers $N_\text{A}$ and $N_\text{B}$. It is worth noting that the expressions in \Eqn{dmu-excess} come from a finite difference approximation to the derivative of the free energy with respect to concentration; hence, the leading order error is $O(N^{-1})$. In practical calculations, we can combine both expressions in \Eqn{dmu-excess}, weighted by the concentrations, to evaluate
\begin{widetext}
  \begin{equation}
    \mu_\text{B}^\text{excess}-\mu_\text{A}^\text{excess}
    = -\kB T\frac{N_\text{A}}{N_\text{A}+N_\text{B}} \ln\left\langle\exp\left( -\beta \DUAB \right) \right\rangle
    +\kB T \frac{N_\text{B}}{N_\text{A}+N_\text{B}}\ln\left\langle\exp\left( -\beta \DUBA \right) \right\rangle.
    \label{eqn:dmu-excess-comp}
  \end{equation}
\end{widetext}
Thus, at regular intervals, we compute the energy from changing each single A atom into a B atom, and each B atom into an A atom, and return the averages of $\exp(-\beta\DUAB)$ and $\exp(-\beta\DUBA)$. For our ternary, there are six averages computed. After multiple samples taken in this way, we can then use \Eqn{dmu-excess-comp} to find the difference in excess chemical potential; combined with \Eqn{dmu-ideal}, we have our computed differences in chemical potential.

The differences in chemical potentials can be used to parameterize an expression for the Gibbs free energy as a function of composition using Redlich-Kister polynomials\cite{Lupis1983}. The Gibbs free energy per atom for our ternary liquid is written in terms of the concentrations $\xFe$, $\xCu$, $\xNi$ (which sum to 1) as
\begin{equation}
  G = \kB T\sum_i x_i\ln x_i + \sum_i x_i \mu_i^* + G^\text{excess}
  \label{eqn:gibbs-basic}
\end{equation}
with the free energy of mixing, $\mu_i^*$ the pure phase chemical potential, and $G^\text{excess}$ the excess Gibbs free energy. Note that the ideal gas chemical potential, \Eqn{dmu-ideal}, contains both the contribution to the free energy of mixing ($\kB T\ln x_\text{A}/x_\text{B}$) \textit{and} a contribution to $\mu_i^*$ from the mass ratios, and that the excess chemical potential \Eqn{dmu-excess} will also contain contributions to $\mu_i^*$. The excess Gibbs free energy are deviations from the ideal solution model; our ternary liquid is a subregular solution model, where we parameterize
\begin{equation}
  \begin{split}
    G^\text{excess}
    &= \xFe\xCu\left(A_{210}\xFe + A_{120}\xCu + A_{220}\xFe\xCu\right) \\
    &+ \xCu\xNi\left(A_{021}\xCu + A_{012}\xNi + A_{022}\xCu\xNi\right) \\
    &+ \xNi\xFe\left(A_{102}\xNi + A_{201}\xFe + A_{202}\xCu\xNi\right) \\
    &+ \xFe\xCu\xNi\left(A_{211}\xFe + A_{121}\xCu + A_{112}\xNi\right).
  \end{split}
  \label{eqn:gibbs-excess}
\end{equation}
The first three lines describe the binaries, with a subregular solution model. The final line is a ternary correction. \Eqn{gibbs-excess} has 12 parameters, plus three parameters for $\mu_i^*$ (which, following our schema, would be $A_{100}$, $A_{010}$, and $A_{001}$). To find the parameters, we relate chemical potential differences to derivatives of the Gibbs free energy, that is
\begin{equation}
  \Delta\mu_{\text{B}-\text{A}} = \frac{\partial G}{\partial\conc{B}}
  - \frac{\partial G}{\partial\conc{A}}
  \label{eqn:diffG}
\end{equation}
and thus, all of our parameters can be determined from a linear least-squares fit \textit{except} the sum $\mu_\text{Fe}^* + \mu_\text{Cu}^* + \mu_\text{Ni}^*$, which is the absolute Gibbs free energy. This requires a separate calculation---described below---but does not affect the determination of the phase diagram. It should be noted that the expansion in \Eqn{gibbs-excess} is not unique; one can use fewer or more terms in the Redlich-Kister expansion. The form chosen here is determined by fitting to the numerical data for our Fe-Cu-Ni EAM system. Moreover, all of the parameters are found at each temperature, and then the temperature dependence of the parameters can also be fit, to construct a general model of the Gibbs free energy with temperature.

\section{Results}
\subsection{Chemical potentials}
\begin{figure*}[tbh]
  \includegraphics[width=0.32\wholefigwidth]{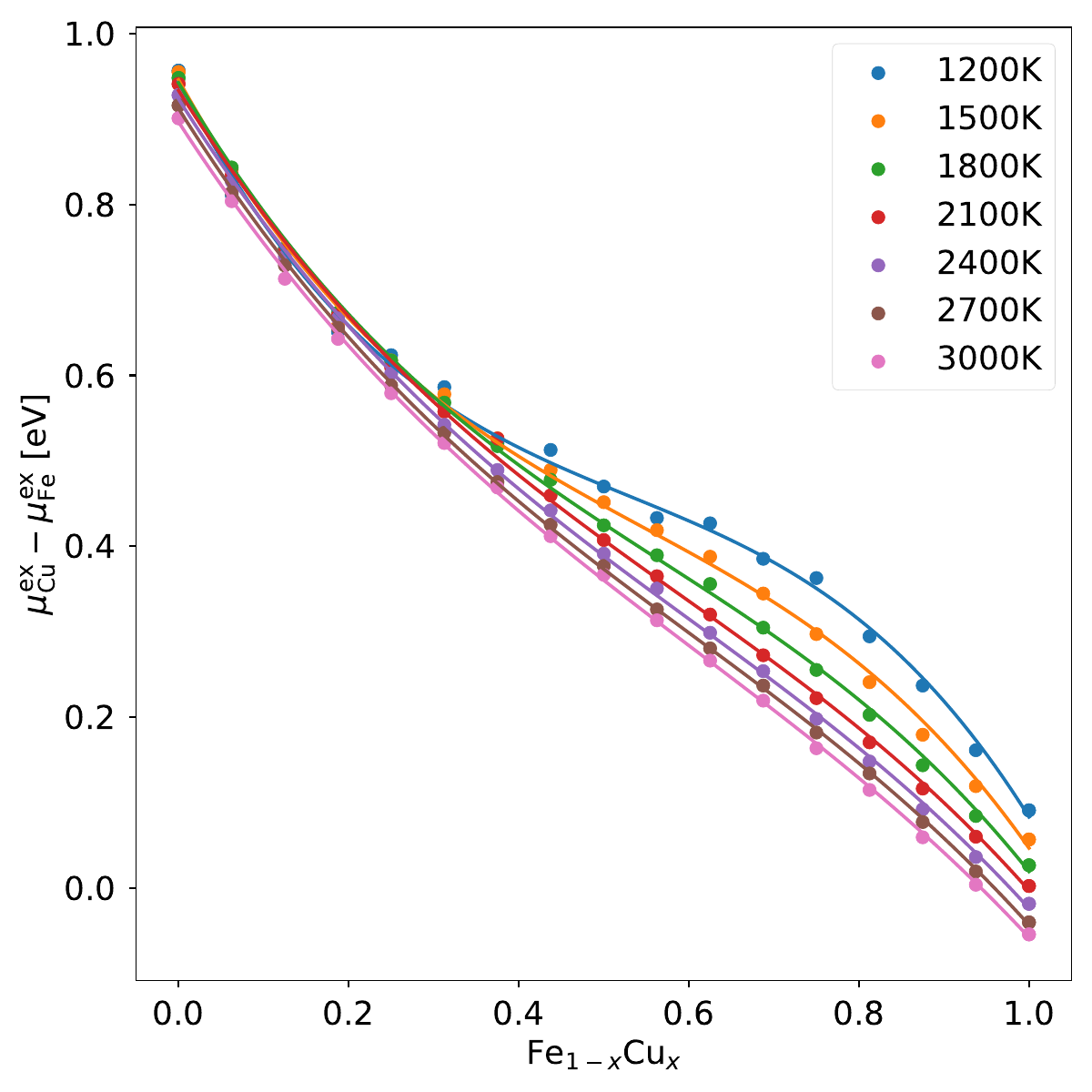}
  \includegraphics[width=0.32\wholefigwidth]{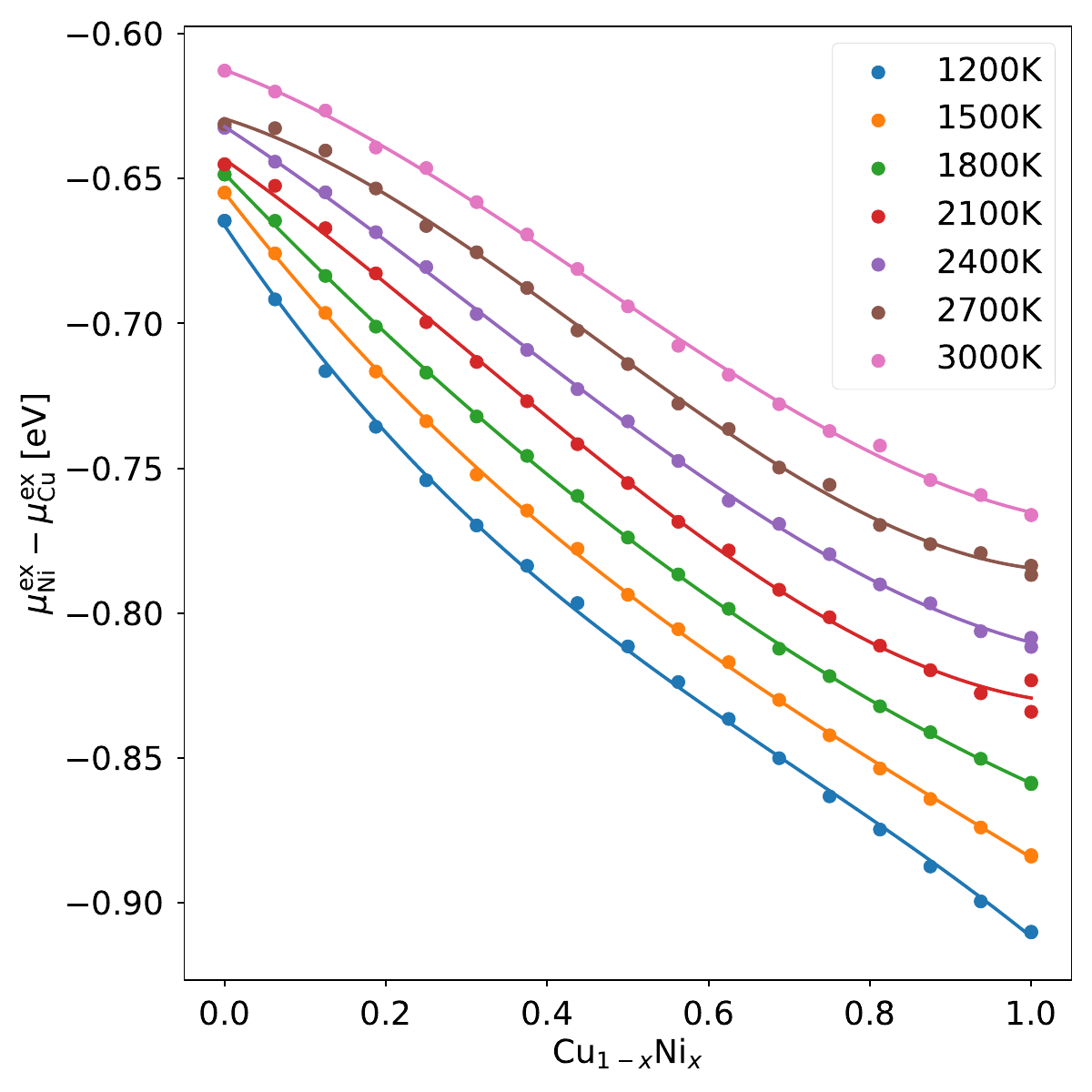}
  \includegraphics[width=0.32\wholefigwidth]{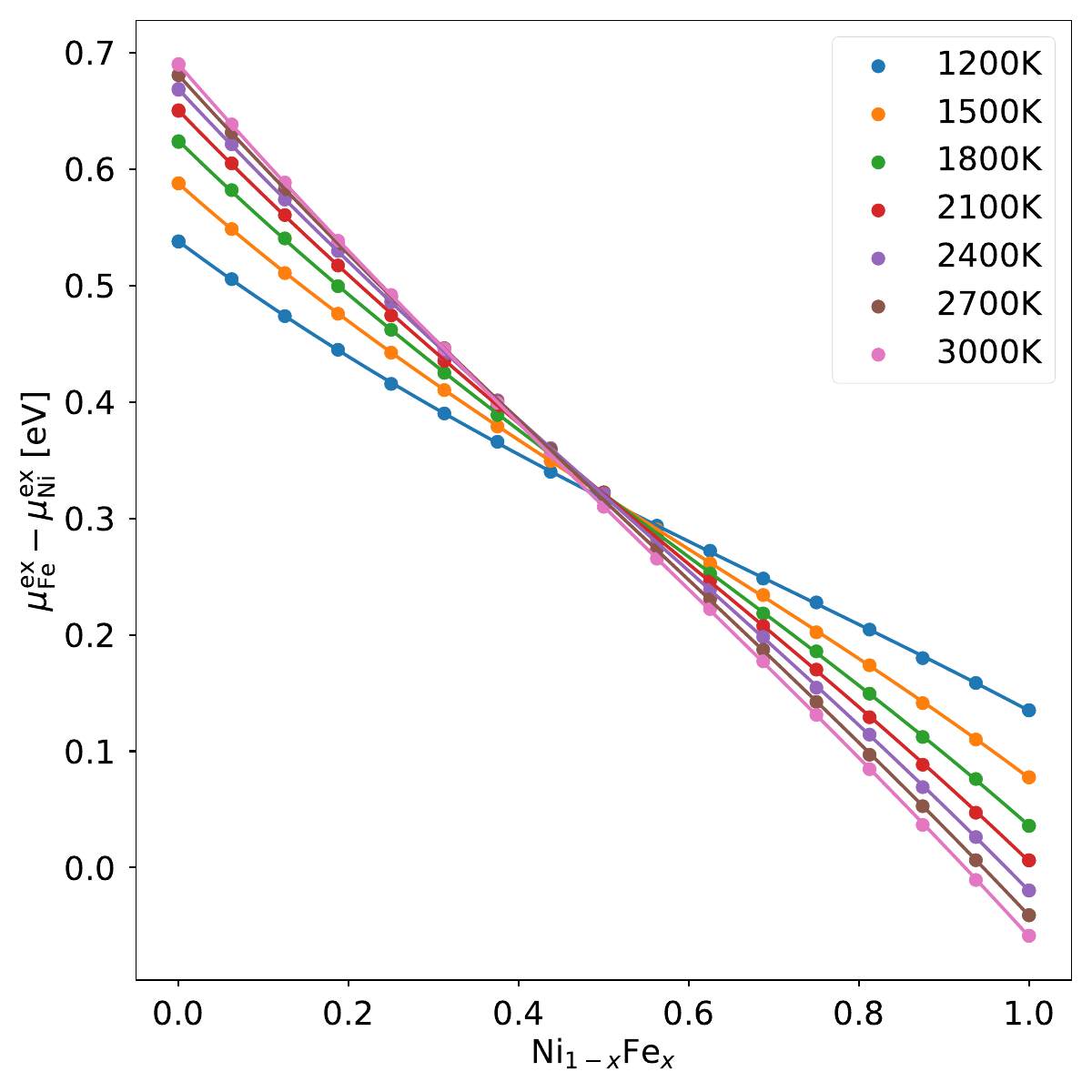}
  \caption{Difference in excess chemical potentials for binary Fe-Cu, Cu-Ni, and Ni-Fe directly from difference method and Redlich-Kister polynomial fits. The excess chemical potential lacks the ideal mixing contribution ($\kB T\ln x_\text{B}/x_\text{A}$), and can be extracted from changes in energy of virtual swaps in \Eqn{dmu-excess-comp}.}
  \label{fig:mu-diff}
\end{figure*}

\Fig{mu-diff} shows the difference in excess chemical potential for the three binary subsystems calculated directly from the LAMMPS semigrand canonical Widom approach, and the fit to Redlich-Kister polynomials. We analyze the statistical errors in the excess chemical potential from the standard error of the averages in \Eqn{dmu-excess-comp}, and from the sum of the three differences. If the average $f := \left\langle\exp(-\beta\DUAB)\right\rangle$ has standard error $\delta f$, the error estimate in chemical potential difference is $\kB T \delta f/f$, weighted similarly to \Eqn{dmu-excess-comp}. In addition, the sum of the three excess chemical potentials should be exactly 0. We find that the average standard error in excess chemical potential is below 3 meV, and that the mean absolute error in the sum is 5 meV. The largest standard error is 33 meV at 3000K for the Ni-Cu difference; the Fe-Cu largest difference is 24 meV at 2100K and 37.5 at.\%\ Cu, which is the closest sample to the critical point. Still, all of these errors are small fractions of the thermal energy $\kB T$. Only the unary and binary terms in \Eqn{gibbs-excess} contribute to the fit; each curve is determined by three binary parameters and one difference in unary parameters. The curvature requires the use of $x_\text{A}^2x_\text{B}^2$ terms, and the fits achieve chemical accuracy for all temperatures. The largest root mean-squared (RMS) fitting error is 12.5 meV at 1200K for the Fe-Cu system, which is 1/8 $\kB T$. The remaining RMS fitting errors are below 10 meV for Fe-Cu, below 3 meV for Cu-Ni, and below 1 meV for Ni-Fe. These are all similar in magnitude to the statistical errors in the excess chemical potential difference. The relatively larger error seen in the Fe-Cu system is largest below the miscibility gap; this suggests that as the liquid is phase separating, higher order corrections in \Eqn{gibbs-excess} are becoming larger, but still sufficiently small to be ignored in our fits.

As a benchmark test of our chemical potential values for the ternary compositions, we compare against semi-grand canonical Monte Carlo simulations\cite{Sadigh2012}. Taking the differences in excess chemical potentials extracted from our virtual semi-grand canonical Widom calculations at the ternary compositions, we performed hybrid molecular dynamics / semi-grand canonical MC simulations in LAMMPS, where every 2048 steps, 2048 trial moves are attempted to swap atomic identities. After $2^{17}$ fs, the chemical composition is averaged, and compared with the nominal composition. For all temperatures above 2100K, the disagreement is less than 5\%; starting at 2100K and below, there is disagreement in compositions in the two-phase regime, as the SGCMC method will favor one side of the two-phase separation.

\begin{figure}[tbh]
  \includegraphics[width=\figwidth]{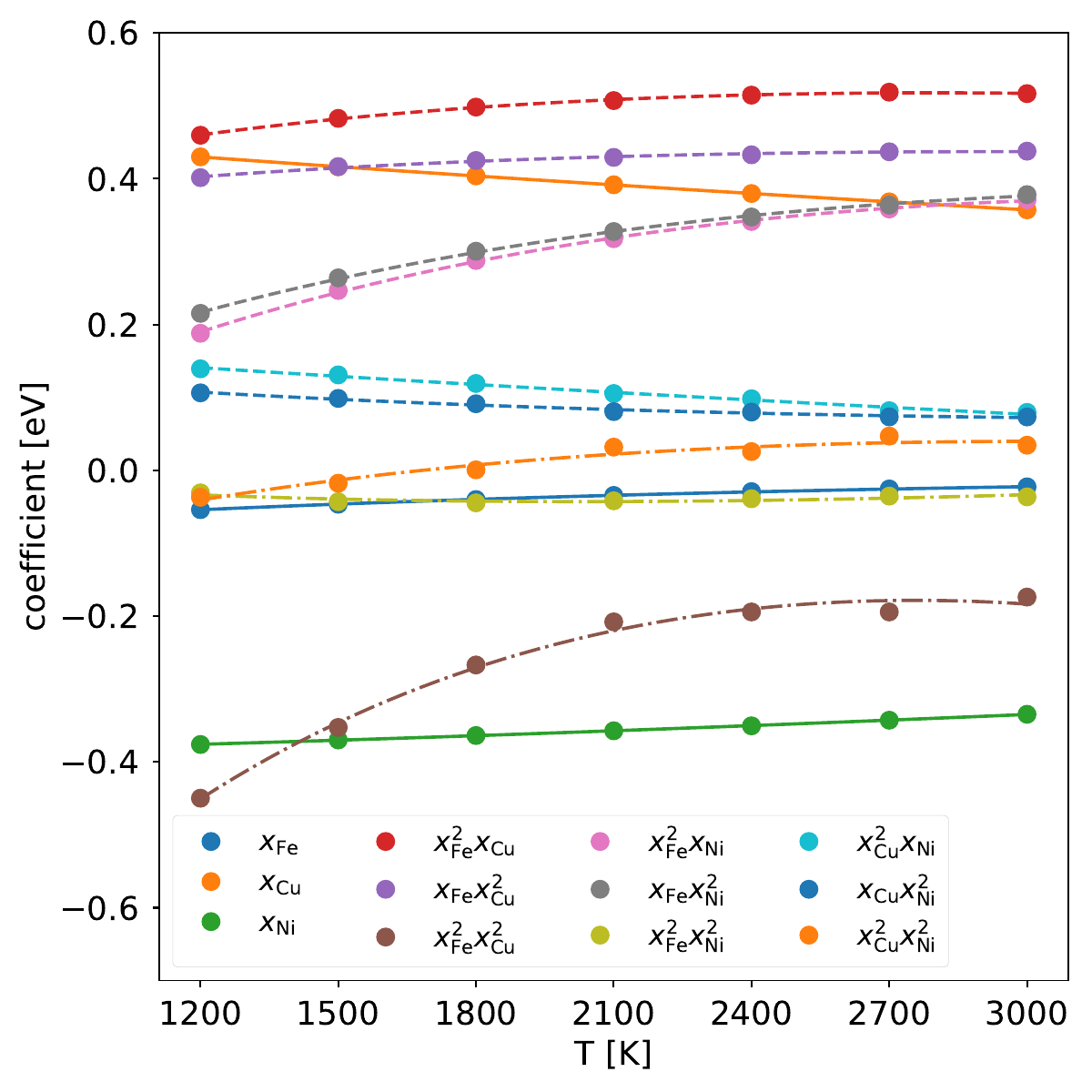}
  \caption{Binary coefficients for Redlich-Kister polynomial fits at different temperatures, along with fits of temperature dependence. Solid lines are unary coefficients (such as $A_{100} \xFe$), dashed lines are linear terms in the binary (such as $\xFe\xCu(A_{210}\xFe + A_{120}\xCu)$), and dashed-dotted lines are quadratic terms in the binary (such as $A_{220}\xFe^2\xCu^2$). Note that the unary coefficients are not absolute, as $A_{100}+A_{010}+A_{001}=0$; they represent the \textit{differences} in unary chemical potentials $\mu^*_i$. The absolute quantities have been computed in \Fig{gibbs-absolute}. The temperature dependence is described as $c_0 + c_1 T +c_2 T\ln T$.}
  \label{fig:binary-coeff}
\end{figure}

\Fig{binary-coeff} shows the temperature dependence of the unary and binary Redlich-Kister polynomials as well as a simple parameterized fit. The majority of the coefficients show weak temperature dependence; the primary outliers are the three coefficients in the Fe-Cu subsystem. This is to be expected, as Fe-Cu has a miscibility gap at approximately 2300K for the EAM potential. The Fe-Ni binary coefficients show some temperature dependence, which can be deduced also from \Fig{mu-diff}. To parameterize the temperature dependence, we note that the average energy of our pure liquid systems are closely linear with the temperature; if a linear temperature dependence $U(T) = U_0 + C_V T$ is integrated to get a difference in Gibbs free energy (as $\partial (\beta G)/\partial \beta = U$), then the Gibbs free energy would have the form $c_0 + c_1 T + c_2 T\ln T$. This functional form with temperature is also used in the CALPHAD fitting of the Fe-Cu system\cite{Chen1995}, among others. The RMS error from this form is below 10 meV for all coefficients, and below 3 meV for all but two; again, the errors are fractions of $\kB T$, and similar to the underlying small statistical errors.

\begin{figure}[tbh]
  \includegraphics[width=\figwidth]{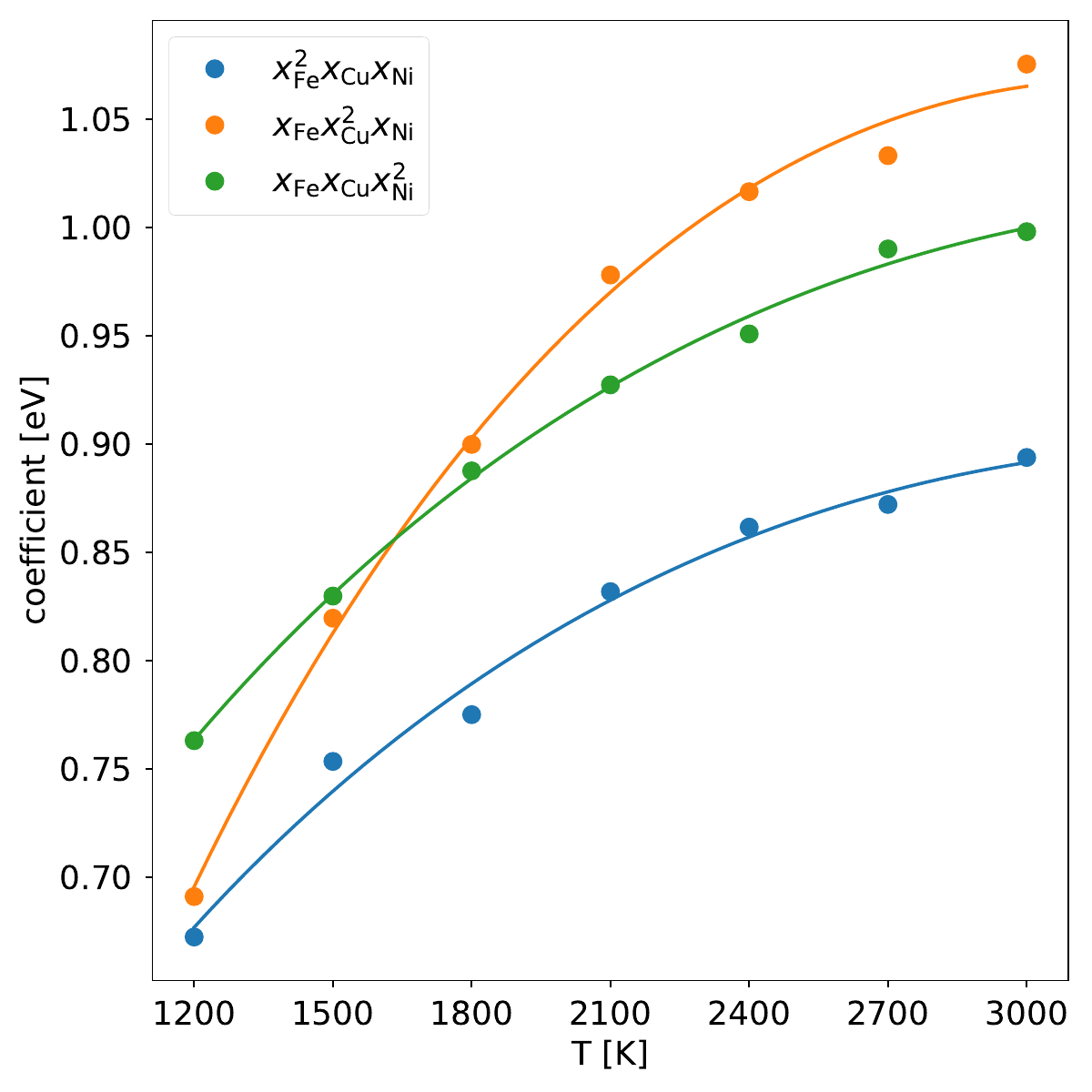}
  \caption{Ternary coefficients for Redlich-Kister polynomial fits at different temperatures, along with fits of temperature dependence. The temperature dependence is described as $c_0 + c_1 T +c_2 T\ln T$.}
  \label{fig:ternary-coeff}
\end{figure}

\Fig{ternary-coeff} shows the temperature dependence of the ternary coefficients in \Eqn{gibbs-excess}. A similar approach is used to fit the ternary coefficients as the binary coefficients; however, the binary contributions are first subtracted from the excess chemical potential differences. The ternary concentrations are used as well as the binary ``endpoints'' where the third species excess chemical potential difference has also been evaluated, but not used in the fitting. The statistical errors in the ternary field are no different than the binaries; if anything, their maximum errors are lower than for the binaries. The mean absolute error in the chemical potentials if the ternary terms are \textit{not} included range from 75 meV at 1200K up to 100 meV at 3000K, which is much larger than any error statistical or fitting error. After including the ternary terms, the fitting error in 13 meV at 1200K, dropping to 8 meV at 3000K. After fitting to the same temperature dependence as the binary coefficients, we have fitting root mean square errors of 8 meV. It might be possible to reduce this error with additional ternary terms, but our errors are all well below $\kB T$ for all temperatures and compositions.

\subsection{Phase diagrams}

\begin{figure}[tbh]
  \includegraphics[width=\figwidth]{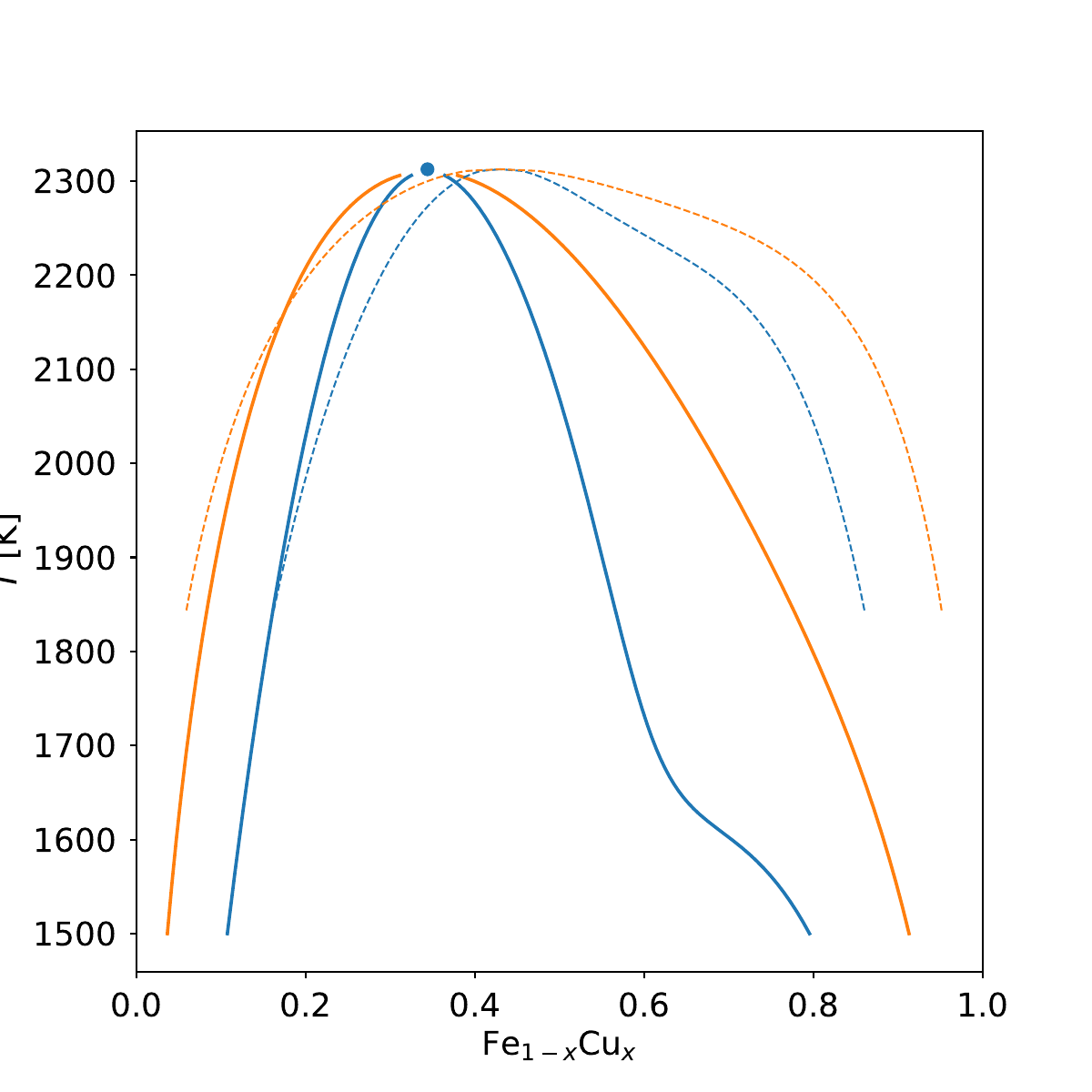}
  \caption{Liquid binodal (orange) and spinodal (blue) decomposition points for Fe-Cu, compared with experiment (dashed). The curves are computed directly from the analytic fits to free energy. The EAM potential critical temperature is 2286.7K and critical Cu concentration is 33.775 at.\%; the experimental curves come from CALPHAD\cite{Chen1995} where the temperature has been scaled to match the EAM critical temperature.}
  \label{fig:FeCu-phasediagram}
\end{figure}

With our parameterized, analytic form for the Gibbs free energy we can find the miscibility gap for our binary liquid in \Fig{FeCu-phasediagram}. The binodal decomposition lines identify where we should see phase separation into Fe-rich and Cu-rich liquid phases. This phase separation requires nucleation, while the spinodal decomposition range should show near spontaneous separation due to instabilities against local chemical fluctuations. The spinodal points are easily identified where the curvature $d^2G/dx^2$ goes to zero at a given temperature using numerical root-finding. The binodal points require a common tangent construction, which is more complex. We can find these points numerically by starting with the spinodal points, and moving to the more Fe- and Cu-rich endpoints; then, we use the values, slope, and curvature of $G$ at the two points to parameterize two parabolas, from which we can identify new guesses for the common tangent points. This is iterated until the compositions are determined within $10^{-4}$. The critical point where the binodal and spinodal meet is a temperature and composition where $d^2G/dx^2$ is zero \textit{and} is local minimum ($d^3G/dx^3 = 0$), which can be found via root-finding. The EAM potential puts the critical point at a higher temperature compared with experimental phase diagram assessments\cite{Chen1995}, where $T_\text{c}=1694 \text{K}$ and $x_\text{c} \approx 43 \text{at.\%\ Cu}$. The scaled CALPHAD binodal and spinodal experimental results show that the lower temperature shoulder in the Cu-rich side of the spinodal is captured by EAM. Despite there being large uncertainties in the experimental values for the metastable liquid miscibility gap, this indicates that the Fe-Cu repulsion in the EAM potential would need to be reduced to correctly predict the miscibility gap. The stronger asymmetry in the spinodal and binodal compared with experiments would also need correction.

\begin{figure*}[tbh]
  \includegraphics[width=\PDfigwidth]{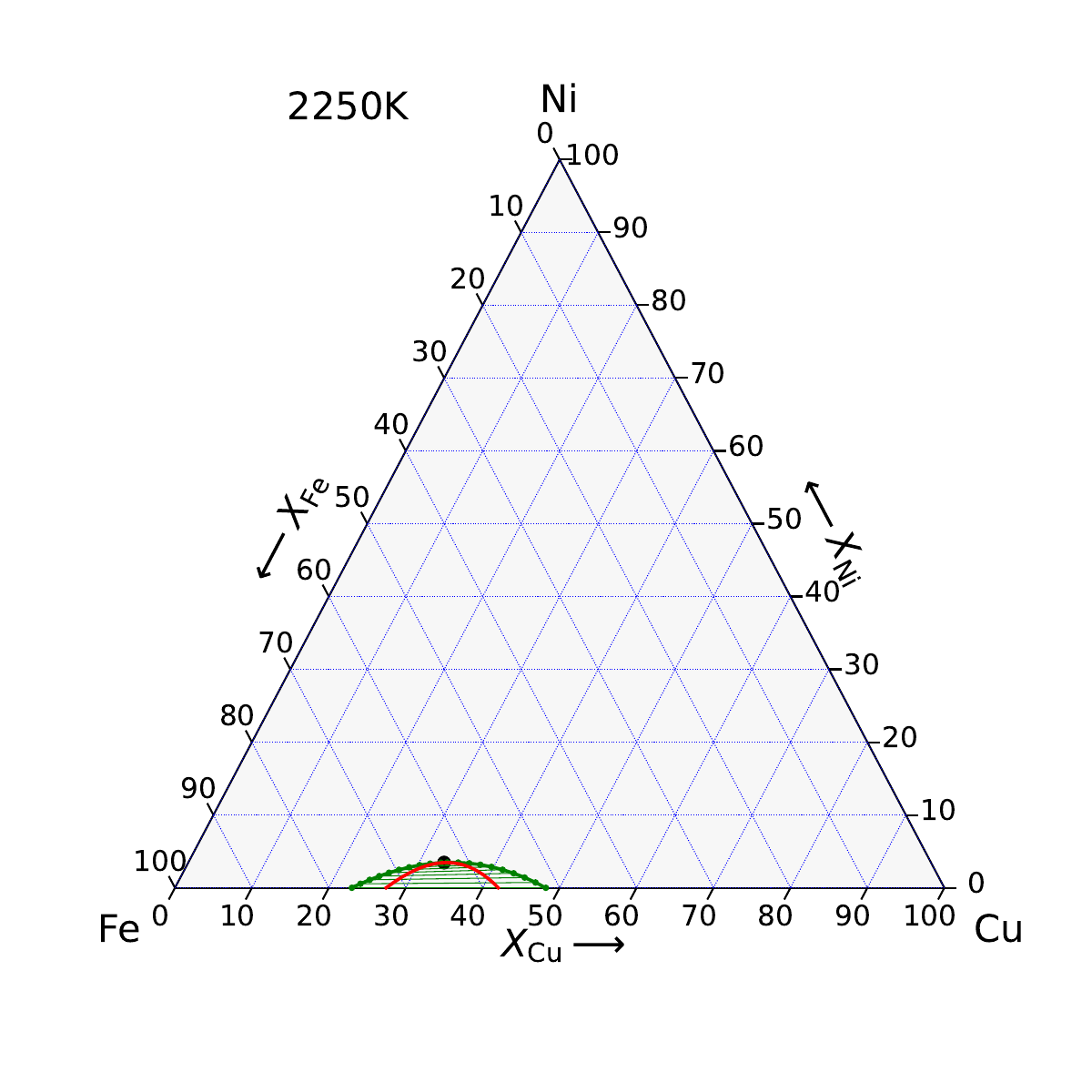}\!
  \includegraphics[width=\PDfigwidth]{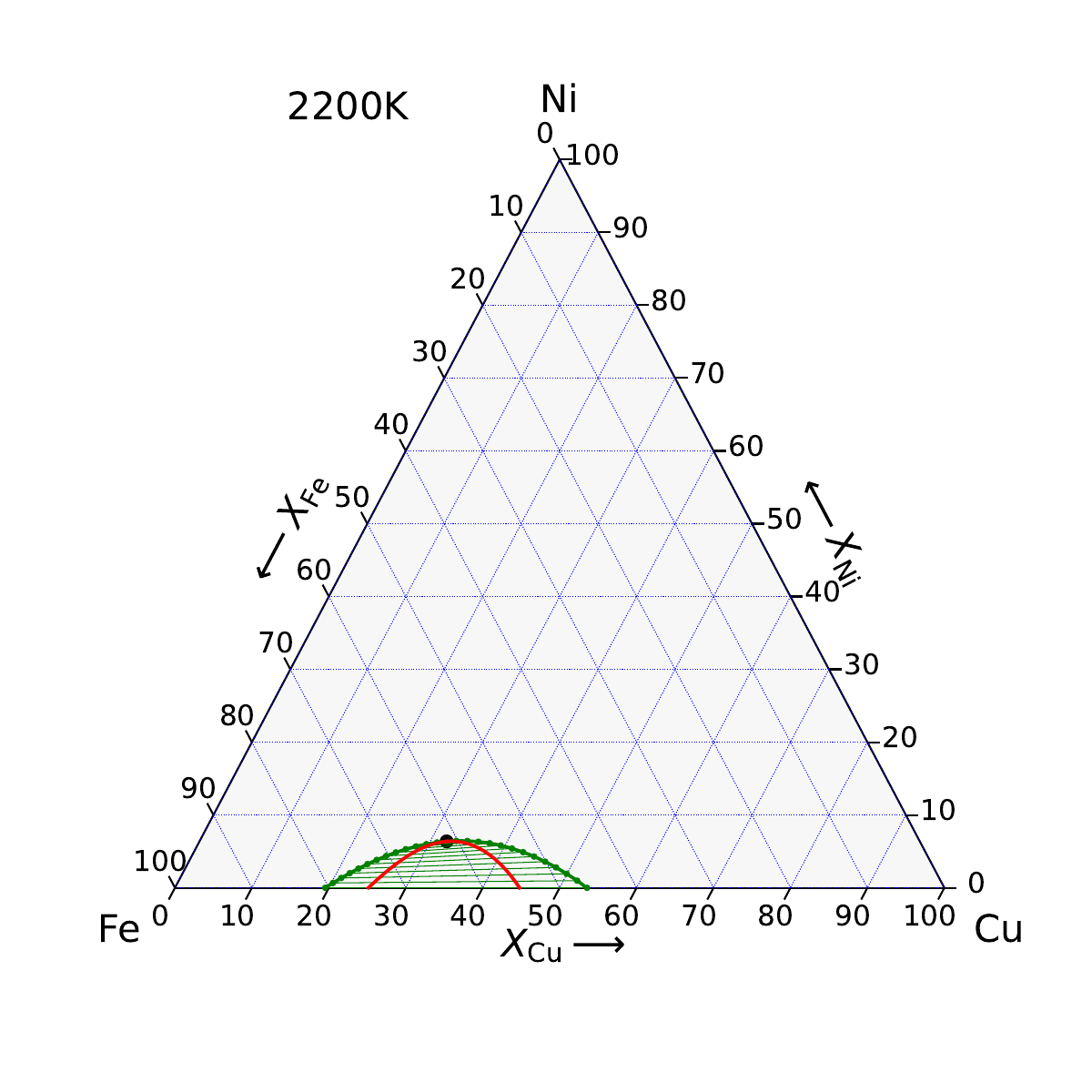}\!
  \includegraphics[width=\PDfigwidth]{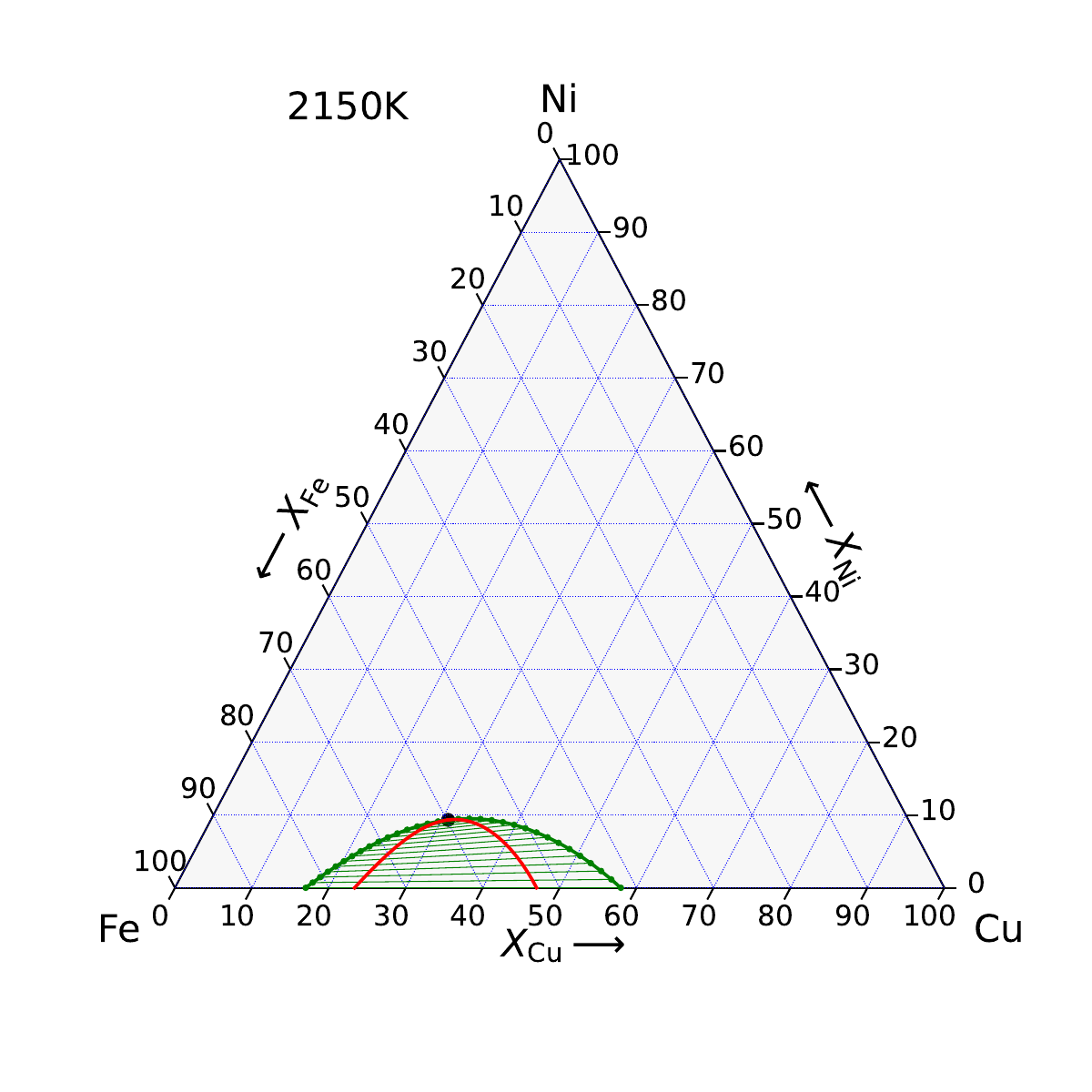}\!
  \includegraphics[width=\PDfigwidth]{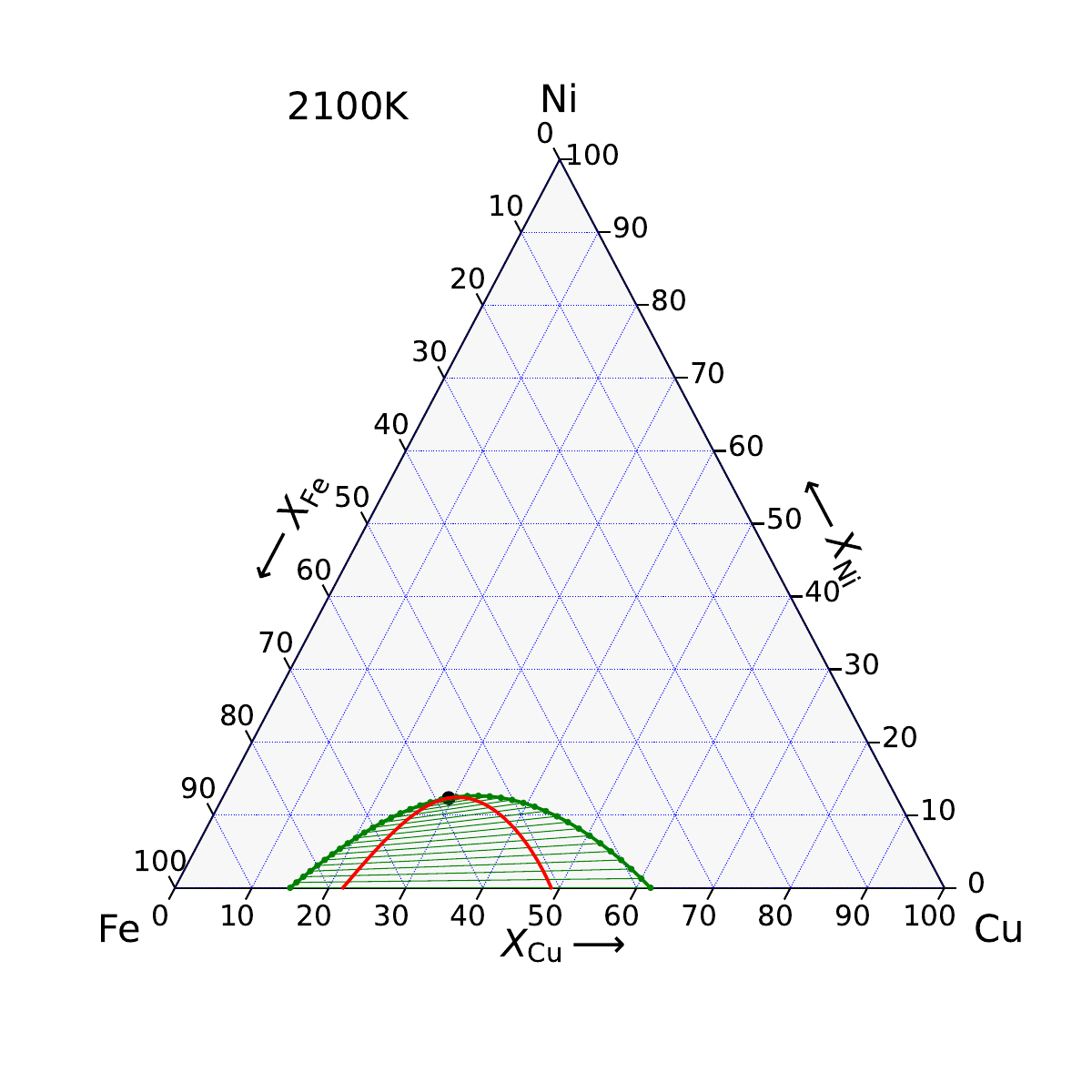}\\
  \includegraphics[width=\PDfigwidth]{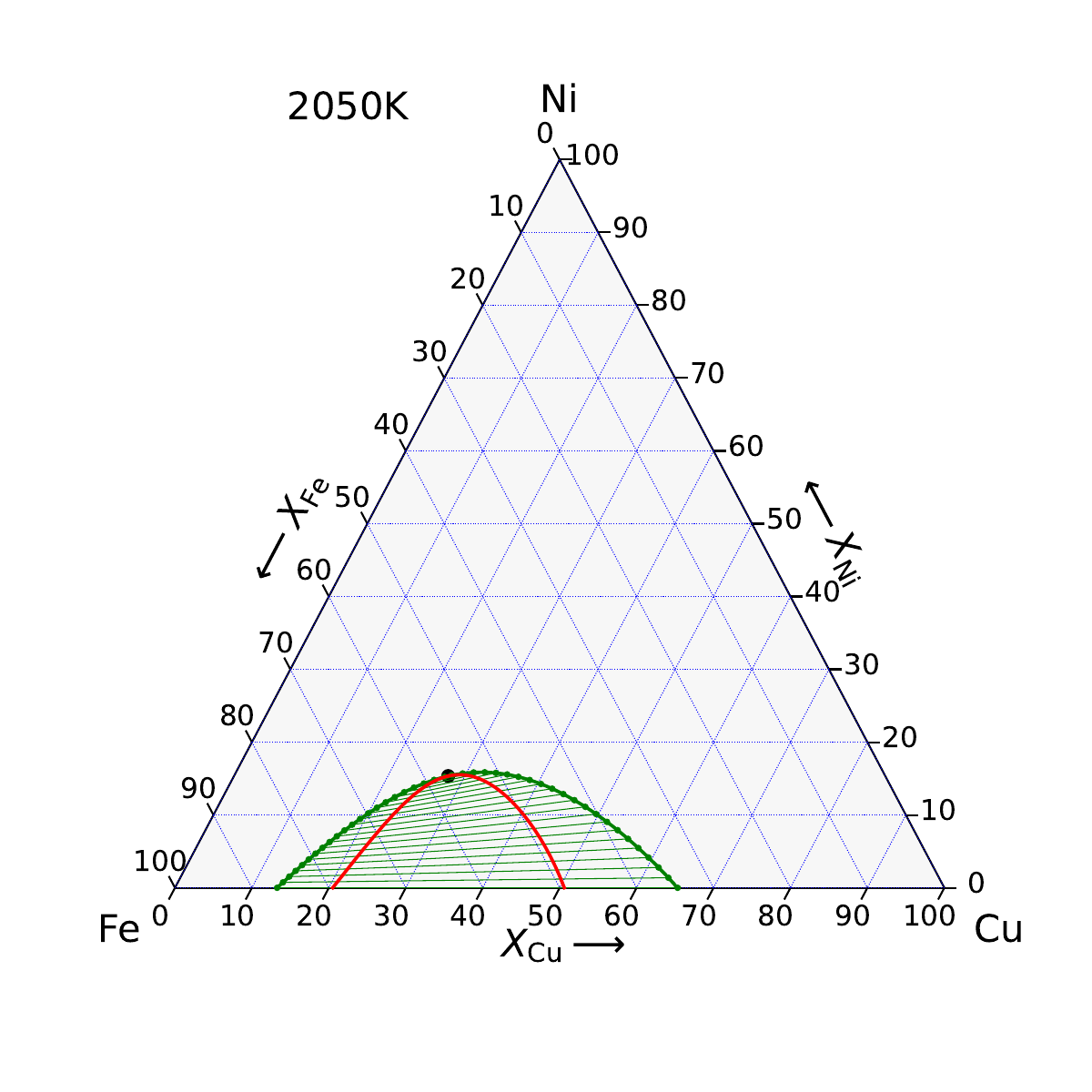}\!
  \includegraphics[width=\PDfigwidth]{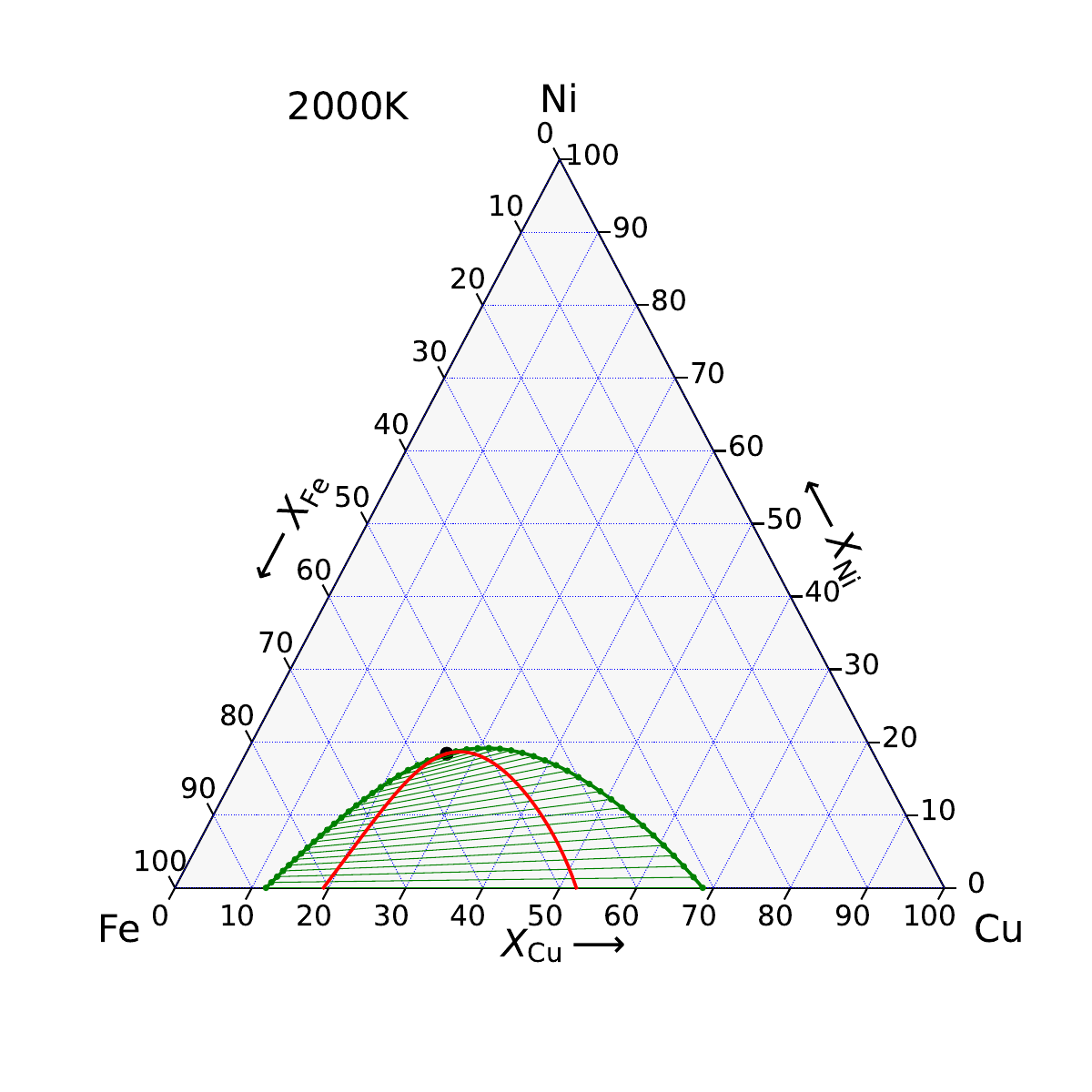}\!
  \includegraphics[width=\PDfigwidth]{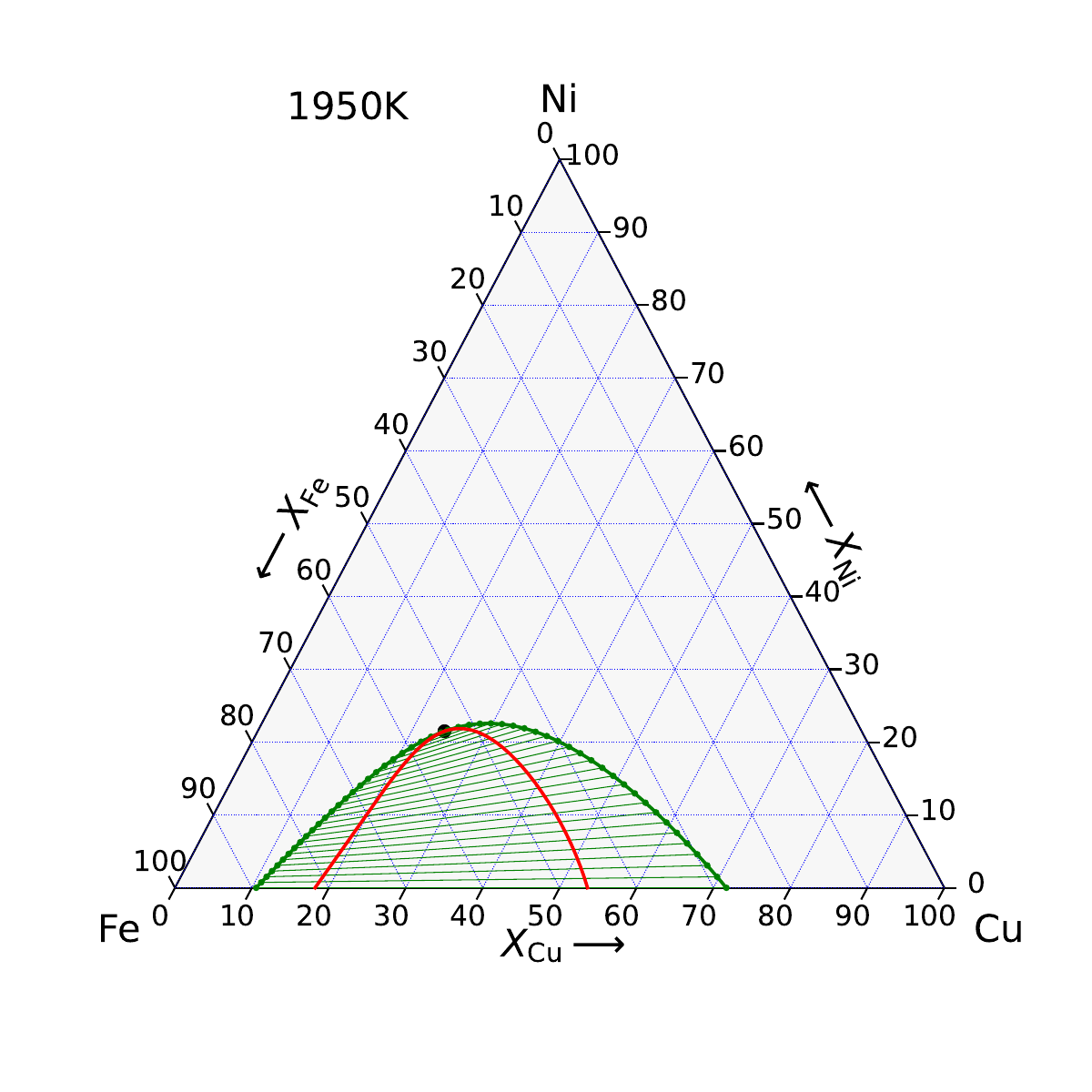}\!
  \includegraphics[width=\PDfigwidth]{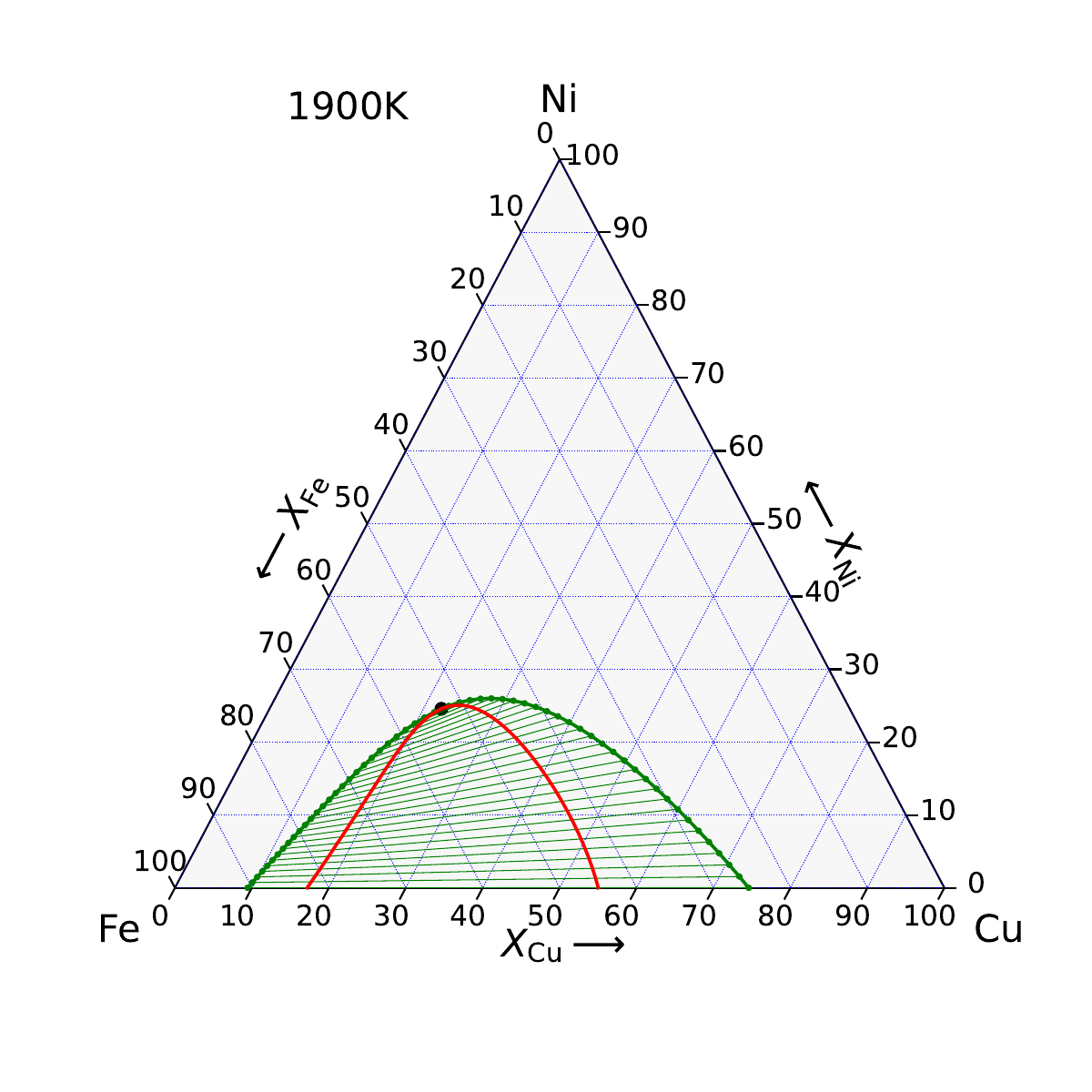}\\
  \includegraphics[width=\PDfigwidth]{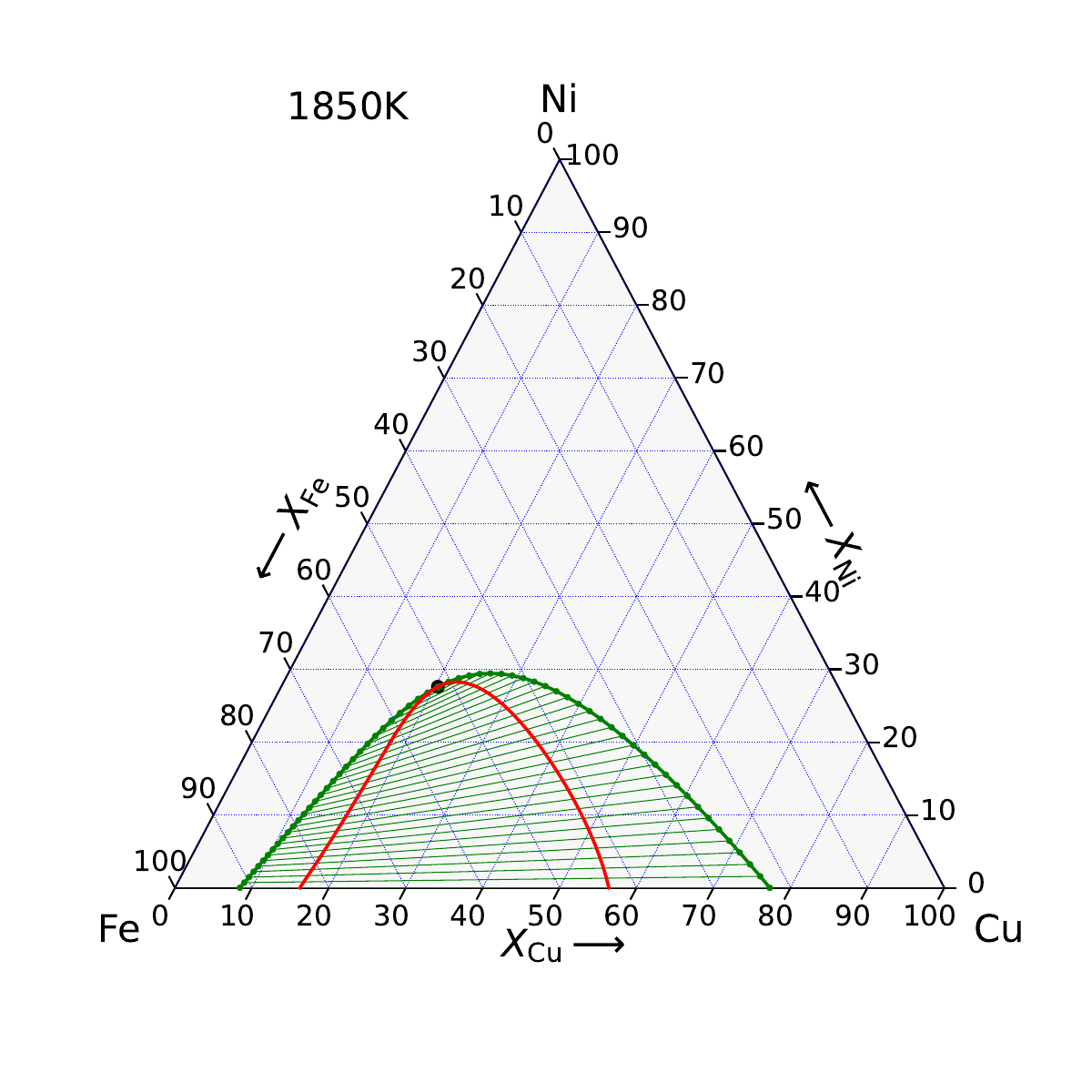}\!
  \includegraphics[width=\PDfigwidth]{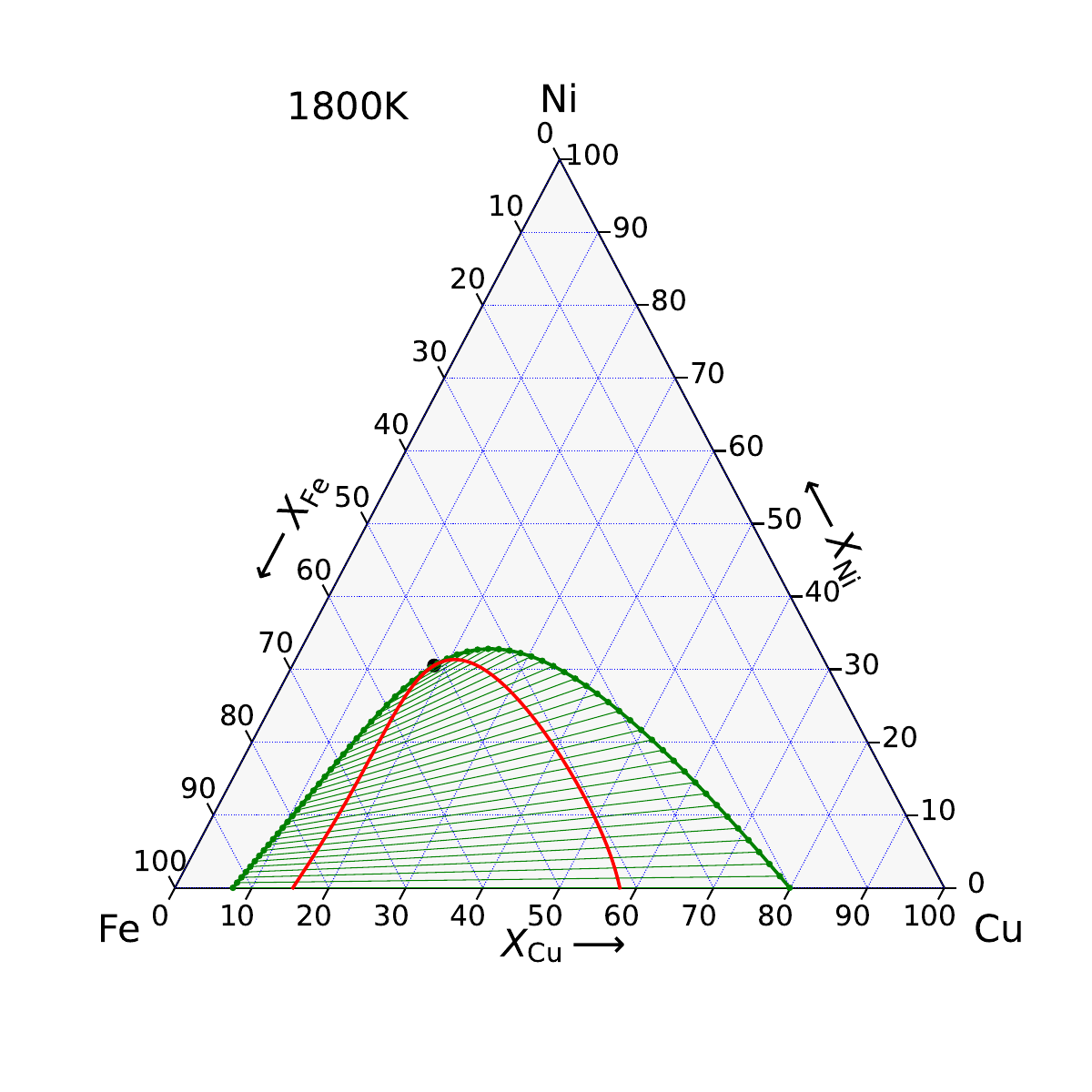}\!
  \includegraphics[width=\PDfigwidth]{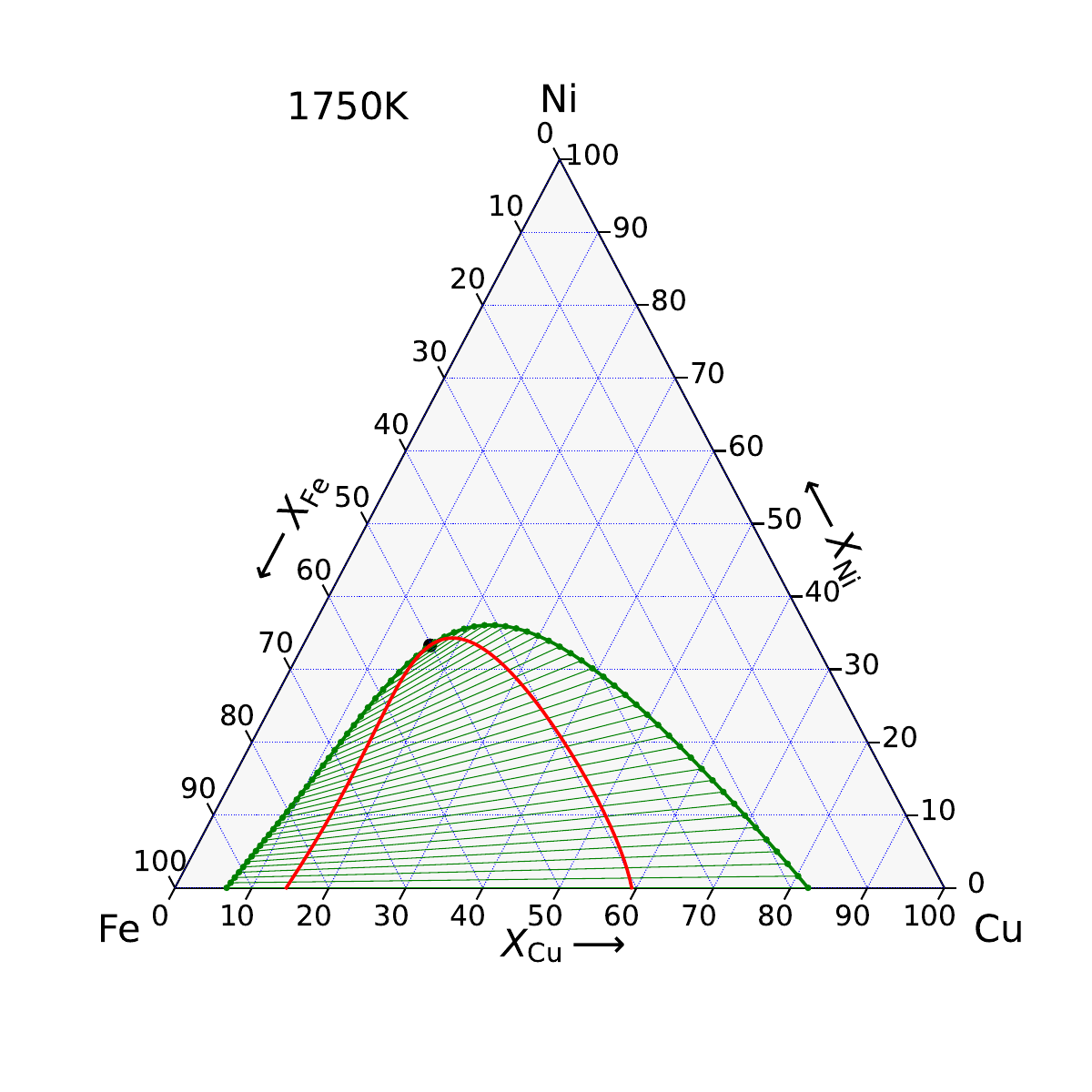}\!
  \includegraphics[width=\PDfigwidth]{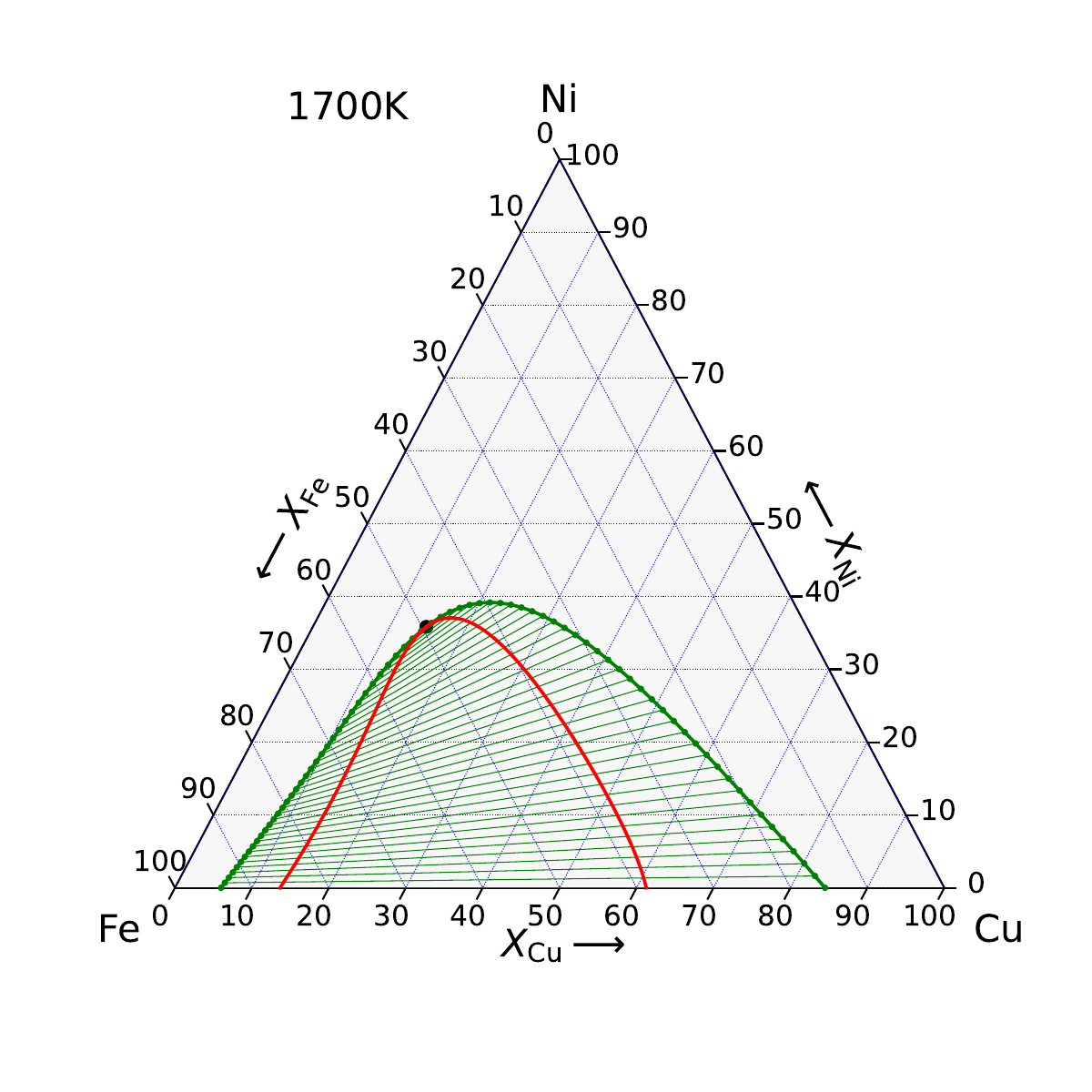}\\
  \includegraphics[width=\PDfigwidth]{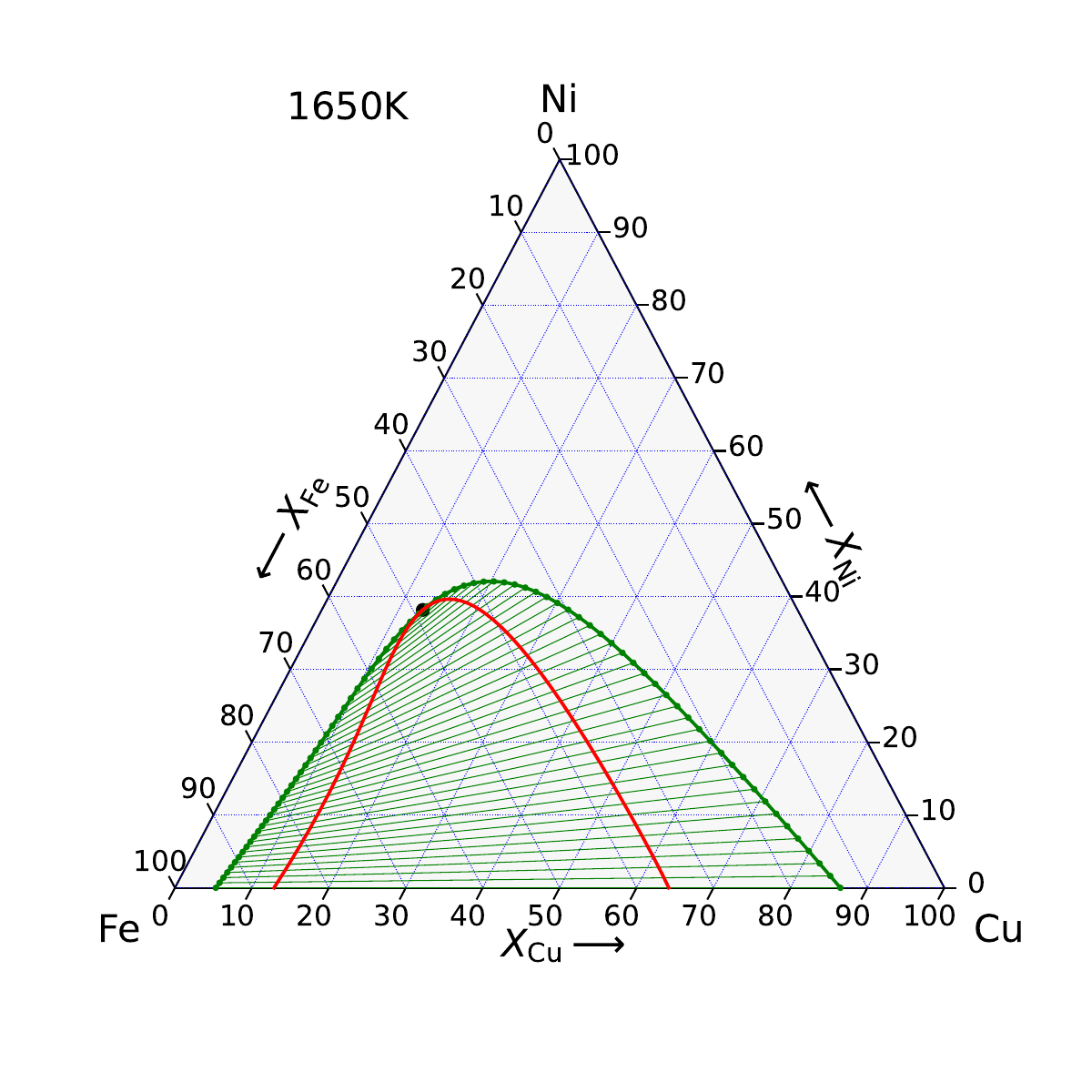}\!
  \includegraphics[width=\PDfigwidth]{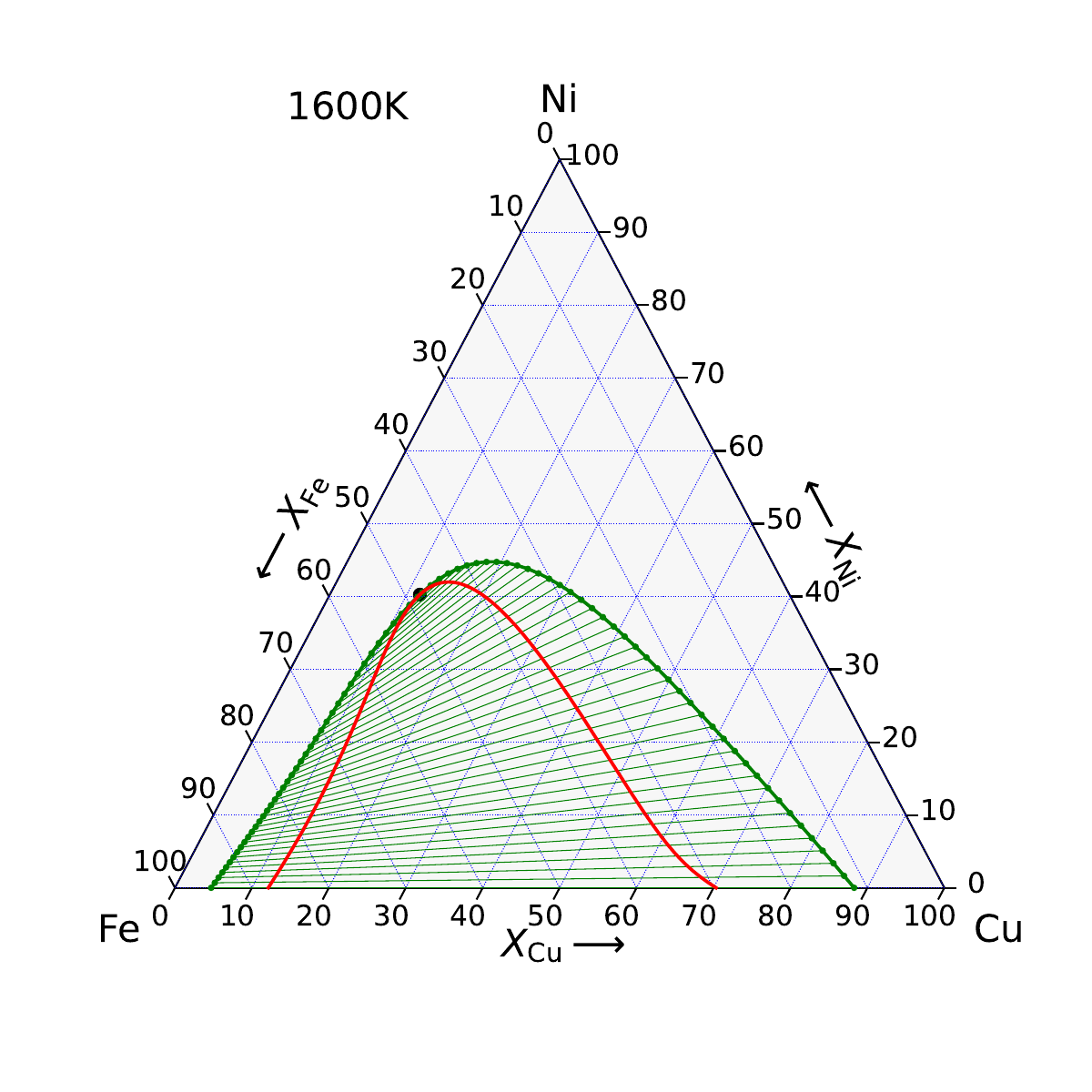}\!
  \includegraphics[width=\PDfigwidth]{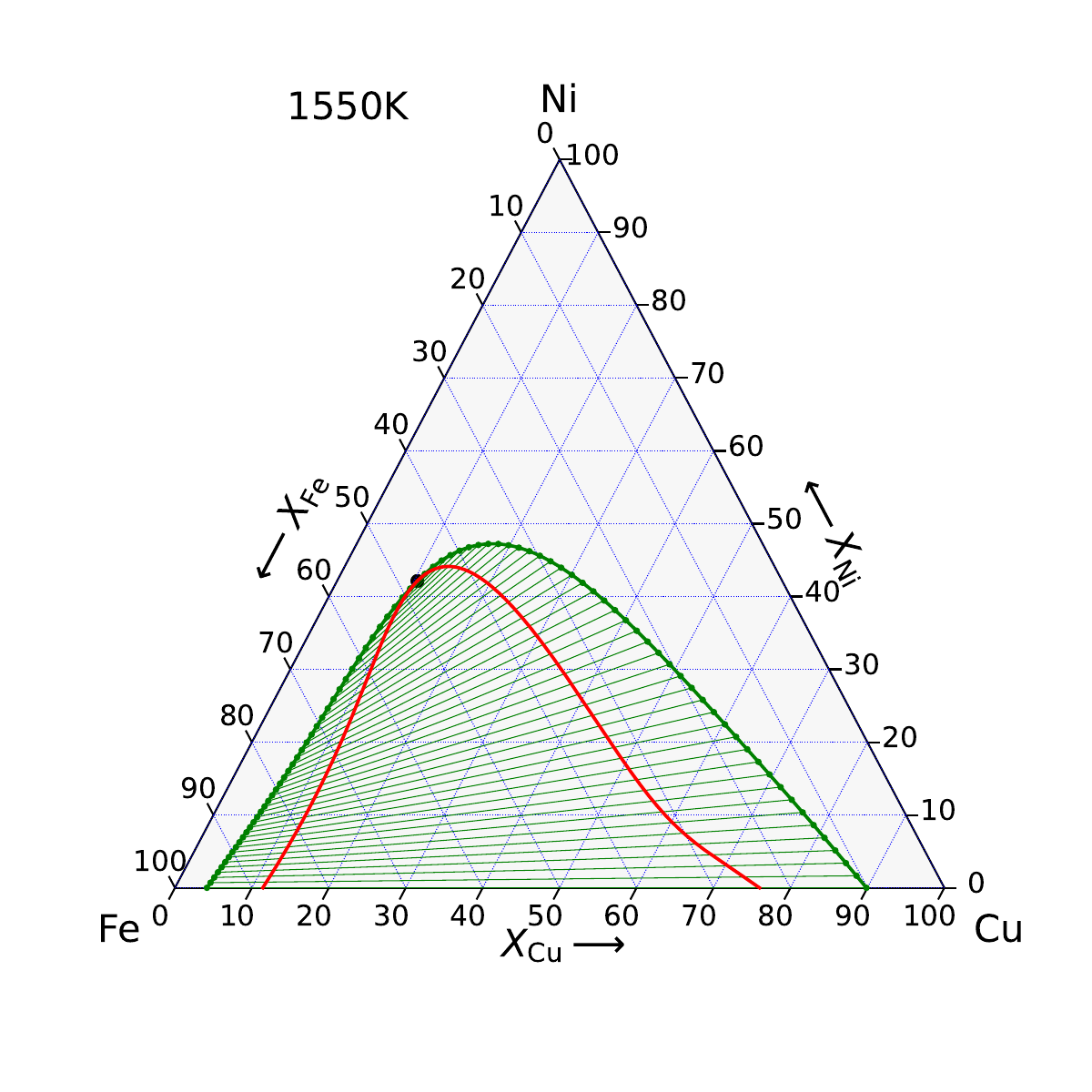}\!
  \includegraphics[width=\PDfigwidth]{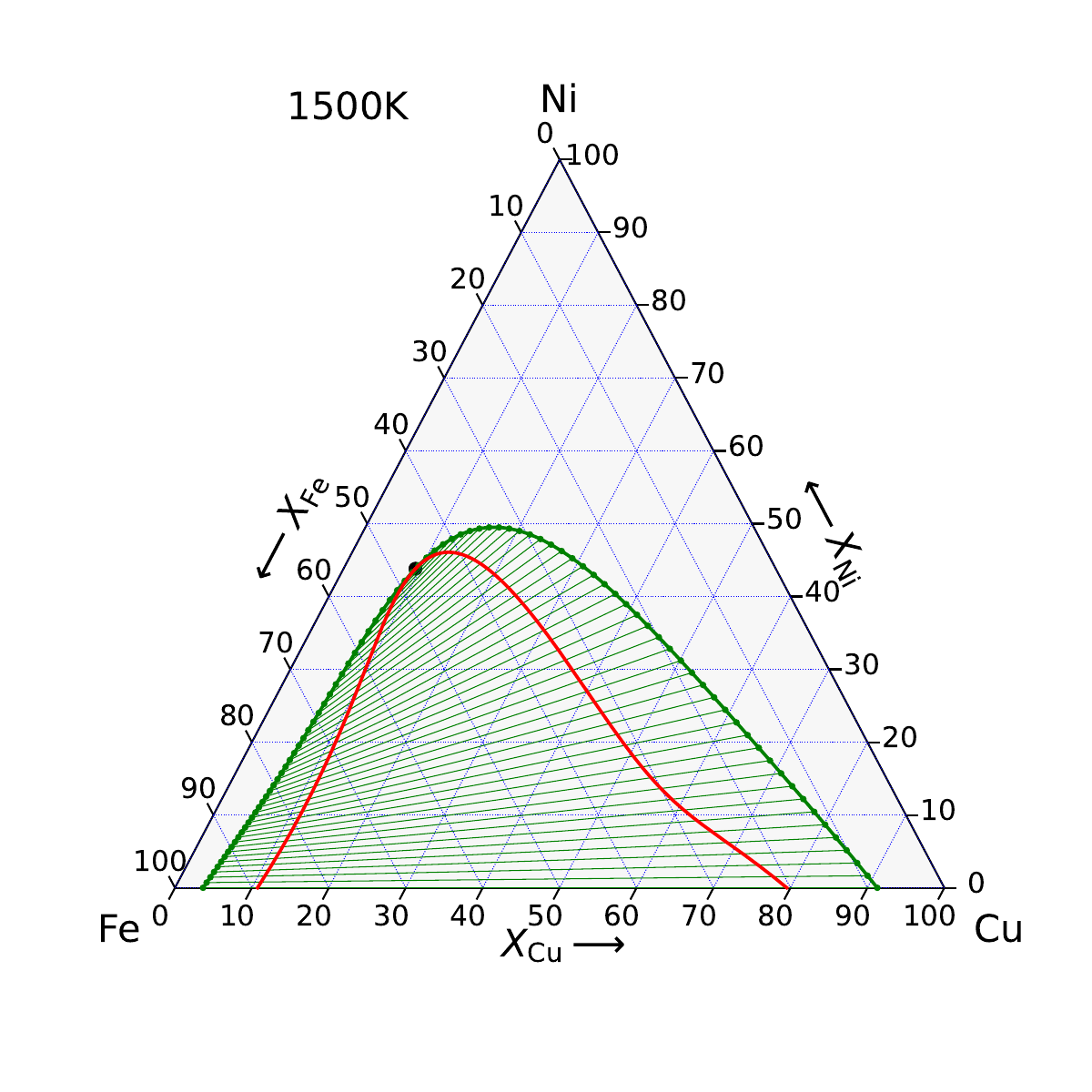}  
  \caption{Temperature slices of the Fe-Cu-Ni ternary liquid phase diagram. The binodal decomposition is shown in green with tie lines, and the spinodal decomposition is in red, with the critical point identified as a black dot.}
  \label{fig:FeCuNi-Tslices}
\end{figure*}

Computing the phase diagram for a ternary is more complex than the binary; we map out the boundary iteratively at each temperature, by proceeding from the Fe-Cu binary into the ternary phase field. We take the two binodal points, and given the Gibbs free energy, chemical potential (first derivatives) and derivatives of the chemical potentials, we make a small change in each concentration to (a) keep the chemical potentials equal to first order, and (b) maintain a common tangent to first order. This requires solving a coupled quadratic equation. This procedure identifies a tie line, and we continue stepping along until the binodal collapses at the critical point. The full algorithm is described in \App{phase-diagram}; alternate algorithms to map the phase boundaries from a parameterized free-energy function are available in CALPHAD software such as \textsc{OpenCalphad}\cite{Sundman2015}. Along each tie-line, the spinodal points are identified as the points where the second derivative of $G$ is zero via numerical root-finding.

\begin{figure}[tbh]
  \includegraphics[width=\figwidth]{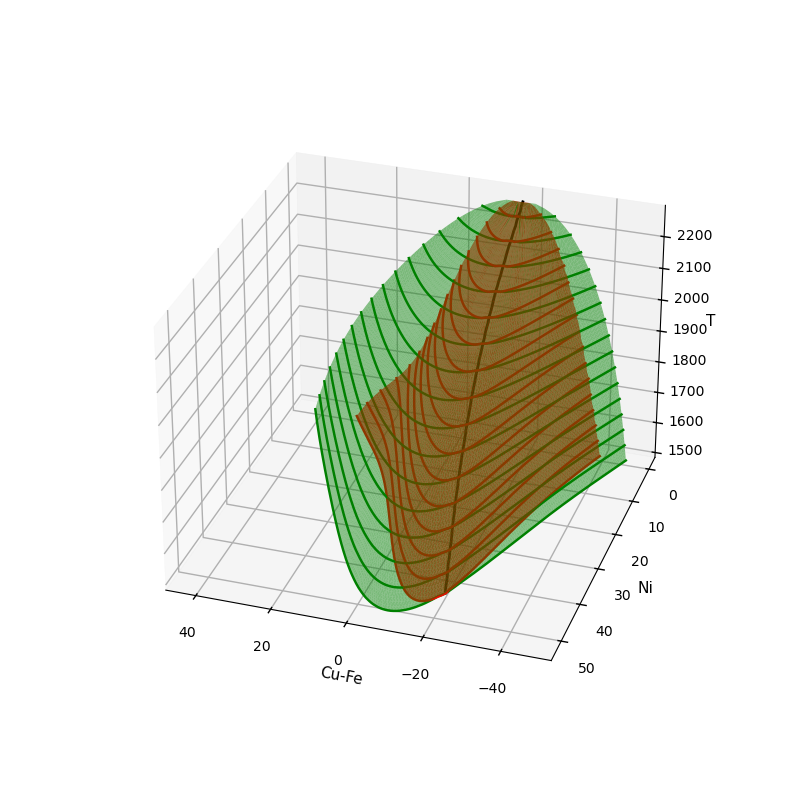}
  \caption{3D rendering of the Fe-Cu-Ni ternary liquid phase diagram. The binodal (green) and spinodal (red) decompositions are demarcated in 50K slices, and the critical curve is shown in black.}
  \label{fig:FeCuNi-3D}
\end{figure}

\Fig{FeCuNi-Tslices} and \Fig{FeCuNi-3D} shows how Ni affects the Fe-Cu liquid miscibility gap. 
In general, adding Ni makes Fe and Cu more miscible, shrinking the binodal and spinodal regions. This suggests that Ni additions may be used to tune the liquid miscibility gap in real alloys. In addition, from the angle of the tie-lines, we see that Ni preferentially segregates to the Cu-rich phase over the Fe-rich phase. This would improve the mechanical properties of the alloy microstructures, as the Cu rich alloy can benefit more from solid-solution strengthening. The addition of Ni also increases the Fe content in the Cu-rich phase. The 3D rendering of the phase boundaries shows more clearly how the spinodal separates more from the binodal on the Cu side of the Gibbs triangle, and that the critical point remains on the Fe side. To our knowledge, this ternary phase diagram has not been experimentally assessed; we would expect this phase diagram to be qualitatively correct, but not predictive, given that the Fe-Cu critical point was shown to be too high.

\subsection{Structural analysis}

\begin{figure*}[tbh]
  \includegraphics[width=0.32\wholefigwidth]{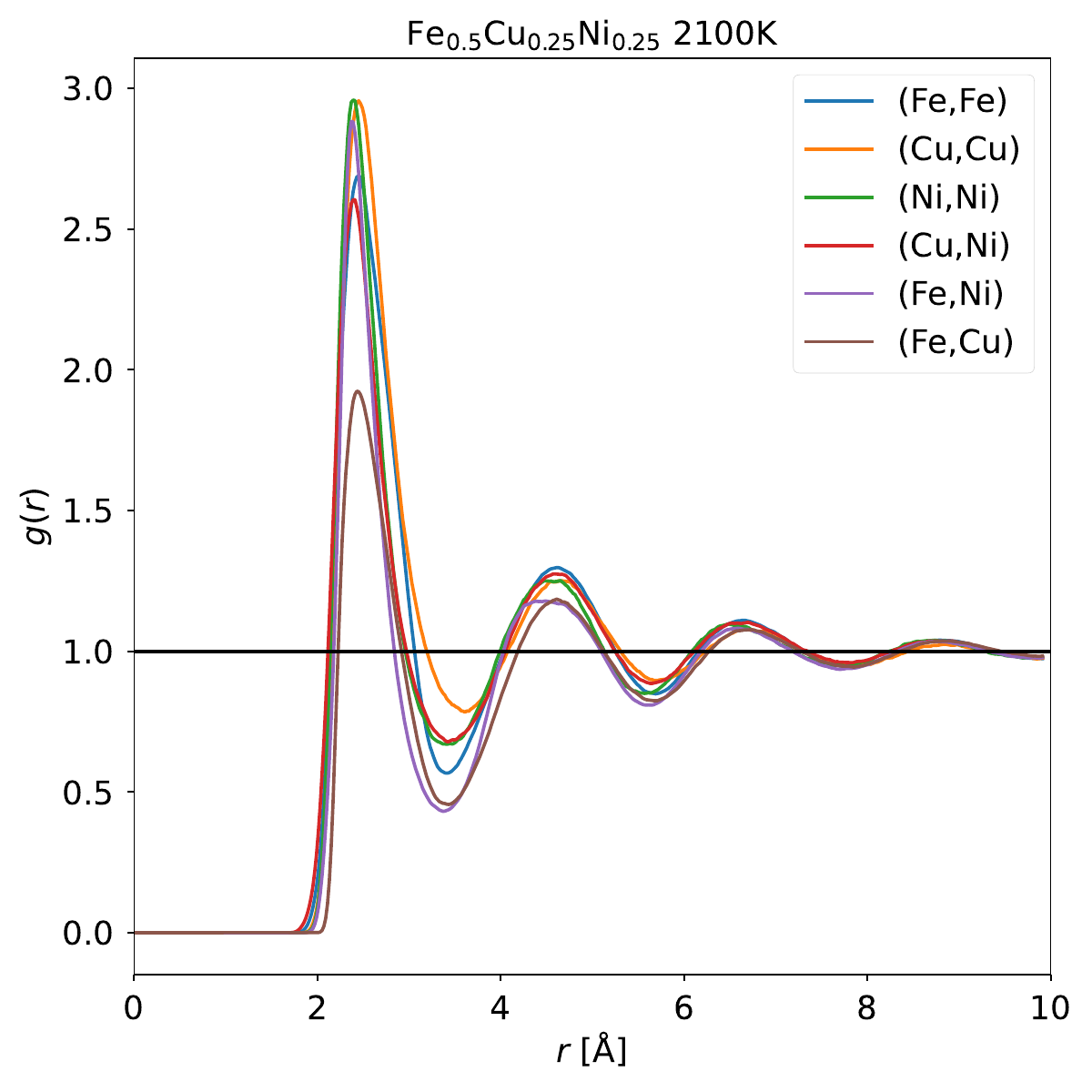}
  \includegraphics[width=0.32\wholefigwidth]{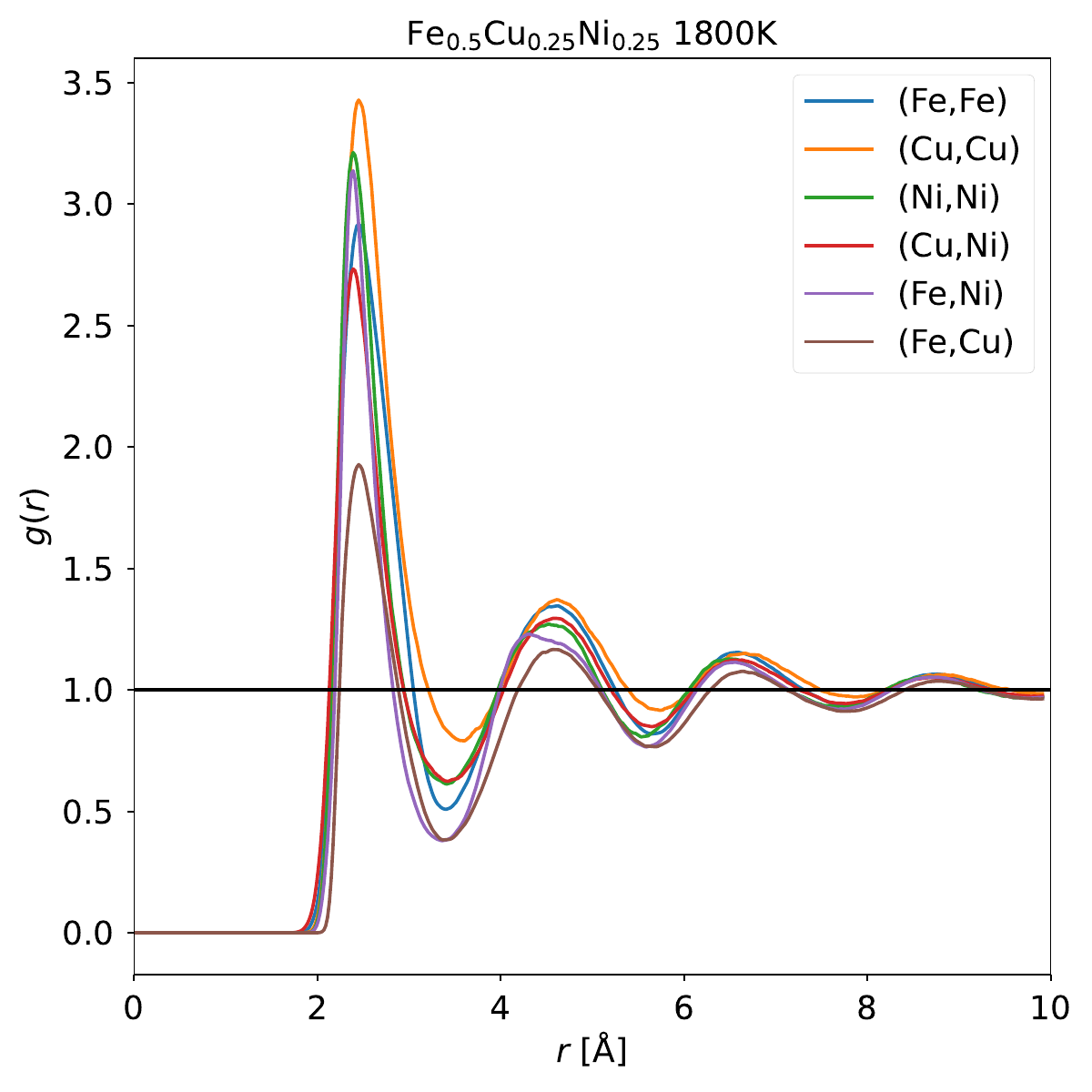}
  \includegraphics[width=0.32\wholefigwidth]{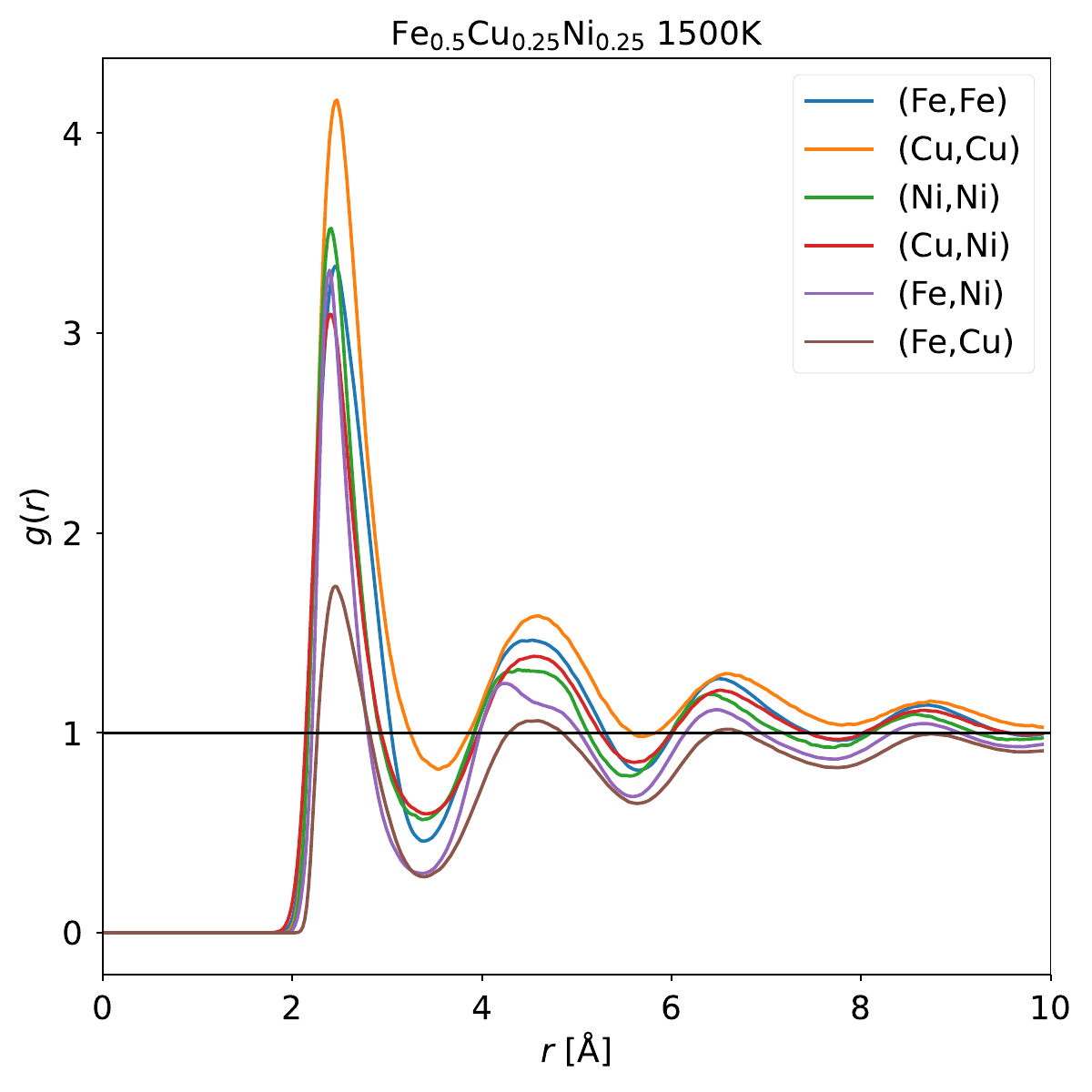}
  \caption{Pair correlation functions for Fe$_{50}$Cu$_{25}$Ni$_{25}$ above, just below, and far below the spinodal decomposition temperature.}
  \label{fig:gr-422}
\end{figure*}

In addition to computing the chemical potential differences, molecular dynamics allows us to determine our liquid structure; in \Fig{gr-422}, we examine pair correlation functions in Fe$_{50}$Cu$_{25}$Ni$_{25}$ above and below the spinodal decomposition. Pair correlation functions $g_{\text{A}\text{B}}(r)$ give the probability of finding a pair of A-B atoms at a distance $r$ relative to a uniform distribution given by the concentration. Above the spinodal decomposition temperature (2100K), we see very similar $g(r)$ for all combination of chemistries; all of the shells are at similar distances. The peaks are similar, except Fe-Cu has a lower first peak of approximately 2 compared with 2.5--3. This corresponds with some short-range ordering as a precursor to the spinodal decomposition. At the second peak, both the Fe-Cu and Fe-Ni peaks are smaller. The lower Fe-Ni second peak is likely due to depletion, as it appears Ni neighbors are substituting for the missing Cu in the first shell. These trends grow just below the spinodal temperature (1800K); in addition, the Cu-Cu first and second peaks grow. The Fe-Ni second peak starts to show some structure; this may be a precursor to liquid phase separation. As the temperature is lowered further (1500K), the Fe-Cu peak drops further. Moreover, the positions of the first, second, and third peaks shift between Cu-Cu, Fe-Fe, and Ni-Ni. In the large $r$ limit, the $g(r)$ begins to deviate from the expected limit of 1, indicating the start of liquid phase separation.

\begin{figure*}[tbh]
  
  \includegraphics[width=0.32\wholefigwidth]{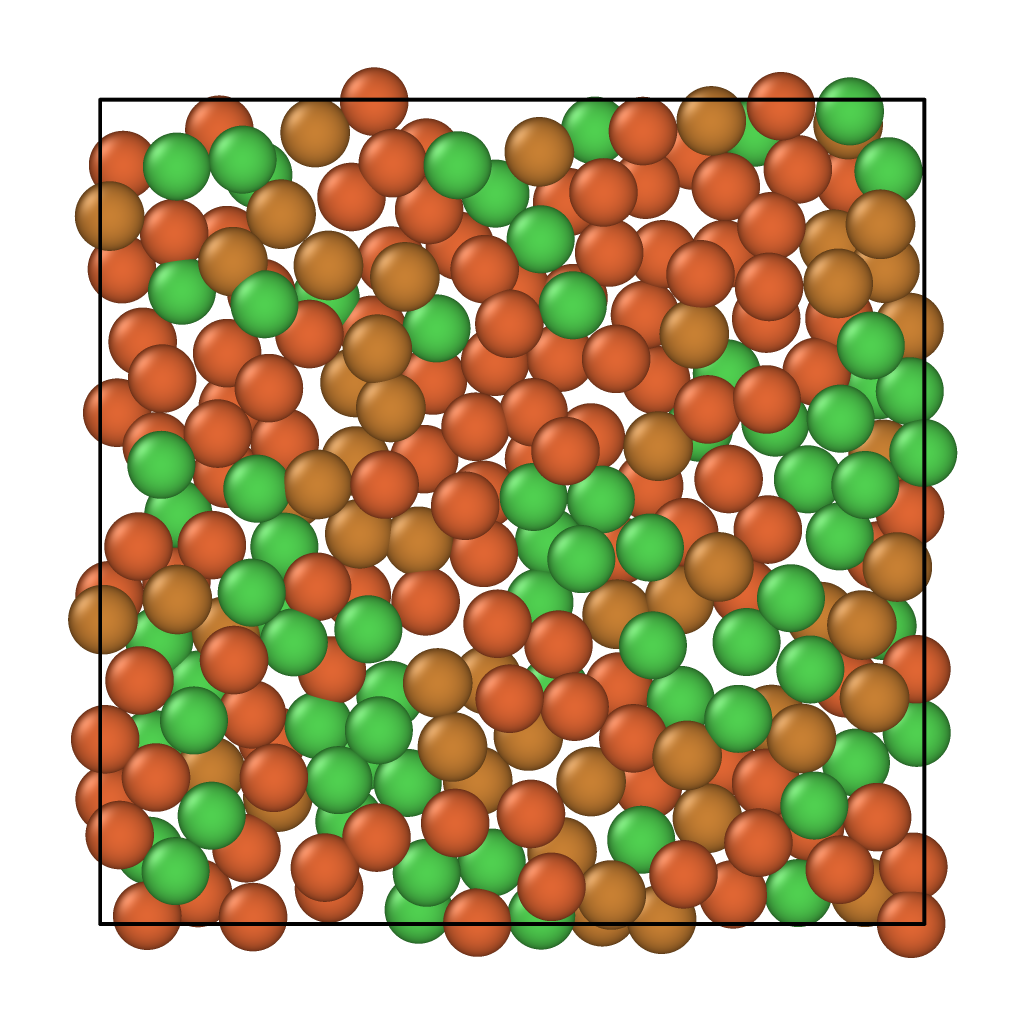}
  \includegraphics[width=0.32\wholefigwidth]{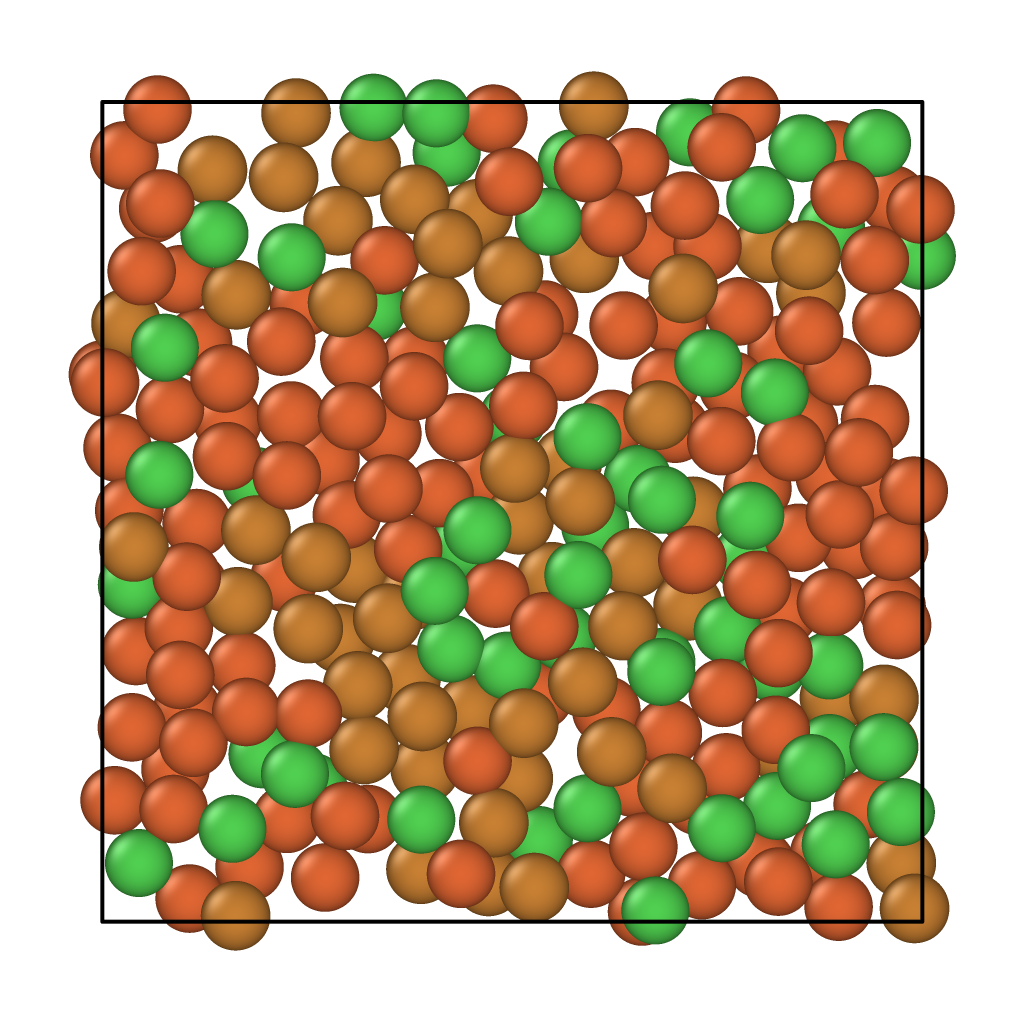}
  \includegraphics[width=0.32\wholefigwidth]{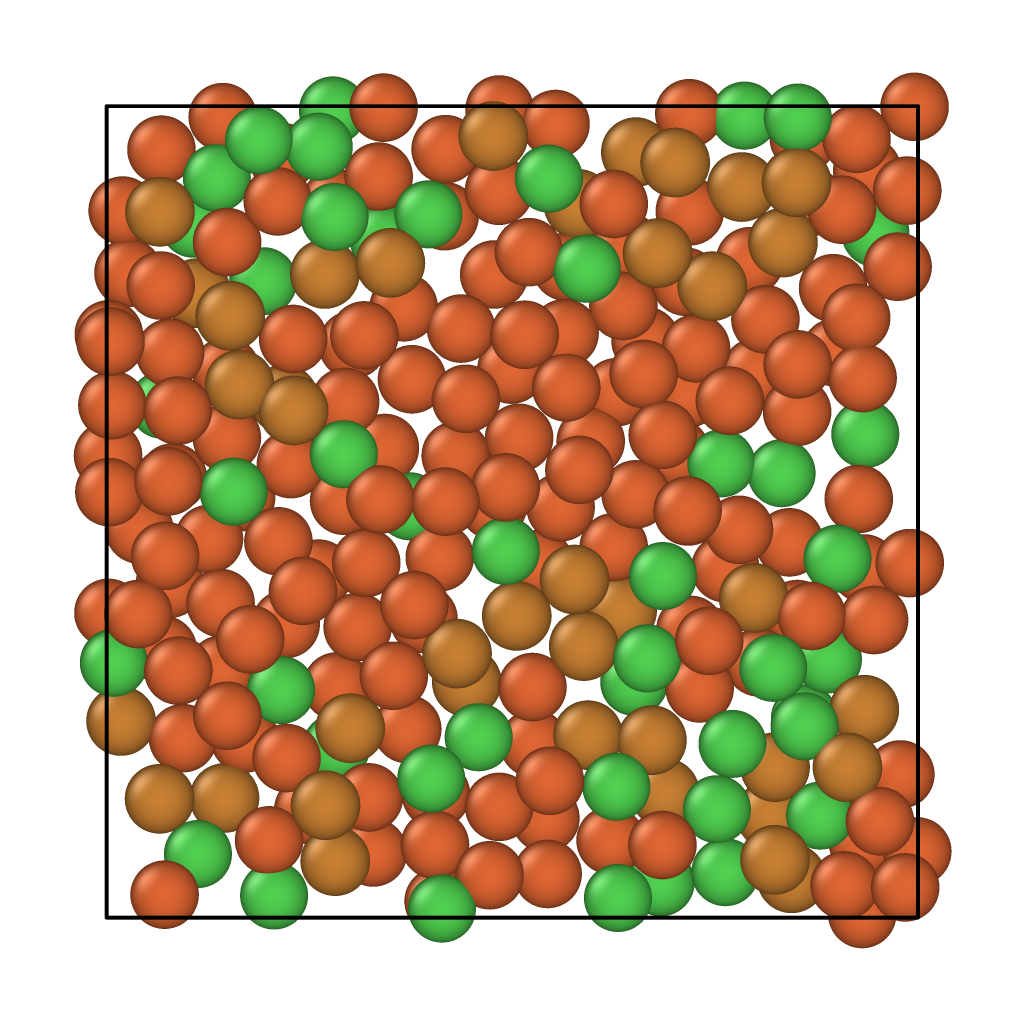}
  \caption{Molecular dynamics snapshots Fe$_{50}$Cu$_{25}$Ni$_{25}$ at 2100K (left), 1800K (middle), 1500K (right). For clarity, a slice of thickness 7.8\AA\ (corresponding to the dip in $g(r)$ between 3rd and 4th neighbor peaks) is shown with Fe (red), Cu (copper), and Ni (green). }
  \label{fig:snapshots}
\end{figure*}

\Fig{snapshots} shows molecular dynamics snapshots of Fe$_{50}$Cu$_{25}$Ni$_{25}$ at the same temperatures as the pair correlation functions. As temperature is lowered, from 1800K down to 1500K, the partial separation into Cu-rich and Fe-rich phases can be seen in the snapshots. In all cases, Ni is intermixed between both phases, as one would expect from the phase diagram. Due to the size of the simulation cell, full phase separation is not possible, even at low temperatures. The limited phase separation is a benefit to our computation of chemical potential differences, as it allows the calculation of driving forces in metastable and even partially unstable regimes. The lack of full phase separation allows the computation of the Gibbs free energy in the metastable regions of the phase diagram, and to predict the phase boundaries.

\begin{figure}[tbh]
  \includegraphics[width=\figwidth]{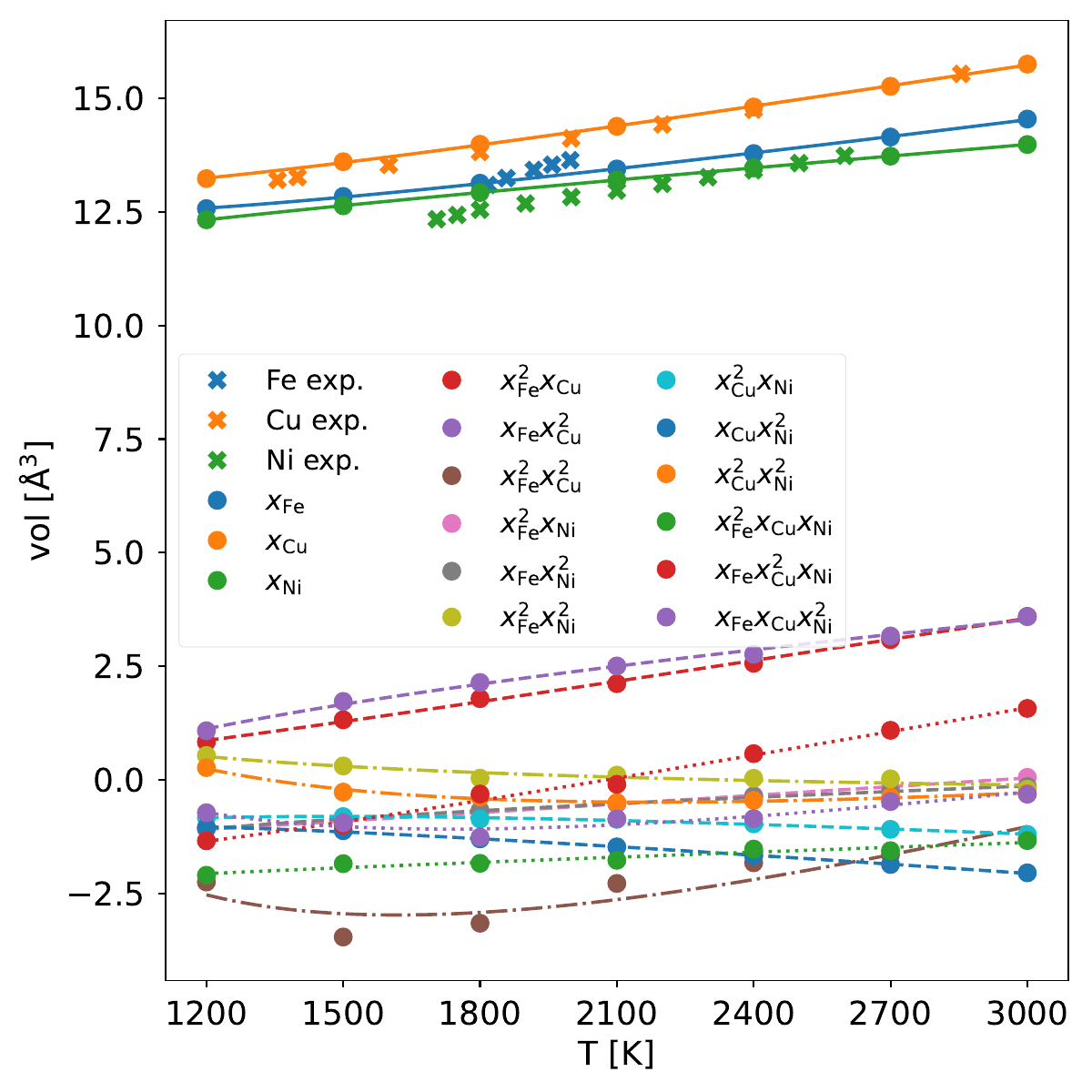}\\
  \includegraphics[width=\figwidth]{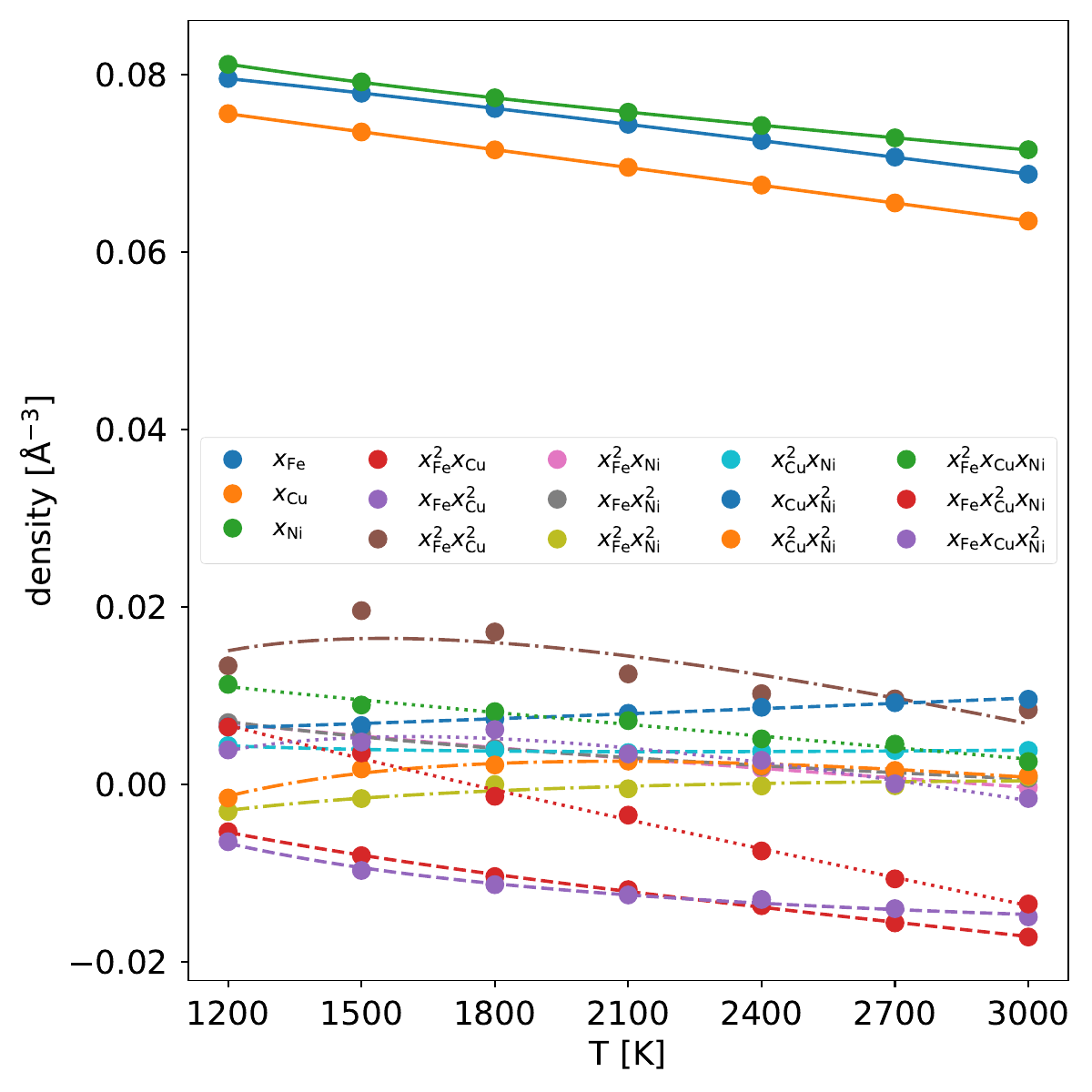}
  \caption{Redlich-Kister polynomial fits for volume and density with temperature. Solid lines are unary coefficients (such as $V_{100} x_1$), dashed lines are linear terms in the binary (such as $x_1x_2(V_{210}x_1 + V_{120}x_2)$), dashed-dotted lines are quadratic terms in the binary (such as $V_{220}x^2_1x^2_2$), and dotted lines are ternary coefficients (such as $V_{211} x_1^2x_2x_3$). The temperature dependence is described as $c_0 + c_1 T +c_2 T^{-1}$. Experimental liquid volume data for pure Fe\cite{Kamiya2021}, Cu\cite{Cahill1962}, and Ni\cite{Schmon2015} is shown for comparison.}
  \label{fig:volume-density}
\end{figure}

As the molecular dynamics simulations are performed with a barostat, we can also compute the average volume as a function of temperature and composition; the expansion coefficients appear in \Fig{volume-density}. The volume per atom, or number density, can each expanded as a function of composition, similar to the excess chemical potential with a unary contribution. The pure liquids show highest density for Ni and lowest for Cu, as well as the largest thermal expansion for Cu. From these expansions, the partial molar volume at any composition can be determined. In the Fe-Cu system, the intermediate volumes are all larger than a simple Vegard's law would predict. The individual chemical expansion coefficients at each temperature are also fit to a simple temperature dependence of the form $c_0 + c_1 T +c_2 T^{-1}$; over the 1200--3000K temperature range, this better captured the curvature than quadratics in temperature. The comparison with experimental measurements of pure Fe\cite{Kamiya2021}, Cu\cite{Cahill1962}, and Ni\cite{Schmon2015} liquid atomic volumes is quite good, despite other shortcomings of the EAM potential.

\subsection{$S^0$ method}
\begin{figure*}[tbh]
  \includegraphics[width=0.32\wholefigwidth]{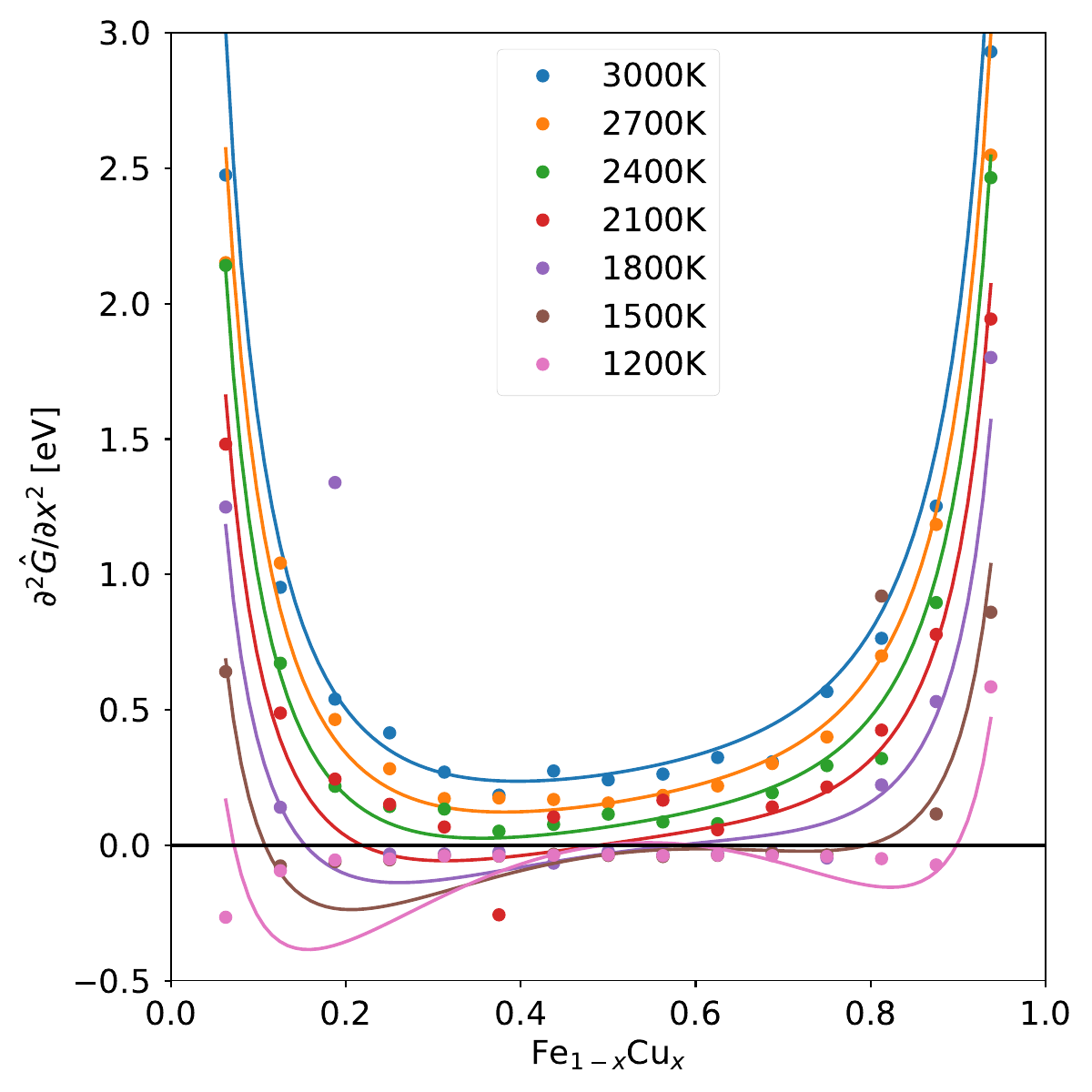}
  \includegraphics[width=0.32\wholefigwidth]{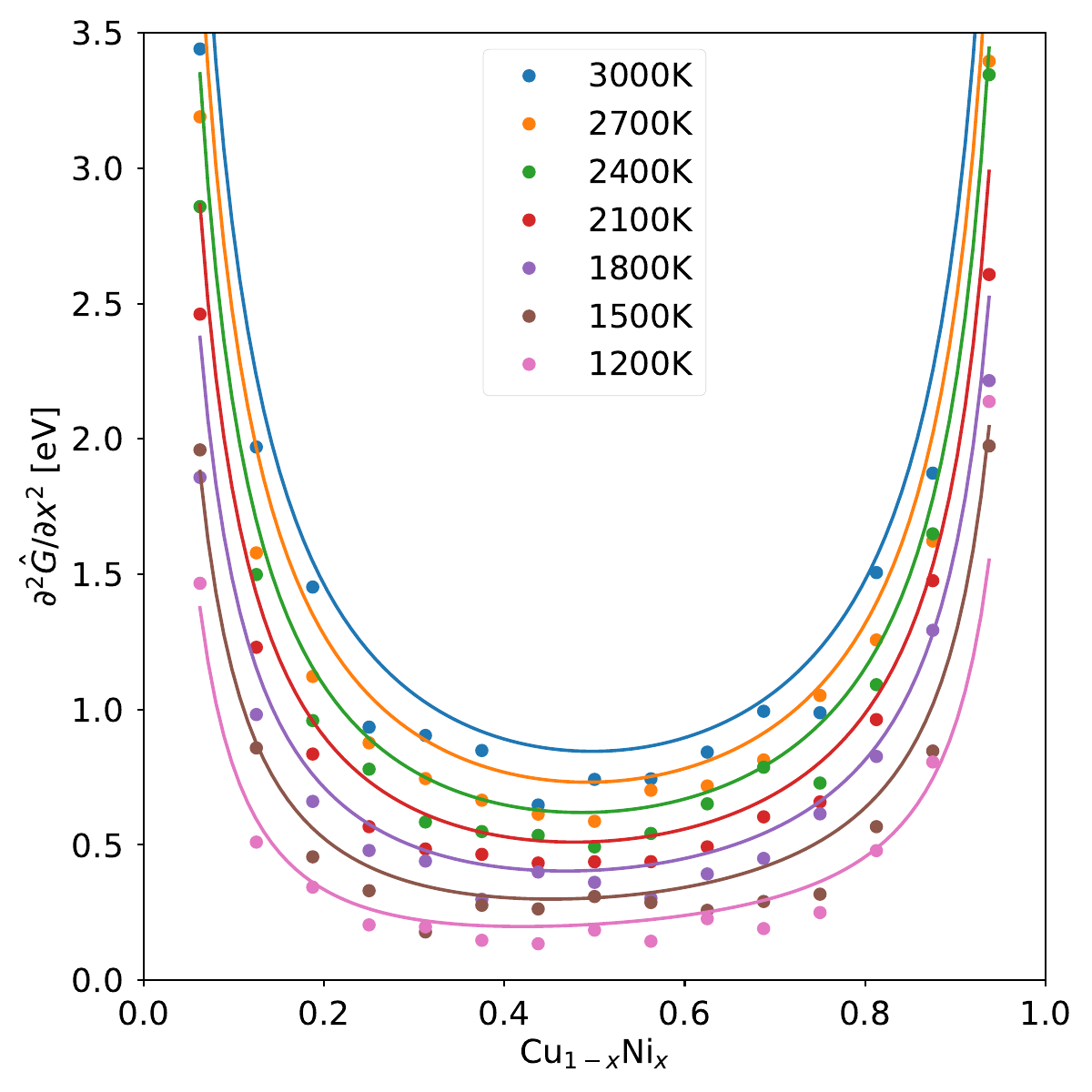}
  \includegraphics[width=0.32\wholefigwidth]{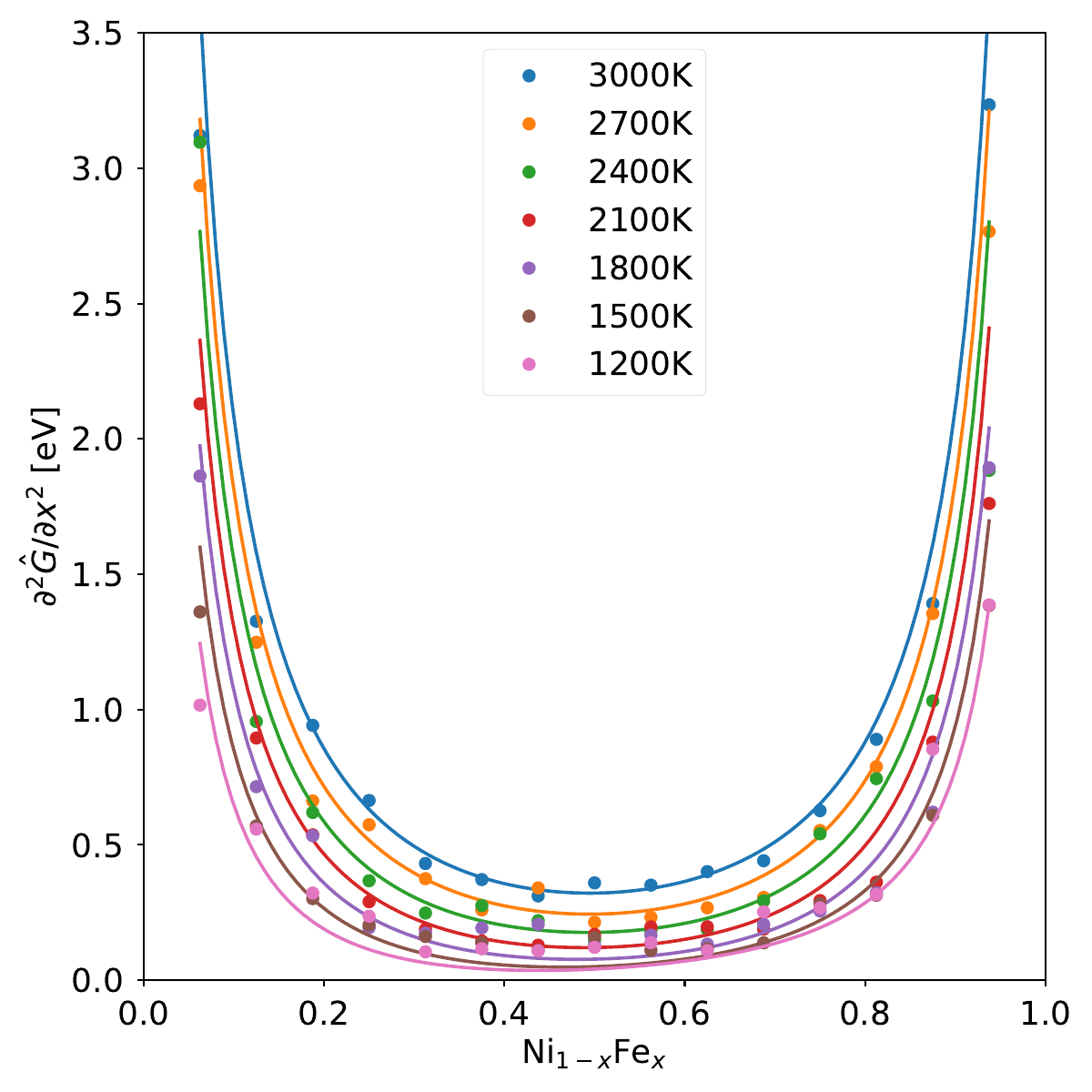}
  \caption{Analytic (curves) vs. $S^0$ (points) evaluation of $d^2G/dx^2$ for Fe-Cu, Cu-Ni, and Ni-Fe. The $S^0$ method extracts the derivative of the chemical potential with respect to concentration---the curvature of the free energy---from the structure factor computed in molecular dynamic calculations. While there is a qualitative agreement between the analytic curvature and the $S^0$ results, the quantitative disagreement makes $S^0$ results insufficient to compute accurate free energy expressions. In addition, as the Fe-Cu system cools and shows phase separation (2100K and below), the $S^0$ method is unable to be used in the metastable and unstable regimes.}
  \label{fig:S0-d2Gdx2}
\end{figure*}

In addition to pair correlation functions, molecular dynamic simulations can compute static structure factors, which can be related to derivatives of chemical potential\cite{Cheng2022}. While the pair correlation functions capture short-range structure, the static structure factor captures long-range structure. For wave-vector $\kv$, the static structure factor for atoms of chemistry A and B can be computed as
\begin{equation}
  S_{\text{A}\text{B}}(\kv) = \frac{1}{\sqrt{N_\text{A}N_\text{B}}}
  \left\langle \sum_{i=1}^{N_\text{A}}\sum_{j=1}^{N_\text{B}}
  \exp\left(i\kv\cdot\left(\rhv_i^\text{A} - \rhv_j^\text{B}\right)\right) \right\rangle
  \label{eqn:Sk}
\end{equation}
where $\rhv^\text{A}_i := \rv^\text{A}_i \langle\ell\rangle/\ell$ and $\rv^\text{B}_j := \rv^\text{B}_j \langle\ell\rangle/\ell$ are the rescaled positions of atoms A and B to the average cell dimension $\langle\ell\rangle$. The structure factors contain information about density correlations over larger distances. As $k\to 0$,
\begin{equation*}
  S^0_{\text{A}\text{B}} := \lim_{\kv\to 0} S_{\text{A}\text{B}}(\kv)
\end{equation*}
is the ratio of particle number fluctuations in A and B to geometric mean of $N_\text{A}$ and $N_\text{B}$. This can be related to chemical potential derivatives; following Cheng\cite{Cheng2022},
\begin{equation}
  \left(\frac{\partial\mu_\text{A}}{\partial x_\text{A}}\right)_{T,P} =
  \frac{\kB T}{x_\text{A}} \left[\frac{1}%
    {S^0_{\text{A}\text{A}} - S^0_{\text{A}\text{B}} \sqrt{x_\text{A}/x_\text{B}} }
    \right]
  \label{eqn:S0}
\end{equation}
This is the $S^0$ method. To find $S^0$ from $S(\kv)$, we fit the small $\kv$ behavior to the functional form $S^0/(1+k^2\xi^2)$ for length scale $\xi$. We can directly compare the calculation of the second derivative of the Gibbs free energy from our fits to that determined via the $S^0$ method in \Fig{S0-d2Gdx2}. As can be seen for the three binary subsystems, the $S^0$ calculations of the second derivatives qualitatively agree with the analytic forms, but show significant quantitative disagreement. While the disagreement is most significant for the Fe-Cu system, especially as it undergoes partial phase separation where $S^0$ will predict $\partial^2G/\partial x^2=0$, there are disagreements for all temperatures in the Cu-Ni and Ni-Fe systems as well. While we did not attempt to fit analytic functions to the $S^0$ data, it seems unlikely to reproduce the binary coefficients in \Fig{binary-coeff}. It is possible that this may be remedied by significantly longer molecular dynamics trajectories and larger cells for a more accurate calculation of the small $k$-limit. In addition, as the $S^0$ method only produces \textit{derivatives} of the chemical potentials, additional computation would be required to provide a baseline chemical potential, which is necessary to compute a phase diagram.

\subsection{Absolute Gibbs free energy}

\begin{figure}[tbh]
  \includegraphics[width=\figwidth]{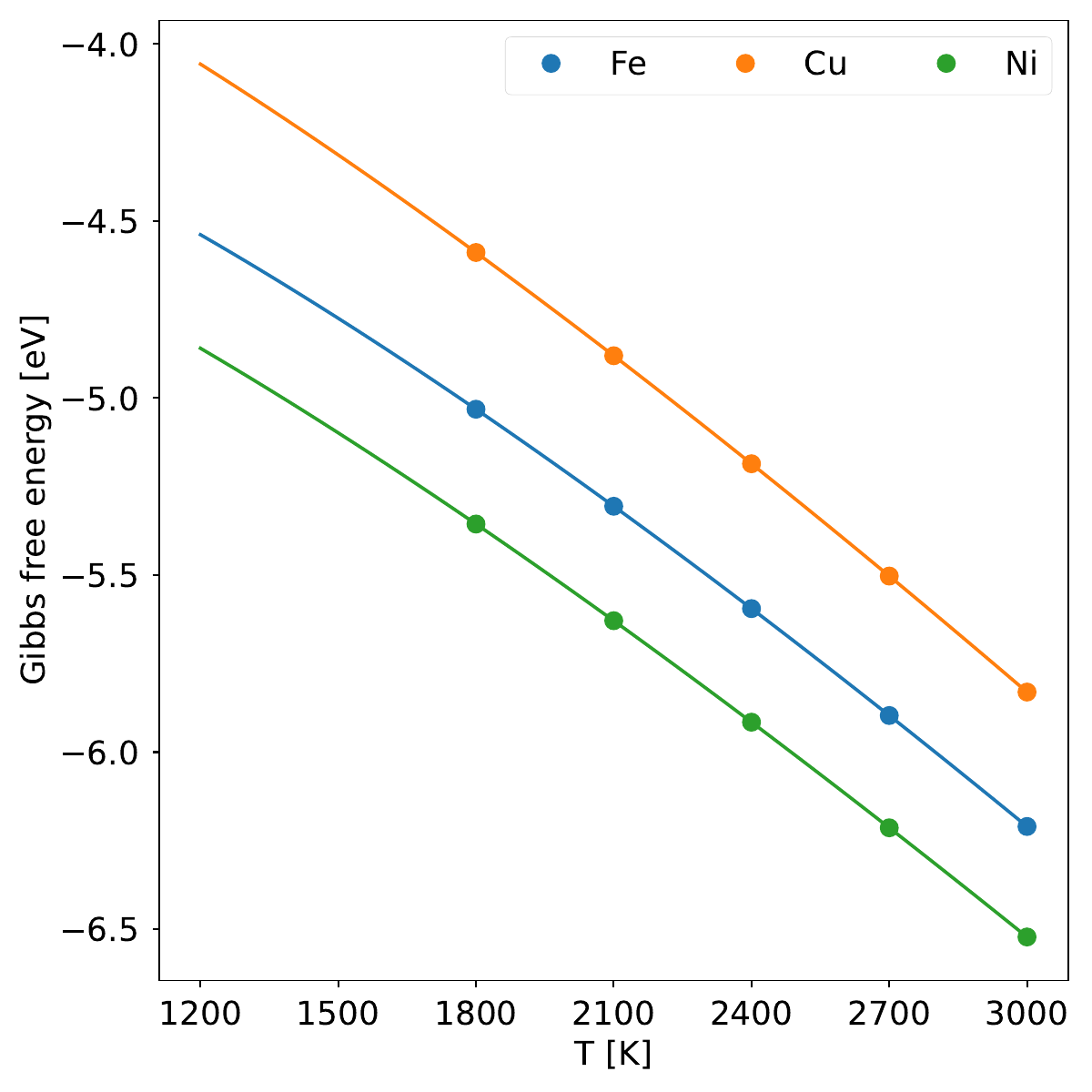}
  \caption{Absolute Gibbs free energies of pure liquid Fe, Cu, and Ni. The temperature dependence is described as $c_0 + c_1 T +c_2 T\ln T$.}
  \label{fig:gibbs-absolute}
\end{figure}

\Fig{gibbs-absolute} shows the absolute Gibbs free energies of the pure liquid states, calculated using non-equilibrium Hamiltonian integration. While all of the data needed to compute the liquid phase diagram is available from a virtual semigrand canonical Widom approach, computing the absolute Gibbs free energy for the full phase diagram requires additional reference data. The differences in chemical potentials allows us to determine the Gibbs free energy up to an additive constant for each temperature; in order to compute equilibria against other phases, we would need to find a reference. \App{NEHI} gives the details of our computation of the absolute Gibbs free energy for pure liquid Fe, Cu, and Ni, performed at 3000K, 2700K, 2400K, 2100K, and 1800K. The standard error in the Gibbs free energy from integration is less than 0.6 meV for all of the simulations. Similarly, the Gibbs free energies all fit our temperature dependence with errors below 0.1 meV. When we compare the difference in absolute Gibbs free energies for our pure liquids to that given by our coefficients in \Fig{binary-coeff}, we find differences of 1 meV or lower. If we extrapolate the fit down to 1200K, we find a difference of 10 meV for Fe-Cu. In addition, we can take the thermal average energy as a function of temperature from our runs, and integrate to get the difference in $G(T)$ at different temperatures; compared with the non-equilibrium Hamiltonian integration, the error is 1 meV over the 1800--3000K temperature range. So while we have benchmarked our results, in a practical calculation, one would need to do a single absolute Gibbs free energy calculation at one reference temperature, and then use the energy integration and chemical potential differences to have absolute Gibbs free energies across the entire chemical and temperature space. Note also that each non-equilibrium Hamiltonian integration calculation required 2 million energy evaluations for each composition and temperature, compared with approximately 260,000 energy evaluations for the virtual semigrand canonical Widom method including the trajectory and the calculation of energy changes.

\section{Conclusions}
We have demonstrated the prediction of phase separation in a ternary liquid system using  a virtual semigrand canonical Widom approach, sampled with molecular dynamics. In addition the determination of the Gibbs free energy for the ternary Fe-Cu-Ni system as a function of composition and temperature over the range 1200--3000K, we mapped the binodal and spinodal limits in the Gibbs triangle. Using molecular dynamics, we could also determine density, pair correlation functions, and static structure factors. The latter was used to provide benchmark computation of the second derivative of the Gibbs free energy via the $S^0$ method; it appears that longer simulations with larger cells may be necessary to provide sufficient accuracy to be useful for fitting of Gibbs free energies. The virtual semigrand canonical Widom approach has overhead of roughly a factor of 2 with a regular molecular dynamics run; that said, the standard error estimates shown here indicate that shorter trajectories or fewer samples could be used. In addition, a hybrid Monte Carlo / molecular dynamics approach may further increase the efficiency of sampling, albeit at the expense of full trajectory information. We also benchmarked our computation against absolute Gibbs free energy calculations for the pure liquid; while these were computationally intensive, they show that the computed free energy matches to well small fractions of the thermal energy. The method is not restricted to liquids, and can be applied to multicomponent solid systems as well. While the numerical results hinge on the accuracy of the underlying EAM potential to describe the interaction of Fe, Cu, and Ni, the trends identified should be useful for the design of ternary alloys in additive manufacturing. Moreover, this serves as a useful benchmark for future, more accurate, computation of phase diagrams.

\begin{acknowledgments}
Part of this research was performed while the author was visiting the Institute for Mathematical and Statistical Innovation (IMSI), which is supported by the National Science Foundation (Grant No. DMS-1929348).
The author thanks Marie A.\ Charpagne and Baron G.\ Peters for helpful conversations. The data is available at the Materials Data Facility\cite{Blaiszik2016, Blaiszik2019} doi:10.18126/p0gx-tt30 \cite{FeCuNi2025Data}.
\end{acknowledgments}

\appendix
\section{Iterative algorithm for ternary binodal points and tie-lines}
\label{sec:phase-diagram}
We begin our identification of the binodal points in the ternary from the known binodal points in the Fe-Cu binary. Initially, our two concentrations obey the common tangent construction and have equal chemical potentials \textit{only} along the Fe-Cu axis, as $\xNi=0$. As a first step, we introduce a small $\xNi=10^{-4}$ concentration, but split between the two binodal points to make the full chemical potentials equal. Then, let our initial binodal concentration vectors be $\xv_1$ and $\xv_2$ with Gibbs free energies $G_1$ and $G_2$, chemical potentials $\muv_1$ and $\muv_2$, and second derivative matrices $\hh_1:=\nablav\nablav G_1$ and $\hh_2 := \nablav\nablav G_2$. We define $\dxv := \xv_2-\xv_1$ as the tie-line direction, and identify two sets of (normalized) directions: $\hpv_i := (\hh_i\cdot\dxv)/|\hh_i\cdot\dxv|$ and $\htv_i$ that is perpendicular to $\hpv_i$ in the Gibbs triangle plane. We change our compositions with $\xv_1 \to \xv_1 + (s-\alpha)\htv_1 + \beta_1\hpv_1$ and $\xv_2 \to \xv_2 + (s+\alpha)\htv_2 + \beta_2\hpv_2$ with a step size $s$. The three unknowns, $\alpha$, $\beta_1$, and $\beta_2$ are chosen to guarantee that the chemical potentials are equal to first order and that the new points lie on the common tangent to second order. The chemical potential difference is
\begin{widetext}
\begin{equation}
  0 = \muv_2-\muv_1 + \hh_2\cdot((s+\alpha)\htv_2 + \beta_2\hpv_2)
  - \hh_1\cdot((s-\alpha)\htv_1 + \beta_1\hpv_1)
\end{equation}
\end{widetext}
whose equality we check by dotting into $\dxv$ and a direction $\nv$ that is perpendicular in the Gibbs triangle plane. Defining $\dmuv := \muv_2-\muv_1$ (the small error in chemical-potential equality before changing composition), we have
\begin{equation}
  \begin{split}
    \dxv\cdot\dmuv =&\ (\hh_1\dxv)\cdot((s-\alpha)\htv_1 + \beta_1\hpv_1)\\
    &- (\hh_2\dxv)\cdot((s+\alpha)\htv_2 + \beta_2\hpv_2)\\
    =&\ (\dxv\hh_1\hpv_1)\beta_1 - (\dxv\hh_2\hpv_2)\beta_2
  \end{split}
  \label{eqn:mu_dx}
\end{equation}
and
\begin{equation}
  \begin{split}
    \nv\cdot\dmuv =&\ (\hh_1\nv)\cdot((s-\alpha)\htv_1 + \beta_1\hpv_1)\\
    &- (\hh_2\nv)\cdot((s+\alpha)\htv_2 + \beta_2\hpv_2)\\
    =&\ (\nv\hh_1\htv_1-\nv\hh_2\htv_2)s - (\nv\hh_1\htv_1+\nv\hh_2\htv_2)\alpha\\
    &+ (\nv\hh_1\hpv_1)\beta_1 - (\nv\hh_2\hpv_2)\beta_2
  \end{split}
  \label{eqn:mu_n}
\end{equation}
These two equations are linear in our three unknowns.

The final equation is the common-tangent requirement. We define $\Delta G_i := G_2-G_1-\muv_i\cdot\dxv$, which is the small error in the common-tangent before changing composition. The common-tangent equation is written most compactly with the composition changes $\dxv_i := (s\mp\alpha)\htv_i + \beta_i\hpv_i$ as
\begin{widetext}
\begin{equation}
  G_2 + \muv_2\cdot\dxv_2 + \frac12 \dxv_2\hh_2\dxv_2 - G_1 - \muv_1\cdot\dxv_1
  -\frac12 \dxv_1\hh_1\dxv_1 =
  (\muv_i + \hh_i\dxv_i)\cdot(\xv_2 - \xv_1 + \dxv_2 - \dxv_1)
  \label{eqn:common-tangent}
\end{equation}
\end{widetext}
which holds for $i=1,2$. With some algebraic manipulation and the requirement that $\muv_1+\hh_1\dxv_1=\muv_2+\hh_2\dxv_2$, \Eqn{common-tangent} becomes
\begin{equation}
  \Delta G_i - \dxv\cdot\hh_i \dxv_i = \frac12 \dxv_2\hh_2\dxv_2 - \frac12 \dxv_1\hh_1\dxv_1
\end{equation}
which can be summed over $i=1,2$ to give
\begin{widetext}
\begin{equation}
  \Delta G_1 + \Delta G_2 =  \dxv\cdot\hh_1 \dxv_1 + \dxv\cdot\hh_2\dxv_2 + \dxv_2\hh_2\dxv_2 - \dxv_1\hh_1\dxv_1.
\end{equation}
\end{widetext}
Expanding this equation in terms of $\alpha$ and $\beta_i$ gives the final equation
\begin{widetext}
\begin{multline}
  (2\htv_1\hh_1\htv_1 + 2\htv_2\hh_2\htv_2)s\alpha
  + (-2\hpv_1\hh_1\htv_1 s + \dxv\hh_1\hpv_1)\beta_1
  + (2\hpv_2\hh_2\htv_2 s + \dxv\hh_2\hpv_2)\beta_2 
  = \Delta G_1 + \Delta G_2 \\
  + (\htv_1\hh_1\htv_1 - \htv_2\hh_2\htv_2)(s^2+\alpha^2)
  - 2\hpv_1\hh_1\htv_1 \beta_1\alpha
  - 2\hpv_2\hh_2\htv_2 \beta_2\alpha
  + \hpv_1\hh_1\hpv_1\beta_1^2
  - \hpv_2\hh_2\hpv_2\beta_2^2
  \label{eqn:quadratic}
\end{multline}
\end{widetext}
The three equations, \Eqn{mu_dx}, \Eqn{mu_n}, and \Eqn{quadratic} can be written as a linear system in our variables $\alpha$ and $\beta_i$ equal to constants plus quadratic terms from \Eqn{quadratic}. This system can be solved numerically via iteration to self-consistency. With a few iterations, the solution converges, and the tie-line is determined. Because the deviations from the common-tangent and equal chemical potentials are included in the next iteration, the algorithm shows numerical stability for small values of $s$. The algorithm continues until the binodal points converge on each other near the critical point.

\section{Non-equilibrium Hamiltonian integration}
\label{sec:NEHI}
We perform a nonequilibrium free-energy calculation for pure liquid Fe, Cu, and Ni following the approach of R.\ Paula Leite and M.\ de Koning\cite{PaulaLeite2019}; the methodology is similar to that of R.\ Freitas et al.\ for solids\cite{Freitas2016}. Thermodynamic integration is performed via a non-equilibrium Hamiltonian integration to a reference state with a known free energy. In the case here, our reference system is the Uhlenbeck-Ford (UF) fluid phase, whose free energy is computable via its virial coefficients\cite{PaulaLeite2016}. Similar to \cite{PaulaLeite2019}, we do a constant volume molecular dynamics simulation at the determined atomic volume from our previous calculations. We use a Langevin thermostat\cite{Schneider1978} with a timestep of 1 fs, and a damping parameter of 0.1 ps. We equilibrate for $5\times10^4$ steps, then over $5\times10^4$ steps, switch to a UF potential with $p=50$, $\sigma=1.5$, and $\varepsilon = p\kB T$. We then reequilibrate, and run the process in reverse. We use the smooth switching function $\lambda(\tau) = \tau^5(70\tau^4-315\tau^3+540\tau^2-420\tau+126)$ where $\tau$ goes from 0 to 1 over the length of the switching simulation. This switching function results in a less dissipative switching process\cite{deKoning1996}. We integrate the dynamical work to get the difference in free energy between our EAM system and our reference UF system; this is done over 10 passes to ensure good statistics, and the forward and backwards directions are compared to ensure smooth switching and no intermediate phase transitions.


\begin{thebibliography}{40}%
\makeatletter
\providecommand \@ifxundefined [1]{%
 \@ifx{#1\undefined}
}%
\providecommand \@ifnum [1]{%
 \ifnum #1\expandafter \@firstoftwo
 \else \expandafter \@secondoftwo
 \fi
}%
\providecommand \@ifx [1]{%
 \ifx #1\expandafter \@firstoftwo
 \else \expandafter \@secondoftwo
 \fi
}%
\providecommand \natexlab [1]{#1}%
\providecommand \enquote  [1]{``#1''}%
\providecommand \bibnamefont  [1]{#1}%
\providecommand \bibfnamefont [1]{#1}%
\providecommand \citenamefont [1]{#1}%
\providecommand \href@noop [0]{\@secondoftwo}%
\providecommand \href [0]{\begingroup \@sanitize@url \@href}%
\providecommand \@href[1]{\@@startlink{#1}\@@href}%
\providecommand \@@href[1]{\endgroup#1\@@endlink}%
\providecommand \@sanitize@url [0]{\catcode `\\12\catcode `\$12\catcode
  `\&12\catcode `\#12\catcode `\^12\catcode `\_12\catcode `\%12\relax}%
\providecommand \@@startlink[1]{}%
\providecommand \@@endlink[0]{}%
\providecommand \url  [0]{\begingroup\@sanitize@url \@url }%
\providecommand \@url [1]{\endgroup\@href {#1}{\urlprefix }}%
\providecommand \urlprefix  [0]{URL }%
\providecommand \Eprint [0]{\href }%
\providecommand \doibase [0]{https://doi.org/}%
\providecommand \selectlanguage [0]{\@gobble}%
\providecommand \bibinfo  [0]{\@secondoftwo}%
\providecommand \bibfield  [0]{\@secondoftwo}%
\providecommand \translation [1]{[#1]}%
\providecommand \BibitemOpen [0]{}%
\providecommand \bibitemStop [0]{}%
\providecommand \bibitemNoStop [0]{.\EOS\space}%
\providecommand \EOS [0]{\spacefactor3000\relax}%
\providecommand \BibitemShut  [1]{\csname bibitem#1\endcsname}%
\let\auto@bib@innerbib\@empty
\bibitem [{\citenamefont {Ratke}\ and\ \citenamefont
  {Diefenbach}(1995)}]{Ratke1995}%
  \BibitemOpen
  \bibfield  {author} {\bibinfo {author} {\bibfnamefont {L.}~\bibnamefont
  {Ratke}}\ and\ \bibinfo {author} {\bibfnamefont {S.}~\bibnamefont
  {Diefenbach}},\ }\bibfield  {title} {\bibinfo {title} {Liquid immiscible
  alloys},\ }\href
  {https://doi.org/https://doi.org/10.1016/0927-796X(95)00180-8} {\bibfield
  {journal} {\bibinfo  {journal} {Mater. Sci. Eng. R}\ }\textbf {\bibinfo
  {volume} {15}},\ \bibinfo {pages} {263} (\bibinfo {year} {1995})}\BibitemShut
  {NoStop}%
\bibitem [{\citenamefont {Zafari}\ and\ \citenamefont
  {Xia}(2021{\natexlab{a}})}]{Zafari2021a}%
  \BibitemOpen
  \bibfield  {author} {\bibinfo {author} {\bibfnamefont {A.}~\bibnamefont
  {Zafari}}\ and\ \bibinfo {author} {\bibfnamefont {K.}~\bibnamefont {Xia}},\
  }\bibfield  {title} {\bibinfo {title} {Laser powder bed fusion of ultrahigh
  strength {Fe-Cu} alloys using elemental powders},\ }\href
  {https://doi.org/https://doi.org/10.1016/j.addma.2021.102270} {\bibfield
  {journal} {\bibinfo  {journal} {Additive Manufacturing}\ }\textbf {\bibinfo
  {volume} {47}},\ \bibinfo {pages} {102270} (\bibinfo {year}
  {2021}{\natexlab{a}})}\BibitemShut {NoStop}%
\bibitem [{\citenamefont {Zafari}\ and\ \citenamefont
  {Xia}(2021{\natexlab{b}})}]{Zafari2021b}%
  \BibitemOpen
  \bibfield  {author} {\bibinfo {author} {\bibfnamefont {A.}~\bibnamefont
  {Zafari}}\ and\ \bibinfo {author} {\bibfnamefont {K.}~\bibnamefont {Xia}},\
  }\bibfield  {title} {\bibinfo {title} {Nano/ultrafine grained immiscible
  fe-cu alloy with ultrahigh strength produced by selective laser melting},\
  }\href@noop {} {\bibfield  {journal} {\bibinfo  {journal} {Materials Research
  Letters}\ }\textbf {\bibinfo {volume} {9}},\ \bibinfo {pages} {247} (\bibinfo
  {year} {2021}{\natexlab{b}})}\BibitemShut {NoStop}%
\bibitem [{\citenamefont {Wei}\ \emph {et~al.}(2025)\citenamefont {Wei},
  \citenamefont {Wuu}, \citenamefont {Soh}, \citenamefont {Lau}, \citenamefont
  {Sun}, \citenamefont {Lee}, \citenamefont {Liu}, \citenamefont {Zhang},
  \citenamefont {Wang},\ and\ \citenamefont {Ramamurty}}]{Wei2025}%
  \BibitemOpen
  \bibfield  {author} {\bibinfo {author} {\bibfnamefont {S.}~\bibnamefont
  {Wei}}, \bibinfo {author} {\bibfnamefont {D.}~\bibnamefont {Wuu}}, \bibinfo
  {author} {\bibfnamefont {V.}~\bibnamefont {Soh}}, \bibinfo {author}
  {\bibfnamefont {K.~B.}\ \bibnamefont {Lau}}, \bibinfo {author} {\bibfnamefont
  {Z.}~\bibnamefont {Sun}}, \bibinfo {author} {\bibfnamefont {J.~J.}\
  \bibnamefont {Lee}}, \bibinfo {author} {\bibfnamefont {F.}~\bibnamefont
  {Liu}}, \bibinfo {author} {\bibfnamefont {B.}~\bibnamefont {Zhang}}, \bibinfo
  {author} {\bibfnamefont {P.}~\bibnamefont {Wang}},\ and\ \bibinfo {author}
  {\bibfnamefont {U.}~\bibnamefont {Ramamurty}},\ }\bibfield  {title} {\bibinfo
  {title} {High throughput additive manufacturing and characterization of
  immiscible {Cu-Fe} binary system using compositional gradient approach},\
  }\href {https://doi.org/https://doi.org/10.1016/j.jallcom.2025.178770}
  {\bibfield  {journal} {\bibinfo  {journal} {Journal of Alloys and Compounds}\
  }\textbf {\bibinfo {volume} {1014}},\ \bibinfo {pages} {178770} (\bibinfo
  {year} {2025})}\BibitemShut {NoStop}%
\bibitem [{\citenamefont {Nakagawa}(1958)}]{Nakagawa1958}%
  \BibitemOpen
  \bibfield  {author} {\bibinfo {author} {\bibfnamefont {Y.}~\bibnamefont
  {Nakagawa}},\ }\bibfield  {title} {\bibinfo {title} {Liquid immiscibility in
  copper-iron and copper-cobalt systems in the supercooled state},\ }\href
  {https://doi.org/https://doi.org/10.1016/0001-6160(58)90061-0} {\bibfield
  {journal} {\bibinfo  {journal} {Acta Metallurgica}\ }\textbf {\bibinfo
  {volume} {6}},\ \bibinfo {pages} {704} (\bibinfo {year} {1958})}\BibitemShut
  {NoStop}%
\bibitem [{\citenamefont {Elder}\ \emph {et~al.}(1989)\citenamefont {Elder},
  \citenamefont {Munitz},\ and\ \citenamefont {Abbaschian}}]{Elder1989}%
  \BibitemOpen
  \bibfield  {author} {\bibinfo {author} {\bibfnamefont {S.~P.}\ \bibnamefont
  {Elder}}, \bibinfo {author} {\bibfnamefont {A.}~\bibnamefont {Munitz}},\ and\
  \bibinfo {author} {\bibfnamefont {G.~J.}\ \bibnamefont {Abbaschian}},\
  }\bibfield  {title} {\bibinfo {title} {Metastable liquid immiscibility in
  {Fe-Cu} and {Co-Cu} alloys},\ }in\ \href
  {https://doi.org/10.4028/www.scientific.net/MSF.50.137} {\emph {\bibinfo
  {booktitle} {Materials Processing in Space}}},\ \bibinfo {series} {Materials
  Science Forum}, Vol.~\bibinfo {volume} {50}\ (\bibinfo  {publisher} {Trans
  Tech Publications Ltd},\ \bibinfo {year} {1989})\ pp.\ \bibinfo {pages}
  {137--150}\BibitemShut {NoStop}%
\bibitem [{\citenamefont {Chen}\ and\ \citenamefont {Jin}(1995)}]{Chen1995}%
  \BibitemOpen
  \bibfield  {author} {\bibinfo {author} {\bibfnamefont {Q.}~\bibnamefont
  {Chen}}\ and\ \bibinfo {author} {\bibfnamefont {Z.}~\bibnamefont {Jin}},\
  }\bibfield  {title} {\bibinfo {title} {The {Fe-Cu} system: A thermodynamic
  evaluation},\ }\href {https://doi.org/10.1007/BF02664678} {\bibfield
  {journal} {\bibinfo  {journal} {Metall. Mater. Trans. A}\ }\textbf {\bibinfo
  {volume} {26}},\ \bibinfo {pages} {417} (\bibinfo {year} {1995})}\BibitemShut
  {NoStop}%
\bibitem [{\citenamefont {Norman}\ and\ \citenamefont
  {Filinov}(1969)}]{Norman1969}%
  \BibitemOpen
  \bibfield  {author} {\bibinfo {author} {\bibfnamefont {G.~E.}\ \bibnamefont
  {Norman}}\ and\ \bibinfo {author} {\bibfnamefont {V.~S.}\ \bibnamefont
  {Filinov}},\ }\bibfield  {title} {\bibinfo {title} {Investigations of phase
  transitions by a {Monte-Carlo} method},\ }\href@noop {} {\bibfield  {journal}
  {\bibinfo  {journal} {High Temperature}\ }\textbf {\bibinfo {volume} {7}},\
  \bibinfo {pages} {216} (\bibinfo {year} {1969})}\BibitemShut {NoStop}%
\bibitem [{\citenamefont {Frenkel}\ and\ \citenamefont
  {Smit}(2002)}]{Frenkel2002}%
  \BibitemOpen
  \bibfield  {author} {\bibinfo {author} {\bibfnamefont {D.}~\bibnamefont
  {Frenkel}}\ and\ \bibinfo {author} {\bibfnamefont {B.}~\bibnamefont {Smit}},\
  }\href@noop {} {\emph {\bibinfo {title} {Understanding Molecular Simulation:
  From Algorithms to Applications}}}\ (\bibinfo  {publisher} {Elsevier},\
  \bibinfo {year} {2002})\BibitemShut {NoStop}%
\bibitem [{\citenamefont {Panagiotopoulos}(1987)}]{Panagiotopoulos1987}%
  \BibitemOpen
  \bibfield  {author} {\bibinfo {author} {\bibfnamefont {A.~Z.}\ \bibnamefont
  {Panagiotopoulos}},\ }\bibfield  {title} {\bibinfo {title} {Direct
  determination of phase coexistence properties of fluids by {Monte Carlo}
  simulation in a new ensemble},\ }\href@noop {} {\bibfield  {journal}
  {\bibinfo  {journal} {Molecular Physics}\ }\textbf {\bibinfo {volume} {61}},\
  \bibinfo {pages} {813} (\bibinfo {year} {1987})}\BibitemShut {NoStop}%
\bibitem [{\citenamefont {Mehta}\ and\ \citenamefont
  {Kofke}(1994)}]{Mehta1994}%
  \BibitemOpen
  \bibfield  {author} {\bibinfo {author} {\bibfnamefont {M.}~\bibnamefont
  {Mehta}}\ and\ \bibinfo {author} {\bibfnamefont {D.~A.}\ \bibnamefont
  {Kofke}},\ }\bibfield  {title} {\bibinfo {title} {Coexistence diagrams of
  mixtures by molecular simulation},\ }\href@noop {} {\bibfield  {journal}
  {\bibinfo  {journal} {Chemical engineering science}\ }\textbf {\bibinfo
  {volume} {49}},\ \bibinfo {pages} {2633} (\bibinfo {year}
  {1994})}\BibitemShut {NoStop}%
\bibitem [{\citenamefont {Hitchcock}\ and\ \citenamefont
  {Hall}(1999)}]{Hitchcock1999}%
  \BibitemOpen
  \bibfield  {author} {\bibinfo {author} {\bibfnamefont {M.~R.}\ \bibnamefont
  {Hitchcock}}\ and\ \bibinfo {author} {\bibfnamefont {C.~K.}\ \bibnamefont
  {Hall}},\ }\bibfield  {title} {\bibinfo {title} {Solid--liquid phase
  equilibrium for binary {Lennard-Jones} mixtures},\ }\href@noop {} {\bibfield
  {journal} {\bibinfo  {journal} {J. Chem. Phys.}\ }\textbf {\bibinfo {volume}
  {110}},\ \bibinfo {pages} {11433} (\bibinfo {year} {1999})}\BibitemShut
  {NoStop}%
\bibitem [{\citenamefont {Kofke}\ and\ \citenamefont
  {Glandt}(1988)}]{Kofke1988}%
  \BibitemOpen
  \bibfield  {author} {\bibinfo {author} {\bibfnamefont {D.~A.}\ \bibnamefont
  {Kofke}}\ and\ \bibinfo {author} {\bibfnamefont {E.~D.}\ \bibnamefont
  {Glandt}},\ }\bibfield  {title} {\bibinfo {title} {{Monte Carlo} simulation
  of multicomponent equilibria in a semigrand canonical ensemble},\ }\href@noop
  {} {\bibfield  {journal} {\bibinfo  {journal} {Molecular Physics}\ }\textbf
  {\bibinfo {volume} {64}},\ \bibinfo {pages} {1105} (\bibinfo {year}
  {1988})}\BibitemShut {NoStop}%
\bibitem [{\citenamefont {Ghazisaeidi}(2021)}]{Ghazisaeidi2021}%
  \BibitemOpen
  \bibfield  {author} {\bibinfo {author} {\bibfnamefont {M.}~\bibnamefont
  {Ghazisaeidi}},\ }\bibfield  {title} {\bibinfo {title} {Alloy thermodynamics
  via the {Multi-cell} {Monte} {Carlo} ({MC})$^2$ method},\ }\href
  {https://doi.org/https://doi.org/10.1016/j.commatsci.2021.110322} {\bibfield
  {journal} {\bibinfo  {journal} {Comp. Mater. Sci.}\ }\textbf {\bibinfo
  {volume} {193}},\ \bibinfo {pages} {110322} (\bibinfo {year}
  {2021})}\BibitemShut {NoStop}%
\bibitem [{\citenamefont {Hong}\ and\ \citenamefont {{van de
  Walle}}(2012)}]{Hong2012}%
  \BibitemOpen
  \bibfield  {author} {\bibinfo {author} {\bibfnamefont {Q.-J.}\ \bibnamefont
  {Hong}}\ and\ \bibinfo {author} {\bibfnamefont {A.}~\bibnamefont {{van de
  Walle}}},\ }\bibfield  {title} {\bibinfo {title} {Direct first-principles
  chemical potential calculations of liquids},\ }\href
  {https://doi.org/10.1063/1.4749287} {\bibfield  {journal} {\bibinfo
  {journal} {J. Chem. Phys.}\ }\textbf {\bibinfo {volume} {137}},\ \bibinfo
  {pages} {094114} (\bibinfo {year} {2012})}\BibitemShut {NoStop}%
\bibitem [{\citenamefont {Qin}\ and\ \citenamefont {Zhou}(2016)}]{Qin2016}%
  \BibitemOpen
  \bibfield  {author} {\bibinfo {author} {\bibfnamefont {S.}~\bibnamefont
  {Qin}}\ and\ \bibinfo {author} {\bibfnamefont {H.-X.}\ \bibnamefont {Zhou}},\
  }\bibfield  {title} {\bibinfo {title} {Fast method for computing chemical
  potentials and liquid–liquid phase equilibria of macromolecular
  solutions},\ }\href {https://doi.org/10.1021/acs.jpcb.6b01607} {\bibfield
  {journal} {\bibinfo  {journal} {J. Phys. Chem. B}\ }\textbf {\bibinfo
  {volume} {120}},\ \bibinfo {pages} {8164} (\bibinfo {year}
  {2016})}\BibitemShut {NoStop}%
\bibitem [{\citenamefont {Perego}\ \emph {et~al.}(2018)\citenamefont {Perego},
  \citenamefont {Valsson},\ and\ \citenamefont {Parrinello}}]{Perego2018}%
  \BibitemOpen
  \bibfield  {author} {\bibinfo {author} {\bibfnamefont {C.}~\bibnamefont
  {Perego}}, \bibinfo {author} {\bibfnamefont {O.}~\bibnamefont {Valsson}},\
  and\ \bibinfo {author} {\bibfnamefont {M.}~\bibnamefont {Parrinello}},\
  }\bibfield  {title} {\bibinfo {title} {Chemical potential calculations in
  non-homogeneous liquids},\ }\href {https://doi.org/10.1063/1.5024631}
  {\bibfield  {journal} {\bibinfo  {journal} {J. Chem. Phys.}\ }\textbf
  {\bibinfo {volume} {149}},\ \bibinfo {pages} {072305} (\bibinfo {year}
  {2018})}\BibitemShut {NoStop}%
\bibitem [{\citenamefont {Freitas}\ \emph {et~al.}(2016)\citenamefont
  {Freitas}, \citenamefont {Asta},\ and\ \citenamefont {{de
  Koning}}}]{Freitas2016}%
  \BibitemOpen
  \bibfield  {author} {\bibinfo {author} {\bibfnamefont {R.}~\bibnamefont
  {Freitas}}, \bibinfo {author} {\bibfnamefont {M.}~\bibnamefont {Asta}},\ and\
  \bibinfo {author} {\bibfnamefont {M.}~\bibnamefont {{de Koning}}},\
  }\bibfield  {title} {\bibinfo {title} {Nonequilibrium free-energy calculation
  of solids using {LAMMPS}},\ }\href
  {https://doi.org/https://doi.org/10.1016/j.commatsci.2015.10.050} {\bibfield
  {journal} {\bibinfo  {journal} {Comp. Mater. Sci.}\ }\textbf {\bibinfo
  {volume} {112}},\ \bibinfo {pages} {333} (\bibinfo {year}
  {2016})}\BibitemShut {NoStop}%
\bibitem [{\citenamefont {{Paula Leite}}\ and\ \citenamefont {{de
  Koning}}(2019)}]{PaulaLeite2019}%
  \BibitemOpen
  \bibfield  {author} {\bibinfo {author} {\bibfnamefont {R.}~\bibnamefont
  {{Paula Leite}}}\ and\ \bibinfo {author} {\bibfnamefont {M.}~\bibnamefont
  {{de Koning}}},\ }\bibfield  {title} {\bibinfo {title} {Nonequilibrium
  free-energy calculations of fluids using {LAMMPS}},\ }\href
  {https://doi.org/https://doi.org/10.1016/j.commatsci.2018.12.029} {\bibfield
  {journal} {\bibinfo  {journal} {Comp. Mater. Sci.}\ }\textbf {\bibinfo
  {volume} {159}},\ \bibinfo {pages} {316} (\bibinfo {year}
  {2019})}\BibitemShut {NoStop}%
\bibitem [{\citenamefont {Sindzingre}\ \emph {et~al.}(1987)\citenamefont
  {Sindzingre}, \citenamefont {Ciccotti}, \citenamefont {Massobrio},\ and\
  \citenamefont {Frenkel}}]{Sindzingre1987}%
  \BibitemOpen
  \bibfield  {author} {\bibinfo {author} {\bibfnamefont {P.}~\bibnamefont
  {Sindzingre}}, \bibinfo {author} {\bibfnamefont {G.}~\bibnamefont
  {Ciccotti}}, \bibinfo {author} {\bibfnamefont {C.}~\bibnamefont
  {Massobrio}},\ and\ \bibinfo {author} {\bibfnamefont {D.}~\bibnamefont
  {Frenkel}},\ }\bibfield  {title} {\bibinfo {title} {Partial enthalpies and
  related quantities in mixtures from computer simulation},\ }\href
  {https://doi.org/https://doi.org/10.1016/0009-2614(87)87294-9} {\bibfield
  {journal} {\bibinfo  {journal} {Chem. Phys. Lett.}\ }\textbf {\bibinfo
  {volume} {136}},\ \bibinfo {pages} {35} (\bibinfo {year} {1987})}\BibitemShut
  {NoStop}%
\bibitem [{\citenamefont {Anwar}\ \emph {et~al.}(2020)\citenamefont {Anwar},
  \citenamefont {Leitold},\ and\ \citenamefont {Peters}}]{Peters2020}%
  \BibitemOpen
  \bibfield  {author} {\bibinfo {author} {\bibfnamefont {J.}~\bibnamefont
  {Anwar}}, \bibinfo {author} {\bibfnamefont {C.}~\bibnamefont {Leitold}},\
  and\ \bibinfo {author} {\bibfnamefont {B.}~\bibnamefont {Peters}},\
  }\bibfield  {title} {\bibinfo {title} {Solid–solid phase equilibria in the
  {NaCl}–{KCl} system},\ }\href {https://doi.org/10.1063/5.0003224}
  {\bibfield  {journal} {\bibinfo  {journal} {J. Chem. Phys.}\ }\textbf
  {\bibinfo {volume} {152}},\ \bibinfo {pages} {144109} (\bibinfo {year}
  {2020})}\BibitemShut {NoStop}%
\bibitem [{\citenamefont {Cheng}(2022)}]{Cheng2022}%
  \BibitemOpen
  \bibfield  {author} {\bibinfo {author} {\bibfnamefont {B.}~\bibnamefont
  {Cheng}},\ }\bibfield  {title} {\bibinfo {title} {Computing chemical
  potentials of solutions from structure factors},\ }\href
  {https://doi.org/10.1063/5.0107059} {\bibfield  {journal} {\bibinfo
  {journal} {The Journal of Chemical Physics}\ }\textbf {\bibinfo {volume}
  {157}},\ \bibinfo {pages} {121101} (\bibinfo {year} {2022})}\BibitemShut
  {NoStop}%
\bibitem [{\citenamefont {Bonny}\ \emph {et~al.}(2009)\citenamefont {Bonny},
  \citenamefont {Pasianot}, \citenamefont {Castin},\ and\ \citenamefont
  {Malerba}}]{Bonny2009}%
  \BibitemOpen
  \bibfield  {author} {\bibinfo {author} {\bibfnamefont {G.}~\bibnamefont
  {Bonny}}, \bibinfo {author} {\bibfnamefont {R.}~\bibnamefont {Pasianot}},
  \bibinfo {author} {\bibfnamefont {N.}~\bibnamefont {Castin}},\ and\ \bibinfo
  {author} {\bibfnamefont {L.}~\bibnamefont {Malerba}},\ }\bibfield  {title}
  {\bibinfo {title} {Ternary {Fe–Cu–Ni} many-body potential to model
  reactor pressure vessel steels: First validation by simulated thermal
  annealing},\ }\href {https://doi.org/10.1080/14786430903299824} {\bibfield
  {journal} {\bibinfo  {journal} {Philos. Mag.}\ }\textbf {\bibinfo {volume}
  {89}},\ \bibinfo {pages} {3531} (\bibinfo {year} {2009})}\BibitemShut
  {NoStop}%
\bibitem [{\citenamefont {Thompson}\ \emph {et~al.}(2022)\citenamefont
  {Thompson}, \citenamefont {Aktulga}, \citenamefont {Berger}, \citenamefont
  {Bolintineanu}, \citenamefont {Brown}, \citenamefont {Crozier}, \citenamefont
  {in~'t Veld}, \citenamefont {Kohlmeyer}, \citenamefont {Moore}, \citenamefont
  {Nguyen}, \citenamefont {Shan}, \citenamefont {Stevens}, \citenamefont
  {Tranchida}, \citenamefont {Trott},\ and\ \citenamefont {Plimpton}}]{LAMMPS}%
  \BibitemOpen
  \bibfield  {author} {\bibinfo {author} {\bibfnamefont {A.~P.}\ \bibnamefont
  {Thompson}}, \bibinfo {author} {\bibfnamefont {H.~M.}\ \bibnamefont
  {Aktulga}}, \bibinfo {author} {\bibfnamefont {R.}~\bibnamefont {Berger}},
  \bibinfo {author} {\bibfnamefont {D.~S.}\ \bibnamefont {Bolintineanu}},
  \bibinfo {author} {\bibfnamefont {W.~M.}\ \bibnamefont {Brown}}, \bibinfo
  {author} {\bibfnamefont {P.~S.}\ \bibnamefont {Crozier}}, \bibinfo {author}
  {\bibfnamefont {P.~J.}\ \bibnamefont {in~'t Veld}}, \bibinfo {author}
  {\bibfnamefont {A.}~\bibnamefont {Kohlmeyer}}, \bibinfo {author}
  {\bibfnamefont {S.~G.}\ \bibnamefont {Moore}}, \bibinfo {author}
  {\bibfnamefont {T.~D.}\ \bibnamefont {Nguyen}}, \bibinfo {author}
  {\bibfnamefont {R.}~\bibnamefont {Shan}}, \bibinfo {author} {\bibfnamefont
  {M.~J.}\ \bibnamefont {Stevens}}, \bibinfo {author} {\bibfnamefont
  {J.}~\bibnamefont {Tranchida}}, \bibinfo {author} {\bibfnamefont
  {C.}~\bibnamefont {Trott}},\ and\ \bibinfo {author} {\bibfnamefont {S.~J.}\
  \bibnamefont {Plimpton}},\ }\bibfield  {title} {\bibinfo {title} {{LAMMPS} -
  a flexible simulation tool for particle-based materials modeling at the
  atomic, meso, and continuum scales},\ }\href
  {https://doi.org/10.1016/j.cpc.2021.108171} {\bibfield  {journal} {\bibinfo
  {journal} {Comp. Phys. Comm.}\ }\textbf {\bibinfo {volume} {271}},\ \bibinfo
  {pages} {108171} (\bibinfo {year} {2022})}\BibitemShut {NoStop}%
\bibitem [{\citenamefont {Shinoda}\ \emph {et~al.}(2004)\citenamefont
  {Shinoda}, \citenamefont {Shiga},\ and\ \citenamefont
  {Mikami}}]{Shinoda2004}%
  \BibitemOpen
  \bibfield  {author} {\bibinfo {author} {\bibfnamefont {W.}~\bibnamefont
  {Shinoda}}, \bibinfo {author} {\bibfnamefont {M.}~\bibnamefont {Shiga}},\
  and\ \bibinfo {author} {\bibfnamefont {M.}~\bibnamefont {Mikami}},\
  }\bibfield  {title} {\bibinfo {title} {Rapid estimation of elastic constants
  by molecular dynamics simulation under constant stress},\ }\href
  {https://doi.org/10.1103/PhysRevB.69.134103} {\bibfield  {journal} {\bibinfo
  {journal} {Phys. Rev. B}\ }\textbf {\bibinfo {volume} {69}},\ \bibinfo
  {pages} {134103} (\bibinfo {year} {2004})}\BibitemShut {NoStop}%
\bibitem [{\citenamefont {Martyna}\ \emph {et~al.}(1994)\citenamefont
  {Martyna}, \citenamefont {Tobias},\ and\ \citenamefont
  {Klein}}]{Martyna1994}%
  \BibitemOpen
  \bibfield  {author} {\bibinfo {author} {\bibfnamefont {G.~J.}\ \bibnamefont
  {Martyna}}, \bibinfo {author} {\bibfnamefont {D.~J.}\ \bibnamefont
  {Tobias}},\ and\ \bibinfo {author} {\bibfnamefont {M.~L.}\ \bibnamefont
  {Klein}},\ }\bibfield  {title} {\bibinfo {title} {Constant pressure molecular
  dynamics algorithms},\ }\href {https://doi.org/10.1063/1.467468} {\bibfield
  {journal} {\bibinfo  {journal} {The Journal of Chemical Physics}\ }\textbf
  {\bibinfo {volume} {101}},\ \bibinfo {pages} {4177} (\bibinfo {year}
  {1994})}\BibitemShut {NoStop}%
\bibitem [{\citenamefont {Parrinello}\ and\ \citenamefont
  {Rahman}(1981)}]{Parrinello1981}%
  \BibitemOpen
  \bibfield  {author} {\bibinfo {author} {\bibfnamefont {M.}~\bibnamefont
  {Parrinello}}\ and\ \bibinfo {author} {\bibfnamefont {A.}~\bibnamefont
  {Rahman}},\ }\bibfield  {title} {\bibinfo {title} {Polymorphic transitions in
  single crystals: A new molecular dynamics method},\ }\href
  {https://doi.org/10.1063/1.328693} {\bibfield  {journal} {\bibinfo  {journal}
  {Journal of Applied Physics}\ }\textbf {\bibinfo {volume} {52}},\ \bibinfo
  {pages} {7182} (\bibinfo {year} {1981})}\BibitemShut {NoStop}%
\bibitem [{\citenamefont {Tuckerman}\ \emph {et~al.}(2006)\citenamefont
  {Tuckerman}, \citenamefont {Alejandre}, \citenamefont {L{\'o}pez-Rend{\'o}n},
  \citenamefont {Jochim},\ and\ \citenamefont {Martyna}}]{Tuckerman2006}%
  \BibitemOpen
  \bibfield  {author} {\bibinfo {author} {\bibfnamefont {M.~E.}\ \bibnamefont
  {Tuckerman}}, \bibinfo {author} {\bibfnamefont {J.}~\bibnamefont
  {Alejandre}}, \bibinfo {author} {\bibfnamefont {R.}~\bibnamefont
  {L{\'o}pez-Rend{\'o}n}}, \bibinfo {author} {\bibfnamefont {A.~L.}\
  \bibnamefont {Jochim}},\ and\ \bibinfo {author} {\bibfnamefont {G.~J.}\
  \bibnamefont {Martyna}},\ }\bibfield  {title} {\bibinfo {title} {A
  {Liouville}-operator derived measure-preserving integrator for molecular
  dynamics simulations in the isothermal–isobaric ensemble},\ }\href
  {https://doi.org/10.1088/0305-4470/39/19/S18} {\bibfield  {journal} {\bibinfo
   {journal} {Journal of Physics A: Mathematical and General}\ }\textbf
  {\bibinfo {volume} {39}},\ \bibinfo {pages} {5629} (\bibinfo {year}
  {2006})}\BibitemShut {NoStop}%
\bibitem [{\citenamefont {Lupis}(1983)}]{Lupis1983}%
  \BibitemOpen
  \bibfield  {author} {\bibinfo {author} {\bibfnamefont {C.~H.~P.}\
  \bibnamefont {Lupis}},\ }\href@noop {} {\emph {\bibinfo {title} {Chemical
  Thermodynamics of Materials}}}\ (\bibinfo  {publisher} {Elsevier},\ \bibinfo
  {address} {New York},\ \bibinfo {year} {1983})\BibitemShut {NoStop}%
\bibitem [{\citenamefont {Sadigh}\ \emph {et~al.}(2012)\citenamefont {Sadigh},
  \citenamefont {Erhart}, \citenamefont {Stukowski}, \citenamefont {Caro},
  \citenamefont {Martinez},\ and\ \citenamefont {Zepeda-Ruiz}}]{Sadigh2012}%
  \BibitemOpen
  \bibfield  {author} {\bibinfo {author} {\bibfnamefont {B.}~\bibnamefont
  {Sadigh}}, \bibinfo {author} {\bibfnamefont {P.}~\bibnamefont {Erhart}},
  \bibinfo {author} {\bibfnamefont {A.}~\bibnamefont {Stukowski}}, \bibinfo
  {author} {\bibfnamefont {A.}~\bibnamefont {Caro}}, \bibinfo {author}
  {\bibfnamefont {E.}~\bibnamefont {Martinez}},\ and\ \bibinfo {author}
  {\bibfnamefont {L.}~\bibnamefont {Zepeda-Ruiz}},\ }\bibfield  {title}
  {\bibinfo {title} {Scalable parallel monte carlo algorithm for atomistic
  simulations of precipitation in alloys},\ }\href
  {https://doi.org/10.1103/PhysRevB.85.184203} {\bibfield  {journal} {\bibinfo
  {journal} {Phys. Rev. B}\ }\textbf {\bibinfo {volume} {85}},\ \bibinfo
  {pages} {184203} (\bibinfo {year} {2012})}\BibitemShut {NoStop}%
\bibitem [{\citenamefont {Sundman}\ \emph {et~al.}(2015)\citenamefont
  {Sundman}, \citenamefont {Kattner}, \citenamefont {Palumbo},\ and\
  \citenamefont {Fries}}]{Sundman2015}%
  \BibitemOpen
  \bibfield  {author} {\bibinfo {author} {\bibfnamefont {B.}~\bibnamefont
  {Sundman}}, \bibinfo {author} {\bibfnamefont {U.~R.}\ \bibnamefont
  {Kattner}}, \bibinfo {author} {\bibfnamefont {M.}~\bibnamefont {Palumbo}},\
  and\ \bibinfo {author} {\bibfnamefont {S.~G.}\ \bibnamefont {Fries}},\
  }\bibfield  {title} {\bibinfo {title} {Opencalphad - a free thermodynamic
  software},\ }\bibfield  {journal} {\bibinfo  {journal} {Integ. Mat. Manu.
  Innov.}\ }\textbf {\bibinfo {volume} {4}},\ \href
  {https://doi.org/10.1186/s40192-014-0029-1} {10.1186/s40192-014-0029-1}
  (\bibinfo {year} {2015})\BibitemShut {NoStop}%
\bibitem [{\citenamefont {Kamiya}\ \emph {et~al.}(2021)\citenamefont {Kamiya},
  \citenamefont {Terasaki},\ and\ \citenamefont {Kondo}}]{Kamiya2021}%
  \BibitemOpen
  \bibfield  {author} {\bibinfo {author} {\bibfnamefont {A.}~\bibnamefont
  {Kamiya}}, \bibinfo {author} {\bibfnamefont {H.}~\bibnamefont {Terasaki}},\
  and\ \bibinfo {author} {\bibfnamefont {T.}~\bibnamefont {Kondo}},\ }\bibfield
   {title} {\bibinfo {title} {Precise determination of the effect of
  temperature on the density of solid and liquid iron, nickel, and tin},\
  }\href {https://doi.org/10.2138/am-2021-7509} {\bibfield  {journal} {\bibinfo
   {journal} {American Mineralogist}\ }\textbf {\bibinfo {volume} {106}},\
  \bibinfo {pages} {1077} (\bibinfo {year} {2021})}\BibitemShut {NoStop}%
\bibitem [{\citenamefont {Cahill}\ and\ \citenamefont
  {A.~D.~Kirshenbaum}(1962)}]{Cahill1962}%
  \BibitemOpen
  \bibfield  {author} {\bibinfo {author} {\bibfnamefont {J.~A.}\ \bibnamefont
  {Cahill}}\ and\ \bibinfo {author} {\bibfnamefont {A.~D.}\ \bibnamefont
  {A.~D.~Kirshenbaum}},\ }\bibfield  {title} {\bibinfo {title} {The density of
  liquid copper from its melting point (1356{$^\circ$K}.) to 2500{$^\circ$K}.
  and an estimate of its critical constants},\ }\href
  {https://doi.org/10.1021/j100812a027} {\bibfield  {journal} {\bibinfo
  {journal} {J. Phys. Chem.}\ }\textbf {\bibinfo {volume} {66}},\ \bibinfo
  {pages} {1080} (\bibinfo {year} {1962})}\BibitemShut {NoStop}%
\bibitem [{\citenamefont {Schmon}\ \emph {et~al.}(2015)\citenamefont {Schmon},
  \citenamefont {Aziz},\ and\ \citenamefont {Pottlacher}}]{Schmon2015}%
  \BibitemOpen
  \bibfield  {author} {\bibinfo {author} {\bibfnamefont {A.}~\bibnamefont
  {Schmon}}, \bibinfo {author} {\bibfnamefont {K.}~\bibnamefont {Aziz}},\ and\
  \bibinfo {author} {\bibfnamefont {G.}~\bibnamefont {Pottlacher}},\ }\bibfield
   {title} {\bibinfo {title} {Density determination of liquid copper and liquid
  nickel by means of fast resistive pulse heating and electromagnetic
  levitation},\ }\href {https://doi.org/10.1007/s11661-015-2844-1} {\bibfield
  {journal} {\bibinfo  {journal} {Met. Mater. Trans. A}\ }\textbf {\bibinfo
  {volume} {46}},\ \bibinfo {pages} {2674} (\bibinfo {year}
  {2015})}\BibitemShut {NoStop}%
\bibitem [{\citenamefont {Blaiszik}\ \emph {et~al.}(2016)\citenamefont
  {Blaiszik}, \citenamefont {Chard}, \citenamefont {Pruyne}, \citenamefont
  {Ananthakrishnan}, \citenamefont {Tuecke},\ and\ \citenamefont
  {Foster}}]{Blaiszik2016}%
  \BibitemOpen
  \bibfield  {author} {\bibinfo {author} {\bibfnamefont {B.}~\bibnamefont
  {Blaiszik}}, \bibinfo {author} {\bibfnamefont {K.}~\bibnamefont {Chard}},
  \bibinfo {author} {\bibfnamefont {J.}~\bibnamefont {Pruyne}}, \bibinfo
  {author} {\bibfnamefont {R.}~\bibnamefont {Ananthakrishnan}}, \bibinfo
  {author} {\bibfnamefont {S.}~\bibnamefont {Tuecke}},\ and\ \bibinfo {author}
  {\bibfnamefont {I.}~\bibnamefont {Foster}},\ }\bibfield  {title} {\bibinfo
  {title} {The {Materials Data Facility}: Data services to advance materials
  science research},\ }\href {https://doi.org/10.1007/s11837-016-2001-3}
  {\bibfield  {journal} {\bibinfo  {journal} {JOM}\ }\textbf {\bibinfo {volume}
  {68}},\ \bibinfo {pages} {2045} (\bibinfo {year} {2016})}\BibitemShut
  {NoStop}%
\bibitem [{\citenamefont {Blaiszik}\ \emph {et~al.}(2019)\citenamefont
  {Blaiszik}, \citenamefont {Ward}, \citenamefont {Schwarting}, \citenamefont
  {Gaff}, \citenamefont {Chard}, \citenamefont {Pike}, \citenamefont {Chard},\
  and\ \citenamefont {Foster}}]{Blaiszik2019}%
  \BibitemOpen
  \bibfield  {author} {\bibinfo {author} {\bibfnamefont {B.}~\bibnamefont
  {Blaiszik}}, \bibinfo {author} {\bibfnamefont {L.}~\bibnamefont {Ward}},
  \bibinfo {author} {\bibfnamefont {M.}~\bibnamefont {Schwarting}}, \bibinfo
  {author} {\bibfnamefont {J.}~\bibnamefont {Gaff}}, \bibinfo {author}
  {\bibfnamefont {R.}~\bibnamefont {Chard}}, \bibinfo {author} {\bibfnamefont
  {D.}~\bibnamefont {Pike}}, \bibinfo {author} {\bibfnamefont {K.}~\bibnamefont
  {Chard}},\ and\ \bibinfo {author} {\bibfnamefont {I.}~\bibnamefont
  {Foster}},\ }\bibfield  {title} {\bibinfo {title} {A data ecosystem to
  support machine learning in materials science},\ }\href
  {https://doi.org/10.1557/mrc.2019.118} {\bibfield  {journal} {\bibinfo
  {journal} {MRS Communications}\ ,\ \bibinfo {pages} {1}} (\bibinfo {year}
  {2019})}\BibitemShut {NoStop}%
\bibitem [{\citenamefont {Trinkle}(2025)}]{FeCuNi2025Data}%
  \BibitemOpen
  \bibfield  {author} {\bibinfo {author} {\bibfnamefont {D.~R.}\ \bibnamefont
  {Trinkle}},\ }\href {https://doi.org/10.18126/p0gx-tt30} {\bibinfo {title}
  {Data citation: {Fe-Cu-Ni} liquid ternary calculations with {EAM}}} (\bibinfo
  {year} {2025})\BibitemShut {NoStop}%
\bibitem [{\citenamefont {{Paula Leite}}\ \emph {et~al.}(2016)\citenamefont
  {{Paula Leite}}, \citenamefont {Freitas}, \citenamefont {Azevedo},\ and\
  \citenamefont {{de Koning}}}]{PaulaLeite2016}%
  \BibitemOpen
  \bibfield  {author} {\bibinfo {author} {\bibfnamefont {R.}~\bibnamefont
  {{Paula Leite}}}, \bibinfo {author} {\bibfnamefont {R.}~\bibnamefont
  {Freitas}}, \bibinfo {author} {\bibfnamefont {R.}~\bibnamefont {Azevedo}},\
  and\ \bibinfo {author} {\bibfnamefont {M.}~\bibnamefont {{de Koning}}},\
  }\bibfield  {title} {\bibinfo {title} {The {Uhlenbeck-Ford} model: Exact
  virial coefficients and application as a reference system in fluid-phase
  free-energy calculations},\ }\href {https://doi.org/10.1063/1.4967775}
  {\bibfield  {journal} {\bibinfo  {journal} {J. Chem. Phys.}\ }\textbf
  {\bibinfo {volume} {145}},\ \bibinfo {pages} {194101} (\bibinfo {year}
  {2016})}\BibitemShut {NoStop}%
\bibitem [{\citenamefont {Schneider}\ and\ \citenamefont
  {Stoll}(1978)}]{Schneider1978}%
  \BibitemOpen
  \bibfield  {author} {\bibinfo {author} {\bibfnamefont {T.}~\bibnamefont
  {Schneider}}\ and\ \bibinfo {author} {\bibfnamefont {E.}~\bibnamefont
  {Stoll}},\ }\bibfield  {title} {\bibinfo {title} {Molecular-dynamics study of
  a three-dimensional one-component model for distortive phase transitions},\
  }\href {https://doi.org/10.1103/PhysRevB.17.1302} {\bibfield  {journal}
  {\bibinfo  {journal} {Phys. Rev. B}\ }\textbf {\bibinfo {volume} {17}},\
  \bibinfo {pages} {1302} (\bibinfo {year} {1978})}\BibitemShut {NoStop}%
\bibitem [{\citenamefont {{de Koning}}\ and\ \citenamefont
  {Antonelli}(1996)}]{deKoning1996}%
  \BibitemOpen
  \bibfield  {author} {\bibinfo {author} {\bibfnamefont {M.}~\bibnamefont {{de
  Koning}}}\ and\ \bibinfo {author} {\bibfnamefont {A.}~\bibnamefont
  {Antonelli}},\ }\bibfield  {title} {\bibinfo {title} {Einstein crystal as a
  reference system in free energy estimation using adiabatic switching},\
  }\href {https://doi.org/10.1103/PhysRevE.53.465} {\bibfield  {journal}
  {\bibinfo  {journal} {Phys. Rev. E}\ }\textbf {\bibinfo {volume} {53}},\
  \bibinfo {pages} {465} (\bibinfo {year} {1996})}\BibitemShut {NoStop}%
\end{thebibliography}
\end{document}